\documentstyle[eqsecnum,epsf,aps,preprint]{revtex}
\input epsf.sty
\begin{document}
\draft
\title{Non-Fermi Liquid Behavior in U and Ce Alloys: Criticality, Disorder, 
Dissipation, and Griffiths-McCoy singularities.}
\author{A.~H.~Castro Neto$^1$ and B.~A.~Jones$^2$}
\bigskip
\address
{$^1$ Department of Physics,
University of California,
Riverside, CA 92521 \\
$^2$ IBM Almaden Research Center, San Jose, CA 95120-6099}

\maketitle

\begin{abstract}
In this paper we provide the theoretical basis for the problem
of Griffiths-McCoy singularities close to the quantum critical
point for magnetic ordering in U and Ce intermetallics. We show
that the competition between Kondo effect and RKKY interaction
can be expressed in Hamiltonian form and the dilution effect due to alloying
leads to a quantum percolation problem driven by the number of
magnetically compensated moments. We argue that the exhaustion paradox proposed
by Nozi\`eres is explained when the RKKY interaction is taken
into account. We revisited the one impurity and two impurity Kondo
problem and showed that in the presence of particle-hole
symmetry breaking operators the system flows to a line of fixed
points characterized by coherent (cluster like) motion of the spins.
Moreover, close to the quantum critical point
clusters of magnetic atoms can quantum mechanically tunnel
between different states either
via the anisotropy of the RKKY interaction or by what we call
the {\it cluster Kondo effect}. We calculate explicitly
from the microscopic Hamiltonian the parameters which appear
in all the response functions.
We show that there is a maximum number $N_c$ of spins in
the clusters such that above this number tunneling ceases to occur.
These effects lead to a distribution of cluster Kondo temperatures which
vanishes for finite clusters and therefore leads to strong
magnetic response. From these results we propose a {\it dissipative
quantum droplet model} which describes the critical behavior of
metallic magnetic systems. This model predicts that
in the paramagnetic phase there is a crossover temperature
$T^*$ above which Griffiths-McCoy like singularities with
magnetic susceptibility, $\chi(T) \propto T^{-1+\lambda}$,
and specific heat $C_V(T) \propto T^{\lambda}$ with $\lambda < 1$ appear.
Below $T^*$, however, a new regime dominated by dissipation occurs
with divergences stronger than power law: $\chi(T)
\propto 1/(T \ln(1/T))$ and $C_V(T) \propto 1/\ln^2(1/T)$.
We estimate that $T^*$ is
exponentially small with $N_c$. Our results should be
applicable to a broad class of metallic magnetic systems
which are described by the Kondo lattice Hamiltonian.
\end{abstract}

\bigskip


\narrowtext

\section{Introduction}

In this paper we are interested in the quantum critical behavior
of alloys of actinides and rare earths with metallic atoms.
There is a large number of these alloys and they can be classified
into two main groups accordingly to their chemical composition:
1) {\it Kondo hole} systems which are made out of the substitution
of the rare earth or actinide (R) by a non-magnetic metallic atom (M) with a
chemical formula R$_{1-x}$M$_{x}$ (a typical example is
U$_{1-x}$Th$_x$Pd$_2$Al$_3$); 2){\it Ligand systems} where
one of the metallic atoms (M1) is replaced by another (M2) but the
rare earths or actinides are not touched and thus have the formula
R(M1)$_{1-x}$(M2)$_x$ (as, for instance, UCu$_{5-x}$Pd$_x$).
In most of these alloys one usually has an
ordered magnetic phase at $x=0$ which is destroyed at some
critical value $x=x^*$ with a non-Fermi liquid (NFL) phase in the vicinity 
where strong deviations from the predictions of Landau's theory of the Fermi
liquid are observed (see Fig.\ref{phdi}).
In this paper we propose that the origin for the NFL behavior is
the existence of Griffiths-McCoy singularities at low temperatures.
As we explain below, these singularities have their origin in a quantum
percolation problem driven by the number of magnetically compensated
moments. In this percolation problem magnetic clusters can tunnel
in the presence of a metallic environment which produces dissipation.

On the one hand, at $x=1$ a Kondo hole system becomes an ordinary Fermi liquid
with a temperature independent specific heat coefficient, $\gamma(T)=C_V/T$,
(by low temperature we mean $T \ll  E_F$ where
$E_F$ is the Fermi energy of the metal), the magnetic susceptibility is
paramagnetic and given by Pauli's expression, $\chi(T) = \chi_0$, and
the resistivity has the Fermi liquid form $R(T) = R_0 + A T^2$.
On the other hand, at large $x$ a ligand system is usually a {\it heavy fermion} \cite{greg},
that is, it can also be described by the Fermi liquid expressions but
with coefficients $\gamma$ and $\chi_0$ orders of magnitude
larger than in ordinary metals. The nature of this heavy fermion behavior
can be tracked down to the presence of the rare earths or
actinides in the alloy. In the NFL phase the specific heat coefficient
and the magnetic susceptibility show divergent behavior
as the temperature is lowered.

The root for the understanding of the problem lies on the fact
that in Landau liquids the
thermodynamic and response functions are always well behaved
functions of the temperature.
This is clearly inconsistent with the behavior in the paramagnetic phase
close to a quantum critical point (QCP). Exactly at the QCP divergences are
expected since the system is ordering magnetically, thus, QCP
can generate NFL behavior \cite{hertz,millis}
(albeit with fine tuning of the chemical
composition). It turns out, however, that
many times NFL behavior is observed away from the QCP and inside of
the paramagnetic phase. One possibility is
that the NFL behavior is due to single ion physics and therefore
not at all related to the QCP physics. Another possibility is
that there are still ``traces" of the QCP inside of the
paramagnetic phase. This is possible in the presence of disorder
which can ``pin" pieces
of the ordered phase even in the absence of long range order.
It is even conceivable that single ion physics and critical
behavior co-exist close to a QCP. There is no clear consensus
among researchers on the nature of this NFL behavior and the
scientific debate is intense.
The objective of this work is to shed light into this
controversy.

We organize the paper in the following fashion: Section II briefly reviews
the problem of NFL behavior in U and Ce intermetallics; in Section III
we write an effective Hamiltonian where Kondo effect and RKKY
interaction appear explicitly and we discuss the problem of magnetic
ordering and dilution within this Hamiltonian;
Section IV contains a detailed discussion of the one impurity and
two impurity Kondo problem and the extension of the problem to
$N$ spin clusters where we define a Kondo cluster effect;
in Section V we propose the dissipative quantum droplet model
which is the basis for the description of the problem of magnetic
ordering in metallic magnetic alloys and we show that at low temperatures
dissipation dominates the behavior of the system leading to universal
logarithmic divergences, and that at higher temperatures non-universal 
power law
behavior is expected; in Section VI we study the intermediate temperature
range where quantum Griffiths singularities are expected;
Section VII contains our
conclusions. We also have included a few appendices with detailed
calculations of the results used in the paper.

\section{NFL behavior in U and Ce intermetallics}

The basis for the study of metals in the last 50 years has been
Landau's theory of the interacting Fermi gas \cite{baym}.
As consequences of Landau's theory the thermodynamic and response functions
of the electron fluid are smooth functions of the temperature.
The magnetic susceptibility has the form (we use units such that
$\hbar = k_B =1$)
\begin{eqnarray}
\chi(T) = \frac{(g \mu_B)^2 N(0)}{4 (1+F^a_0)} \, ,
\label{xipauli}
\end{eqnarray}
where $g \approx 2$ is the Land\'e factor for the electron, $\mu_B$ is the
Bohr magneton, $F_0^a$ is an antisymmetric Landau parameter,
\begin{eqnarray}
N(0) = \frac{m^* k_F}{\pi^2} = \frac{3 \rho_e}{2 E_F} \, ,
\label{dsta}
\end{eqnarray}
is the density of states at the Fermi energy,
$m^*$ is the effective mass of the quasiparticles (the effective
mass is related to the bare mass, $m$, by a symmetric Landau parameter:
$m^*/m=1+F^s_1/3$),
\begin{eqnarray}
k_F = (3 \pi^2 \rho_e)^{1/3} \, ,
\label{kf}
\end{eqnarray}
is the Fermi momentum, $\rho_e$ is the number of electrons per unit of volume
and $E_F = k_F^2/(2 m^*)$.
Moreover, the electronic specific heat, $C_V$, is
given by the Fermi liquid expression
\begin{eqnarray}
\frac{C_V}{T}=\gamma(T) = \frac{\pi^2 N(0)}{3}  \, .
\label{cvfermi}
\end{eqnarray}
Naturally these expressions resemble the ones obtained for the free electron
gas where the bare mass $m$ is replaced by the effective mass $m^*$ and
Land\'e factor is renormalized by a factor of $1/\sqrt{1+F^a_0}$. Furthermore,
at low temperatures one expects the electronic resistivity to
behave like \cite{mahan}
\begin{eqnarray}
R(T) = R_0 + A T^2
\label{resis}
\end{eqnarray}
where $R_0$ is the resistivity due to impurities and
$A$ is a constant.
In a magnetic alloy where the magnetic moments are decoupled from
the conduction electrons and do not interact directly among each other
we expect an ordinary paramagnetic behavior for the susceptibility
in the low field limit ($T\gg g \mu_B H$ where $H$ is an applied magnetic field)
\begin{eqnarray}
\chi(T) = \frac{{\cal C}}{T}
\end{eqnarray}
where ${\cal C} = \rho_f S(S+1) (g \mu_B)^2/3$ is the Curie constant
($\rho_f$ is the number of magnetic atoms per unit of volume, $S$ is the
magnetic angular momentum of the atom). At low enough temperatures
($T \ll g \mu_B H$) the susceptibility vanishes.

The effect of disorder in ordinary Fermi liquids has been studied
in great detail \cite{bill} and it was found
that the results quoted above, especially the temperature
dependences of the physical quantities, do not change appreciably
(at least when disorder
is weak enough to be treated in perturbation theory). Therefore,
for a metallic system where Anderson localization \cite{andloc}
does not play a role we still expect Fermi liquid behavior to
be a robust feature.
Thus, it is indeed very surprising
that for such a broad class of U and Ce
alloys (which clearly show
three-dimensional behavior) deviations from Landau's theory
are so abundant.

In NFL systems it is usually observed that even in the
paramagnetic phase the specific heat coefficient and the magnetic
susceptibility
do not saturate as expected from the Landau scenario.
In all the cases studied so far the data for the susceptibility and specific heat have
been fitted to weak power laws or logarithmic functions \cite{staba}.
The resistivity of the systems discussed here can be
fitted with $R(T) = R_0 + A T^{\alpha}$ where $\alpha < 2$.
Neutron scattering experiments
in UCu$_{5-x}$Pd$_x$ \cite{ucupdn} show that the imaginary
part of the frequency dependent susceptibility, $\Im (\chi(\omega))$,
has power law behavior, that is, $\Im (\chi(\omega)) \propto \omega^{1-\lambda}$
with $\lambda \approx 0.7$, over a wide range of frequencies
(for a Fermi liquid one expects $\lambda=1$). Moreover, consistent
with this behavior the static magnetic susceptibility seems
to diverge with $T^{-1+\lambda}$ at low temperatures \cite{staba}.
What is interesting about UCu$_{5-x}$Pd$_x$ is that it has been
shown in recent EXAFS experiments that this compound has a
large amount of disorder \cite{ucupdexaf} consistent
with early NMR and $\mu$SR experiments \cite{ucupdnmr}.

From the magnetic point of view the U and Ce intermetallics
show a rich variety of magnetic ground states: antiferromagnetic
as in the case of UCu$_{5-x}$Pd$_x$ for $0 \leq x \leq 1$ \cite{ucupdn},
ferromagnetic
as in the case of U$_{1-x}$Th$_x$Cu$_2$Si$_2$ for $0.6 \leq x \leq 0.85$
\cite{uthcusi}
and spin glass as in the case of  U$_{1-x}$Y$_x$Pd$_3$ for
$0.6 \leq x \leq 0.8$ \cite{uypd}.
There are many indications that the magnetism in these alloys is
inhomogeneous with strong coupling to the lattice. This seems to be
the case of CeAl$_3$ \cite{ott} or CePd$_2$Al$_3$ \cite{mentink}
where $\mu$SR experiments have shown clear signs of microscopic
inhomogeneity. Moreover, because of the strong spin-orbit coupling
in these systems a myriad of magnetic states with different
ordering configurations are found \cite{miraflores}.

On the theoretical side \cite{miraflores} we can classify the
approaches to NFL
as of {\it single ion} character or {\it cooperative} behavior.
Single impurity approaches for the NFL problem are very
important because the Kondo effect \cite{kondo}
has been suggested as the source of
heavy fermion behavior \cite{greg} in undiluted ligand systems (say, CeAl$_3$)
\cite{klattice}.  Nozi\`eres and Blandin proposed the {\it multichannel
Kondo effect} \cite{over} as a possible source of NFL behavior.
D.~Cox proposed that such a multichannel effect of quadrupolar origin
should exist in these systems \cite{cox}.
Another source of NFL
behavior based on single impurity physics is the so-called {\it Kondo disorder}
approach which postulates an {\it a priori} broad distribution of
single ion Kondo temperatures \cite{ucupdnmr,miranda}.
The cooperative approaches, on the other hand, stress
the proximity to a critical point as a source of NFL behavior
and was pioneered by Hertz \cite{hertz,millis} and has been applied
to many of the systems we discuss here \cite{von,piers,2dcecuau,tsvelik,sgsub}. The problem of disorder close to the QCP has been discussed by
many authors over the years. Hertz studied a classical XY model and
showed that disorder is a relevant perturbation \cite{hertzxy} in dimensions
smaller than $4$. Bhatt and Fisher \cite{bhatt} and more recently
Sachdev \cite{grifsach} have argued that for Hubbard-like models
an instability to a inhomogeneous phase should exist close to the
QCP. A more recent work extending Hertz calculations to the case
of disorder has reached similar conclusions, namely, that close
to the QCP a new type of critical behavior, very different from
the critical behavior of the clean system, should exist \cite{vojta}.

It is clear from the experimental and theoretical point of view
that magnetic interactions and the Kondo effect should be relevant
for the physics of U and Ce intermetallics. Moreover, it has
been claimed for a long time that it is the competition
between Kondo effect and magnetic order that determines the
phase diagram of these systems. This idea was put forward by
Doniach \cite{doni} and we discuss it in more detail below.
In trying to reconciliate the Doniach argument with the existence
of disorder in these systems we proposed recently a new explanation
for the NFL behavior in these systems which is based on the magnetic
inhomogeneity due to this competition. In other words, if the
Kondo effect is responsible for the destruction of long range order
in the magnetically ordered system then at the QCP
the competition between magnetic ordering and magnetic quenching
is strongest \cite{grifeu}.
We have argued that close to the QCP, in the paramagnetic phase,
finite clusters of magnetically ordered atoms can fluctuate quantum
mechanically.
In this case a so-called {\it quantum Griffiths-McCoy singularities}
would be generated and zero temperature divergences of the physical
quantities should be observed. In the next sections we explain exactly
how this process occurs.

Classical Griffiths singularities appear in the context of classical
Ising models as due to the response of rare and large clusters
to an external magnetic field \cite{grif}. A classical model
with columnar disorder was solved exactly by McCoy and Wu \cite{mccoy} in
two dimensions and shows clearly the importance of Griffiths singularities
close to the critical point for magnetic order. There are indeed experimental
indications for the existence of such singularities in  Fe$_{0.47}$Zn$_{0.53}$F$_2$ and other Ising magnets \cite{binek}. Because of the
relationship between the classical statistical mechanics in $d$ dimensions
and the quantum statistical mechanics in $d+1$ dimensions the McCoy-Wu
model can be mapped at $T=0$ into the transverse field Ising model
which has been studied by D.~Fisher \cite{dfisher}. The transverse field
Ising model is supposed to be applicable to insulating magnets since
conduction electron degrees of freedom are not present.
Senthil and Sachdev have studied the transverse
field Ising model in higher dimensions on a percolating lattice and
have found that quantum Griffiths singularities should be present
in the paramagnetic phase in these systems close to the QCP \cite{senthil}.
Moreover,
many physical quantities diverge at zero temperature as a result
of the strong response of the large and rare clusters. Such anomalous
behavior close to the QCP has been confirmed numerically in dimensions
higher than one \cite{tfinum} and there is today strong experimental
indication of such behavior in U and Ce intermetallics \cite{marcio}.

\section{Effective Hamiltonian: RKKY $\times$ Kondo}
\label{uceinter}

In our discussion of the physics of U and Ce intermetallics we are going
to use the so-called Kondo lattice Hamiltonian which describes the exchange
interactions between itinerant electron spins and the localized magnetic
moments:
\begin{eqnarray}
H = \sum_{{\bf k},\kappa} \epsilon_{\kappa}({\bf k})
c^{\dag}_{{\bf k},\kappa} c_{{\bf k},\kappa}
+ \sum_{i} \sum_{a,b,\kappa,\kappa'}
J_{a,b} (i) S^a(i)
c^{\dag}_{\kappa}(i) \tau_{\kappa,\kappa'}^b c_{\kappa'}(i)
\label{kondoarray}
\end{eqnarray}
where $\kappa=1,2$ labels the spin states,
$J_{a,b}(i)$ are the effective exchange constants between the
localized spins and conduction electrons 
at sites ${\bf r}_i$. $S^a(i)$ is the localized
electron spin operator and $\sum_{\kappa,\kappa'}c^{\dag}_{\kappa}(i)
\tau_{\kappa,\kappa'}^b c_{\kappa'}(i)$ the conduction electron spin operator
where $c_{{\bf k},\kappa}$ annihilates an electron with momentum ${\bf k}$
and spin projection $\kappa$ ($\tau^b$ with $a,b = x,y,z$ are Pauli
spin matrices). Observe that we are allowing for the possibility
of the conduction electron dispersion $\epsilon_{\kappa}({\bf k})$ to be
dependent of the electron spin. Indeed, one can show that (\ref{kondoarray})
can be obtained directly from the Anderson Hamiltonian \cite{and} when spin-orbit
effects are included and the hybridization energy between localized and
itinerant electrons is much smaller than the atomic energy scales \cite{next}.
One of the striking features of (\ref{kondoarray})
is the fact that the exchange interaction is not symmetric in spin space
because of spin-orbit coupling.
As we discuss below besides canted magnetism
(\ref{kondoarray}) gives rise to what we call a {\it canted}
Kondo effect.

The main problem in studying the competition of Rudermann-Kittel-Kasuya-Yosida
interaction (RKKY) \cite{rudkit} and Kondo effect in the
Hamiltonian (\ref{kondoarray}) is related to the fact that both
the RKKY and the Kondo effect have origin on the same magnetic coupling
between spins and electrons. What allows us to treat this problem is the
fact that the RKKY interaction is perturbative in $J/E_F$ while the
Kondo effect is non-perturbative in this parameter and requires different
techniques. Moreover, the RKKY interaction depends on electronic
states deep inside of the Fermi sea, in addition to those at $E_F$,
while the Kondo effect is purely a Fermi
surface effect. Thus, it seems to be possible to use perturbation theory
to treat the RKKY interaction.

We will consider, for simplicity the case where the Fermi surfaces for the
two species of electrons are spherical. The local electron operator can
be written in momentum space as
\begin{eqnarray}
c_{\kappa}(i) = \frac{1}{\sqrt{N}} \sum_{{\bf k}} e^{i {\bf k} \cdot {\bf r}} \, c_{{\bf k},\kappa} \, .
\label{cki}
\end{eqnarray}
We now separate the states in momentum space into three different regions of energy
as shown in Fig.\ref{fs}, namely,
$\Omega_0$: $k_{F,\alpha}-\Lambda<k<k_{F,\alpha}+\Lambda$;
$\Omega_1$: $k<k_{F,\alpha}-\Lambda$; and $\Omega_2$: $k>k_{F,\alpha}+\Lambda$ where $\Lambda$ is
an arbitrary cut-off. As we discuss below the value of the cut-off
is related to the number of compensated spins in the system. But
for the moment being we consider it as some arbitrary quantity
we can vary, like in a renormalization group calculation.
Observe that the sum in (\ref{cki}) can also
be split into these three different regions. The problem we want to address
is how the states in region $\Omega_0$ close to the Fermi surface renormalize as one traces out high energy
degrees of freedom which are present in regions $\Omega_1$ and $\Omega_2$. We are going to do this calculation
perturbatively in $J/E_F$ \cite{ian}.
For this purpose it is convenient to use a path
integral representation for the problem and write the quantum partition function as
\begin{eqnarray}
Z= \int D{\bf S}(n,t) D\overline{\psi}({\bf r},t) D\psi({\bf r},t) \exp\left\{
i {\cal S}[{\bf S},\overline{\psi},\psi] \right\}
\label{zspsi}
\end{eqnarray}
in terms of Grassman variables $\overline{\psi}$ and $\psi$ and
where the path integral over the localized spins also contains the constraint that
${\bf S}^2(n,t)=S(S+1)$. The quantum action in (\ref{zspsi}) can be separated into
three different pieces, ${\cal S}={\cal S}_0[{\bf S}]+{\cal S}_0[\overline{\psi},\psi]
+{\cal S}_I[{\bf S},\overline{\psi},\psi]$, where ${\cal S}={\cal S}_0[{\bf S}]$ is
the free actions of the spins (which can be written, for instance, in terms of
spin coherent states \cite{fradkin}),
\begin{eqnarray}
{\cal S}_0[\overline{\psi},\psi] =
\sum_{\alpha,\gamma} \int \frac{d\omega}{2 \pi} \sum_{\bf k}
\overline{\psi}_{\alpha}({\bf k},\omega)
\left(\omega + \mu - \epsilon_{\alpha}({\bf k}) \right) \delta_{\alpha,\gamma} \psi_{\gamma}({\bf k},\omega)
\label{s0psi}
\end{eqnarray}
is the free action for the conduction electrons and
\begin{eqnarray}
{\cal S}_I[{\bf S},\overline{\psi},\psi] = \int dt \sum_{\alpha,\gamma}
\sum_n J_{a,b}(n) S_a(n,t) \tau^b_{\alpha,\gamma} \overline{\psi}_{\alpha}(n,t) \psi_{\gamma}(n,t)
\label{interaction}
\end{eqnarray}
is the exchange interaction between conduction electrons and localized moments.
We now split the Grassman fields into the momentum shells
defined above, that is, we rewrite the path integral as
\begin{eqnarray}
Z= \int D{\bf S}(n,t) \prod_{i=0}^2 D\overline{\psi}_i({\bf r},t) D\psi_i({\bf r},t) \exp\left\{
i {\cal S}[{\bf S},\{\overline{\psi}_i,\psi_i\}] \right\}
\end{eqnarray}
where the indices $0,1,2$ refer to the degrees of freedom which reside in the momentum
regions $\Omega_0$, $\Omega_1$ and $\Omega_2$, respectively. The action
of the problem can be rewritten as ${\cal S} = {\cal S}_0[{\bf S}] + \sum_{i=0}^2 {\cal S}_0[\overline{\psi}_i,\psi_i]
+ {\cal S}_I[{\bf S},\overline{\psi},\psi]$. Notice the free part of the electron action is just a sum
of three terms (essentially by definition since the non-interacting problem is diagonal in momentum
space). Moreover, the exchange part mixes electrons in all three regions defined above:
\begin{eqnarray}
{\cal S}_I =
\sum_n \int dt \sum_{i,i'=0}^2 J_{a,b}({\bf r}_n) S_a({\bf r}_n,t) \tau^b_{\alpha,\gamma}
\overline{\psi}_{\alpha,i}({\bf r}_n,t) \psi_{\gamma,i'}({\bf r}_n,t) \, .
\label{lsplit}
\end{eqnarray}

Since we are interested only on the physics close to the Fermi surface we trace out
the fast electronic modes in the regions $\Omega_1$ and $\Omega_2$ assuming that
$J_{a,b} \ll \mu$. As we show in Appendix (\ref{pertrg}), besides
the renormalization of the parameters in free action of the electrons in the
region $\Omega_0$, we get the RKKY interaction between localized moments.
The effective action of the problem becomes:
\begin{eqnarray}
{\cal S}_{eff}[{\bf S},\overline{\psi}_0,\psi_0] &=& {\cal S}_0[\overline{\psi}_0,\psi_0] +
\sum_n \int dt \sum_{\alpha,\gamma,a,b} J^R_{a,b}({\bf r}_n) S_a({\bf r}_n,t) \tau^b_{\alpha,\gamma}
\overline{\psi}_{\alpha,0}({\bf r}_n,t) \psi_{\gamma,0}({\bf r}_n,t)
\nonumber
\\
&+& \sum_{n,m,a,b} \int dt \, \, \Gamma^R_{a,b}({\bf r}_n-{\bf r}_m,\Lambda) 
S_a({\bf r}_n,t) S_b({\bf r}_m,t)
\label{almosthere}
\end{eqnarray}
where $\Gamma^R_{a,b}({\bf r}_n-{\bf r}_m,\Lambda)$ is the cut-off dependent
RKKY interaction between the local moments and $J^R_{a,b}(n)$ is the {\it Kondo}
electron-spin coupling renormalized by the high energy degrees of freedom.
As we show in Appendix (\ref{pertrg}) this renormalization can be calculated
exactly. For a spherical Fermi surface the exchange interaction between
spins can be written as:
\begin{eqnarray}
\Gamma^R_{a,b}(r,\Lambda) = - \frac{9 c^2 \overline{n}}{E_F}
\sum_c J_{a,c} J_{b,c} {\cal F}(2 k_F r,\Lambda/k_F) e^{-r/\ell}
\label{gammaab}
\end{eqnarray}
where $c$ is the number
of magnetic moments per atom, $\overline{n}$ is the number of electrons per 
atom, $\ell$ is the electron mean-free path and ${\cal F}(x,y)$ is given
in (\ref{rkkylamb}). In Fig.\ref{newrkky} we compare the usual RKKY
with $\Lambda=0$ and the RKKY with finite $\Lambda$. 
At short distances, $r \ll k_F^{-1},\Lambda^{-1}$, 
the interaction is ferromagnetic
and is renormalized by a factor of  $(1-\Lambda/k_F)^3$. Moreover,
as shown in Fig.\ref{reduction} the first zero of the 
RKKY interaction is shifted from  $k_F r \approx 2.2467$ at $\Lambda = 0$ 
to smaller values.  
At intermediate distances, that is, $\Lambda^{-1} \gg r > k_F^{-1}$, 
the RKKY interaction decays like $1/r^3$ as in the case of $\Lambda=0$
but the most striking result is that for large distances, 
$r>>k_F^{-1},\Lambda^{-1}$ the RKKY interactions decays like $1/r^4$ 
instead of the usual $1/r^3$. Thus, for finite
$\Lambda$ the RKKY interaction has a shorter range. Another interesting
result of a finite $\Lambda$ is that at short and intermediate 
distances the RKKY oscillations
are mostly {\it antiferromagnetic}. As one can see directly from Fig.\ref{newrkky} the ferromagnetic part of the RKKY is suppressed after the first
zero. This could perhaps explain why most of the alloys of the type
discussed here have antiferromagnetic ground states. 

We observe further that the perturbation
theory here is well behaved and there are no infrared singularities in the
perturbative expansion. Thus, the limit of $\Lambda \to 0$ is well-defined.
In this limit $\Gamma^R_{a,b}({\bf r}_n-{\bf r}_m,\Lambda \to 0)$ becomes
the usual RKKY interaction one would calculate by tracing all the energy
shells of the problem. Observe that there are no retardation effects
in tracing this high energy degrees of freedom since they are much faster
than the electrons close to the Fermi surface and therefore adapt adiabatically
to their motion.
Hamiltonian (\ref{almosthere}) is the basic starting
point to our approach and contains the basic elements for the discussion
of magnetic order in the system. Observe that the RKKY interaction depends
on the electronic states far away from the Fermi surface while the Kondo
interaction is a pure Fermi surface effect.
While the RKKY interaction leads
to order of the magnetic moments the Kondo coupling induces magnetic
quenching.
It is the interplay of these two interaction which leads to the physics
we discuss here. The action (\ref{almosthere}) has been used as a starting point
for many theoretical discussions of rare earth alloys \cite{coqblin}.
We stress, however, that the action (\ref{almosthere}) describes only the low energy
degrees of freedom of the problem and therefore the exchange constants
that appear there can have strong renormalizations due to the high energy
degrees of freedom. Moreover, as we explain below the cut-off $\Lambda$
depends on the number of compensated moments.

\subsection{Magnetically ordered phase: the role of RKKY}
\label{kondorder}

The existence of local moments is not a sufficient condition
for the existence of long range magnetic order. It is exactly the
interaction between the spins which determines the ordering temperature $T_c$
of the material. There are various ways localized moments can
interact: dipolar-dipolar interactions, kinetic exchange
interactions due to the overlap of the f orbitals and RKKY interaction.
Dipolar interactions are too small to account
for the ordering temperature in these systems (they range from $100$ K down to
$10$ K in the pure compounds) and the direct exchange between f orbitals is very weak
since the spatial extent of the f orbitals is small (with the possible exception
of the 5f orbitals of U).
The RKKY interaction is by far the most important
interactions in metallic rare earth alloys and it will be the only interaction
we will consider in detail.

As it is well-known and as shown in Appendix \ref{pertrg} in the ordered
case ($\Lambda=0$) the RKKY interaction decays like $1/r^3$ and oscillates
in real space with wave-vector $2 k_F$.
The oscillatory terms have to do with the sharpness of the Fermi surface.
Moreover, non-spherical Fermi surfaces will also lead to
an angular dependence on the RKKY interaction with decaying rates
which vary with the direction \cite{roth}. The specific form of the RKKY
interaction is not important in our discussion but the fact that the
RKKY is an interaction which is perturbative in $J/\mu$ and scales like
$N(0) J^2$. The exponential factor due to disorder was obtained originally
by de Gennes \cite{degennes}.

Observe that the spin-orbit coupling generates an RKKY interaction which is
anisotropic and therefore can give rise to canted magnetism.
This effect is the analogue of the
anisotropic spin exchange interaction, or Dzyaloshinsky-Moriya (DM) exchange
interaction \cite{dzmo} in insulating magnets which is obtained via
the kinetic exchange between localized moments in the presence of
spin-orbit coupling. The interaction (\ref{gammaab}) is an indirect
exchange interaction in the presence of spin-orbit coupling. Thus,
as in the case of DM interactions one expects parasitic ferromagnetism
within antiferromagnetic phases. This effect has been observed long ago in
R-Cr0$_3$ systems\cite{yosida,koeler}. The existence of ferromagnetic
and antiferromagnetic coupling creates a very rich situation where many different magnetic phases are possible in the presence of a Lifshitz point \cite{lp}. Recent theoretical approaches for the NFL
problem in CeCu$_{6-y}$Au$_y$ are based on the idea that the Lifshitz point
in these systems is a QCP \cite{piers} and therefore the quantum fluctuations
associated with this point induce NFL behavior in the conduction band.

The critical temperature of the system can be estimated directly from
the mean field theory for (\ref{kondoarray}) and it is given by \cite{mattis}
\begin{eqnarray}
T_c = \frac{2 S (S+1)}{3}
\sum_{{\bf R} \neq 0} \Gamma_M({\bf R}) \cos \left({\bf Q}\cdot {\bf R}\right)
\label{tcmf}
\end{eqnarray}
where ${\bf Q}$ is the ordering vector ($Q=0$ for ferromagnetism) and
$\Gamma_M({\bf R})$ is the largest eigenvalue of $\Gamma_{a,b}({\bf R})$.
Observe that $T_c$ scales with $\Gamma$ and therefore it is proportional
to $N(0) J^2$. This value of $T_c$ gives the order of magnitude of
the transition temperature in Kondo hole or ligand systems,
that is, the magnetically ordered Kondo lattice. In what follows we
discuss the effect of disorder on the magnetic order in these systems.

\subsection{Magnetic dilution}

In Kondo hole systems the transition to the paramagnetic phase happens
because the magnetic sublattice is diluted with non-magnetic atoms.
In this case two main effects occur: (1) the magnetic system loses
its magnetic atoms; (2)
because the non-magnetic atoms
do not have the same size of the magnetic ones there is a local lattice
contraction or expansion. The first effect created by the dilution is
to introduce disorder in the electronic environment and produce a finite
scattering time $\tau$ for the electron. As shown by de Gennes
long ago \cite{degennes}, the RKKY interaction decays exponentially with the electron
mean-free path, $\ell=v_F \tau$.
Notice that this is only true if there is true {\it magnetic long range order}
in the problem. In the case of a spin glass order this is argument
is not correct \cite{elihu}.

The problem of destruction of magnetic order in a ligand system is more
complex since the magnetic atoms are not replaced. In an alloy
like UCu$_{4-x}$Pd$_x$ the Cu atoms are replaced by the somewhat smaller
Pd atoms. This difference between the Pd and the Cu leads to a local lattice
contraction which modifies the local hybridization matrix elements.
Since these matrix elements are exponentially sensitive
to the overlap between different angular momentum orbitals one can have
large local effects in the system. This change of local matrix elements
induces changes in the exchange constants between the conduction band
and the localized moments, $J({\bf r}_i)$, in (\ref{kondoarray}).

Let us start with
(\ref{kondoarray}) and in the homogeneous ordered phase where we
assume that $J_{a,b}({\bf r}_i)=J_{a,b}$ for all
sites. The transition temperature is given by (\ref{tcmf}) with
$\Gamma$ given by (\ref{gammaab}) with the electron mean free path
determined by the extrinsic impurities in the system. As the Kondo lattice
is doped, either by substitution of a magnetic atom by a non-magnetic one
(as in the case of the Kondo hole systems) or just disordered
(as in the case of ligand systems) the local coupling between the localized moments and electrons
is changed from $J_{a,b}$ to a different value $\tilde{J}_{a,b}$ (for simplicity
we will assume just a binary distribution but this assumption can be easily
generalized). In this case can rewrite (\ref{kondoarray}) as $H=H_0 + H_{KL} + H_{imp}$ where:
\begin{eqnarray}
H_0 &=& \sum_{{\bf k},\kappa} \epsilon_{\kappa}({\bf k}) c^{\dag}_{{\bf k},\kappa}
c_{{\bf k},\kappa}
\nonumber
\\
H_{KL} &=& \sum_i \sum_{a,b,\kappa,\kappa'} J_{a,b} S_a({\bf r}_i) c^{\dag}_{\kappa}({\bf r}_i)
\tau^b_{\kappa,\kappa'} c_{\kappa'}(i)
\nonumber
\\
H_{imp}&=& \sum_{a,b,i} p_i \delta J_{a,b} S_a(i) c^{\dag}_{\kappa}(i)
\tau^b_{\kappa,\kappa'} c_{\kappa'}(i)
\label{kondord}
\end{eqnarray}
where $\delta J_{a,b} = \tilde{J}_{a,b} - J_{a,b}$ where the summation over $i$
includes all sites and $p_i =1$ if a particular site $i$ was changed by disorder and
$p_i=0$ otherwise.

Let us consider first the case where the system the ordered state is just slightly
doped. In first order in $x$ (the concentration) the problem reduces
to a single ion problem. If $|\delta J_{a,b}| > |J_{a,b}|$ then one has to treat first
$H_0 + H_{imp}$ and then add $H_{KL}$. It is obvious that we have a
Kondo effect on the sites for which $p_i=1$ with the original conduction band
of the system. In the case of lattice contraction we would have $\delta J_{a,b}>0$ ($\tilde{J}_{a,b}
> J_{a,b}$)
and therefore an antiferromagnetic Kondo effect
and a singlet is formed below a Kondo temperature $T_K \approx E_F \, e^{-1/(N(0) \delta J)}$.
All sites with $p_i=1$
are magnetically quenched. If  $\delta J_{a,b}<0$ ($\tilde{J}_{a,b}
< J_{a,b}$), which is the case of local lattice
dilation, we would have a ferromagnetic
Kondo effect and a local triplet state. Thus, there is
an increase of the local magnetization of the system as a function
of $x$. The effect here is very similar to the problem of enhancement
of the magnetic moment by a highly polarizable metallic environment and creation of ``giant moments''
as it was discussed long ago by Jaccarino and Walker \cite{jaca} and observed
experimentally in Pd$_{1-x}$Ni$_x$ and other similar alloys \cite{pdni}.

As it is well-known in a disordered system,
the electron acquires a life-time, $\tau_K$, due to the
Kondo effect which, at zero temperature, reads \cite{kondo}:
\begin{eqnarray}
\frac{1}{\tau_K} = \frac{3 \pi (\delta J)^2 S(S+1) x}{2 E_F}
\label{ltk}
\end{eqnarray}
where $\delta J$ is the largest eigenvalue of $\delta J_{a,b}$.
This finite lifetime leads to a finite mean-free path, $\ell_K = v_F \tau_K$
for the conduction band motion. After the mean-free
path is taken into account one proceeds as before (described in Appendix \ref{pertrg})
to calculate the RKKY interaction which will be given by (\ref{gammaab}) with $1/\ell$
substituted by $1/\ell + 1/\ell_K$ as in Matthiessen's rule.

In the opposite limit of $|\delta J_{a,b}| < |J_{a,b}|$
we have to consider $H_0 + H_{KL}$ first and then add to it $H_{imp}$. This problem
is more complicated because the ordered Kondo lattice problem has to be solved first.
Here we just consider the simplest mean field theory in which the magnetic moments
order along the $z$ axis. The mean field Hamiltonian can be written as $H = H_{MF} + H_{imp}$
where
 \begin{eqnarray}
H_{MF} &=& \sum_{{\bf k},\kappa} \epsilon_{\kappa}({\bf k}) c^{\dag}_{{\bf k},\kappa}
c_{{\bf k},\kappa} + \sum_i \left[H_S
\left(n_{i,\uparrow}-n_{i,\downarrow}\right) + H_s S_z({\bf r}_i)  \right]
\label{hmf}
\end{eqnarray}
where $H_S = J_{z} \langle S_z({\bf r}_i) \rangle$ and $H_s =  J_{z} \left(\langle n_{i,\uparrow} \rangle -
\langle n_{i,\downarrow} \rangle \right)$ are the molecular fields applied by the localized spins
on the conduction electrons and by the conduction electrons on the localized spins,
respectively.

The problem described by Hamiltonian (\ref{hmf}) can be easily solved
for the case of ferromagnetism as we show in Appendix (\ref{ferromf}).
As a result the electronic degrees of freedom are renormalized
in different ways depending on geometry of the Fermi surface
and the type of ordering one has. For ferromagnetism the main change in
the problem is the change in the density of states
for different electron species. In the
case of antiferromagnetism the situation can be more complicated because
one generates a well defined momentum ${\bf Q}$ and therefore Umklapp
scattering is possible depending on the shape of the Fermi surface.
A spin density wave state (SDW) can be generated and a
gap can open on regions of the Fermi surface. Experimentally, 
the systems we are
discussing here are metallic over the ordered phase which implies that
the whole Fermi surface or perhaps large portions of it would remain gapless.
This is easily understood by the fact that the conditions for
commensurability are hard to obtain in these systems which have
rather complicated Fermi surfaces. Therefore, the conclusions we
reach for the ferromagnetic case can be easily generalized to more
complicated magnetic structures.

Assuming that the system remains metallic
in the magnetically ordered phase we see that $H_K$ describes the Kondo
effect on this new metallic band in the presence of a magnetic field.
We show in Appendix \ref{ferromf} that at $T=0$ the local magnetic
energies are $H_S = - S J_z$ and $H_s = S N(0) J_z^2/\rho_f$.
Observe that while the magnetic field applied on the electron,
$H_S$, by the localized spin
can be positive or negative depending if the local exchange
is antiferromagnetic or ferromagnetic, the local field applied on the local
spin, $H_s$, by the conduction electrons is always ferromagnetic. In the
paramagnetic case ($\langle S_z \rangle = \langle (n_{\uparrow}-n_{\downarrow})
\rangle =0$) the local state of the system is degenerate. That is,
one has a quartet, made out of $|\downarrow,\Uparrow\rangle$,
$|\downarrow,\Downarrow\rangle$,$|\uparrow,\Uparrow\rangle$ and
$|\uparrow,\Downarrow\rangle$  where
$\uparrow,\downarrow$ represents the conduction band spin and $\Downarrow,\Uparrow$ the local moment spin. Since these are all eigenstates of
$S_z$ and $n_{\uparrow}-n_{\downarrow}$ their energies in term
of the molecular fields are
\begin{eqnarray}
|\uparrow,\Downarrow\rangle &\to& E_1 = - S J_z -  \frac{S N(0) J_z^2}{\rho_f}
\nonumber
\\
|\uparrow,\Uparrow\rangle &\to& E_2 = - S J_z +  \frac{S N(0) J_z^2}{\rho_f}
\nonumber
\\
|\downarrow,\Downarrow\rangle &\to& E_3 =  S J_z -  \frac{S N(0) J_z^2}{\rho_f}
\nonumber
\\
|\downarrow,\Uparrow\rangle &\to& E_4 =  S J_z +  \frac{S N(0) J_z^2}{\rho_f} \,  .
\label{fkel}
\end{eqnarray}
Observe that there is level crossing when $J_z = \pm J_c = \pm \rho_f/N(0)$.
Furthermore, $J_c \approx c E_F$ where $c$ is the number of local moments
per atom. Since $J_z \ll E_F$ we can only have $J_z \approx J_c$ when
the density of local moments is very low (dilute limit). In general
we expect $J_z \ll J_c$ in concentrated Kondo lattices.

When $J_z>J_c$ (antiferromagnetic coupling) the local state of
the system is  $|\uparrow,\Downarrow\rangle$ which is
separated from the state
$|\downarrow,\Downarrow\rangle$ by an energy amount $\delta E = 2 S J_z$
and therefore the Kondo effect is suppressed (notice that
$\delta J < \delta E$ and therefore the Kondo effect cannot
bring these two states together). The problem is very
similar to the usual ferromagnetic Kondo effect where quantum
fluctuations are totally suppressed. Observe, however, that
the local moment is compensated in the same way it would be if we
just eliminate the local moment from the lattice. (Note that
the electronic phase shift due to scattering by the impurity 
is zero in both cases.) Thus
doping decreases the magnetization of the system as one
would have in the usual dilution problem (the magnetic
dilution of a ligand system is essentially identical to the dilution
of the Kondo hole system).
In the intermediate coupling regime of
$0<J_z<J_c$ the two lowest energy states are $|\uparrow,\Downarrow\rangle$
and $|\uparrow,\Uparrow\rangle$ which are separated in energy
$\delta E = S N(0) J_z^2/\rho_f$ and no Kondo effect happens.
The ferromagnetic case is somewhat similar with the difference
that the local state of the system is $|\downarrow,\Downarrow\rangle$
but also in this case the Kondo effect does not take place
because the electron spin states are quenched by the local
molecular fields. Therefore, in a ferromagnetically ordered
lattice the Kondo effect is absent independent of the sign
of the Kondo coupling. This state of affairs is very similar
to the one found by Larkin and Mel'nikov in the case of magnetic
impurities in nearly ferromagnetic Fermi liquids \cite{larkin}.
In summary, we conclude that the case of
$|\delta J_{a,b}| < |J_{a,b}|$ there is no real Kondo effect
in a magnetically ordered Kondo lattice. We note, however,
that in the case of antiferromagnetic coupling the magnetization
of the system drops because of a formation of $|\uparrow,\Downarrow\rangle$
states.

We have seen that the ordered state of a Kondo lattice
is destroyed via compensation of the magnetic moments either via
the Kondo quenching or moment compensation. Thus, the percolation
parameter in this problem is the density of quenched moments,
$\rho_Q$. In percolation theory we assign a percolation parameter
$p$ which in the case of the Kondo lattice is essentially $\rho_Q$.
Let us now consider the situation of one of the alloys mentioned
previously where we chemically substitute the atoms by an amount $x$.
In a Kondo hole system the number of magnetic moments decreases
with the alloying because the magnetic moments are replaced by
non-magnetic atoms. At the same time the number of compensated
moments grows because of the changes in the local structure of
the lattice and the increase in the Kondo coupling.
At some particular value of $x$, $\rho_Q$ reaches a maximum since
it cannot grow beyond the actual number of magnetic moments which are left
in the system. Moreover, at percolation threshold, $p_c$, which is
determined by the dilution and lattice changes, the last infinite cluster
disappears and long range order is lost. Beyond this point only
finite clusters can exist.
Eventually both the number of magnetic atoms and the number
of compensated atoms drop to zero. This situation is depicted in Fig.\ref{perco}(a).

In a ligand system the density of magnetic moments is
kept constant with chemical substitution because the magnetic atoms
are not replaced.
The number of compensated moments grows because of the local growth of
the hybridization. At some critical value
of $\rho_Q$ we reach the percolation threshold $p_c$ and long
range order is lost because the last infinite cluster of uncompensated
moments disappears. Moreover, when $\rho_Q$ grows beyond the
threshold it will eventually reach the value $\rho_f$ and all
the magnetic moments in the lattice are compensated.
At this large value of doping one can find a heavy fermion ground state. We depict this situation in
Fig.\ref{perco}(b).

We have to be careful in interpreting the heavy fermion ground
state in light of the Kondo model we are studying.
As we mentioned previously, the dilution in a Kondo hole system
leads to a trivial Fermi liquid state while in a ligand system
it can eventually lead to a Heavy Fermion (HF) state which is
not straightforward to describe. Since the HF state is non-magnetic it is
clear that dilution can drive the system to such a state by increasing
the local hybridization of the conduction band with the localized
f-electrons. If the hybridization becomes of the order of the
local atomic energy scales the Kondo Hamiltonian (\ref{kondoarray})
is not a good starting point for the description of the magnetic
correlations. We should work directly with the Anderson Hamiltonian.
This is usually the route taken by many approaches
to the HF ground state \cite{klattice}. The f-electrons
mix with the conduction band electrons and the Fermi surface is
large since it counts all the electrons in the system. In what
we have discussed we considered only the Kondo Hamiltonian which does
not contain this kind of physics since the occupation of the f-atomic
states is fixed to be $1$. Since we are interested mainly in the
behavior of the system close to the magnetic ordered phase the
Kondo Hamiltonian should give a good description of the problem but
one has to be cautious about the transition from localized to itinerant
behavior in these systems.

Assuming that the Kondo lattice is a good starting problem we immediately
see that the Hamiltonian generated by (\ref{almosthere}) has the
right properties associated with Doniach's famous argument \cite{doni},
namely, there are two main energy scales described in (\ref{almosthere}):
the RKKY strength $\Gamma^R$ and the Kondo temperature $T_K$ generated by
$J^R$. These two energy scales are {\it local} properties of the alloying procedure
and they scale in a very different way with the bare exchange between conduction
electrons and magnetic moments.
The RKKY coupling is proportional to $(N(0) J(i))^2/E_F$ in the weak coupling regime
while $J^R \approx J$ and therefore $T_K(i) \approx E_F e^{-1/(N(0) J(i))}$
(we are going to show below that this $T_K$ is slightly more complicated
if we take anisotropy into account). If we plot these two quantities together
as in Fig.\ref{donifig}
we see that there is a critical value $J_c$ for the local
exchange constant $J(i)$ above which the Kondo temperature is larger
than the RKKY energy scale and therefore as the the system is cooled
from high temperatures the moment is locally compensated and the RKKY
interaction for that particular place becomes irrelevant in the limit
of large $J(i)$. For $J(i)$ below
$J_c$ the RKKY energy scale is larger and therefore as the system is
cooled down the moment can locally order with its environment.
(Note that for just two impurities, a partial Kondo effect occurs
for finite $J(i)$, and the ground state is always a singlet, even
for ferromagnetic RKKY coupling \cite{tikm}. Here, however,
we are treating the case of a larger number of magnetic moments,
tending toward a magnetic ground state for large RKKY.)
We have to stress, however, that in the presence of disorder
this effect is local and represents the a quantum percolation
problem and has nothing to do with the homogeneous change in the
exchange $J$! Indeed, if one interprets the chemical alloying of the
system as a simple change of the exchange over the entire lattice
then the Doniach argument would predict an ordering temperature $T_N$
as in Fig.\ref{donifig} which vanishes at a QCP where a transition
from the ordered state to a fully Kondo compensated state happens.
In this picture the QCP has nothing to do with a percolation problem
where moments are compensated due to local effects. In our picture
this is not possible since a local change in a coupling constant
does not immediately imply a change of the ``average coupling''
constant. Thus, we believe
that interpretations of the Doniach argument based on homogeneous
changes in the couplings are actually erroneous.

We also would like to point out that the same discussion carried out in terms
of Hamiltonian (\ref{kondoarray})
can be done in terms of the effective action (\ref{almosthere}) by introducing
a Hubbard-Stratonovich transformation and studying the saddle point equations
by assuming a homogeneous solution for the spin field.

We have argued that by chemical substitution the order in a Kondo
lattice can be taken away even when the substitution is not on
the magnetic site because individual moments are compensated magnetically
due to the distribution of exchange constants. At some finite
concentration $x^*$ long range order is lost and the
system enters a paramagnetic phase. Since the problem at hand
is a percolative one, the paramagnetic phase can still contain
clusters of atoms in a relatively ordered state. In the rest
of the paper we are going to discuss exactly the physics of these
clusters and how they respond to external probes. We are going
to show how finite clusters give rise to Griffiths-McCoy
singularities in these systems.

\subsection{The value of $\Lambda$}

In our previous discussion the cut-off $\Lambda$ (cf. Fig. (2))
was considered as a free parameter
but now we discuss this problem in more detail. It is clear
from the above discussion that in the magnetically ordered phase we
can set $\Lambda=0$ because all the electrons on the Fermi sea
are participating in the order of the magnetic moments. As one starts
to dilute the Kondo lattice some particular sites with hybridization
larger than average will be magnetically compensated by the Kondo effect.
Therefore $\Lambda$ has to increase in order to accommodate the electrons
which participate in the Kondo coupling. Thus we expect $\Lambda$
to increase with the number of compensated magnetic moments in the system.

Let $\rho_Q$ be the number of compensated moments per unit of volume at
a given chemical concentration. As we have discussed previously this
number is not given only by the Kondo effect alone but by the
competition between the Kondo effect and the RKKY interaction in
(\ref{almosthere}).
If the electrons which participate in the
Kondo effect are the ones in the
region $\Omega_0$ in Fig.\ref{fs} then it is easy to see that
\begin{eqnarray}
\rho_Q &=& \rho_e - 2 \int \frac{d^3 k}{(2 \pi)^3} \Theta(k_F-\Lambda-k)
\nonumber
\\
&=& \rho_e-\frac{(k_F-\Lambda)^3}{3 \pi^2}
\label{rhoulamb}
\end{eqnarray}
which can inverted to give (using (\ref{kf}))
\begin{eqnarray}
\frac{\Lambda}{k_F} = 1 - \left(1-\frac{\rho_Q}{\rho_e}\right)^{1/3} \, .
\label{lkf}
\end{eqnarray}
This equation gives a simple relationship between the number of
compensated moments and the cut-off to be used in (\ref{almosthere}). When
$\rho_Q \ll \rho_e$ we have $\Lambda/k_F \approx \rho_Q/(3 \rho_e)  \ll 1$ and
$\Lambda$ is very small compared with the Fermi momentum. Naturally,
since each magnetic atom in the unit cell gives at least one electron
to the conduction band, we must have $\rho_f \leq \rho_e$. This is true
even when the electronic density changes as a function of the chemical
substitution or percolation parameter $p$ as shown in Fig.\ref{perco}.
Obviously, we have 
$\rho_f \geq \rho_Q$ and therefore $\rho_Q < \rho_e$ from which follows that
$\Lambda \leq k_F$.
The extreme case of  $\rho_Q \approx \rho_e$
($\Lambda \approx k_F$) indicates that Kondo process
involves all electrons in the Fermi sea. Notice that
the perturbative expansion
in terms of $J$ in the effective Hamiltonian (\ref{almosthere})
is still valid but the form of the RKKY interaction
will not be the usual one, that is, (\ref{gammaab}), since it will
depend strongly on $\Lambda/k_F$ (as shown in Appendix \ref{pertrg}).

Our discussion should be contrasted with the well-known {\it exhaustion
paradox} proposed by Nozi\`eres \cite{paradox}: because the characteristic
energy in the Kondo effect is $T_K$ the number of electrons
participating in the Kondo effect is supposed to be
\begin{eqnarray}
\frac{\rho_K}{\rho_f} \approx \frac{N(0)}{\rho_f} T_K = \frac{3 \rho_e}{2 \rho_f} \frac{T_K}{E_F} \, .
\label{rhok}
\end{eqnarray}
So, in order for all moments to be compensated
one has to require this fraction to be $1$.
But because $T_K \ll E_F$ the condition in (\ref{rhok}) is only observed
at extremely small moment concentrations. As one increases $\rho_f$
there are not enough electrons at the Fermi surface to quench
the magnetic moments and one should observe free magnetic moments.
This paradox is based on an energetic argument which takes only the
Kondo effect into account and is correct in the dilute limit.
Indeed when $\rho_F > \rho_Q \to 0$ we would have
$T_K \approx v_F \Lambda$. In
the concentrated limit, where interactions start to play a role, it is
misleading. The energy scales that produce $\Lambda$ involve not
only the Kondo coupling but also the RKKY interaction. Since the RKKY
interaction involves states deep inside of the Fermi sea the energy
scales involved in $\Lambda$ can be rather large, of the order of the
Fermi energy itself. Indeed, recent infrared conductivity measurements
in YbInCu$_4$ indicate that a large fraction of the Fermi sea must
be involved in the Kondo quenching in clear contradiction to the exhaustion
paradox and in agreement with our discussion \cite{garner}.
The calculation of $\rho_Q$ thus involves a self-consistent
calculation of the combined effect of Kondo and RKKY and goes beyond
the scope of this paper.

\section{$1,2,...,N$}
\label{twolevel}

We have argued in the previous section that the local changes in the
hybridization will lead to a finite density $\rho_Q$ of compensated
magnetic moments. The number of compensated moments depends strongly
on lattice structure and the local changes in the electronic wavefunctions
due to alloying. It is obvious, however, that alloying leads to a
percolation problem and the number of magnetic moments drops as
the ordered system moves towards the paramagnetic phase.
Instead of working directly with $\rho_Q$ we define a percolation
parameter $p$. For $p<p_c$ magnetic order exists and for $p>p_c$
long range order is not possible. $p_c$ is therefore the percolation threshold
of the lattice. Thus, above $p_c$ there are only finite clusters of
magnetic moments which are more coupled than the average.
The probability of having $N$ moments together is given in percolation
theory by \cite{perco}
\begin{eqnarray}
P(N,p) = N^{1-\tau_d} f_p\left(N/N_{\xi}\right)
\label{Pn}
\end{eqnarray}
where $N_{\xi}=(\xi(p)/a)^D$ is the number of spins within a correlation
length $\xi$,
$d$ is the spatial dimension of the lattice, $\tau_d$ is a percolation exponent
($\tau_2=187/91 \approx 2.05$, $\tau_3 \approx 2.18$), $D$ is the fractal dimension
of the cluster ($\tau_d = 1+d/D$). Close to
percolation threshold $p_c$ the system is critical and therefore
\begin{eqnarray}
\frac{\xi(p)}{a} \propto \frac{1}{|p-p_c|^{\nu}}
\label{xip}
\end{eqnarray}
where $\nu$ is the correlation length exponent. The scaling function $f_p(x)$ is such
that $f_p(x \to 0) =1$ and for $x\gg 1$ one has
\begin{eqnarray}
f_{p}(x) \approx x^{-\theta_{d,p}+\tau_d} e^{-c_{p} x^{\zeta_p}}
\label{fpx}
\end{eqnarray}
where $\theta_{2,p>p_c}=1$, $\theta_{3,p>p_c}=3/2$, $\theta_{2,p<p_c}=5/4$ and
$\theta_{3,p<p_c}=-1/9$, $c_p$ is a constant of order of unit, $\zeta_{p>p_c}=1$
and $\zeta_{p<p_c} = 1-1/d$. Observe
that the exponential behavior given in (\ref{fpx}) is the dominant part of
the probability.

The above equations are easy to understand in the dilute limit, that is,
when $p>>p_c$. If $c$ is the concentration of magnetic atoms in the systems
then the probability of having $N$ atoms together is
\begin{eqnarray}
P(N,p>>p_c) \approx c^N = e^{-N \ln(1/c)}
\label{nopc}
\end{eqnarray}
and the probability is exponentially small. In particular, the probability of
finding $N$ nearest neighbor atoms in a lattice with coordination number $Z$
is
\begin{eqnarray}
P(N,Z) = \frac{Z!}{N! (Z-N)!} c^N (1-c)^{Z-N}
\label{pnz}
\end{eqnarray}
which reduces to (\ref{nopc}) when $c \to 0$.

It is also important to understand what happens close to percolation threshold
($p-p_c \approx 0^+$)
where from (\ref{Pn}) and (\ref{fpx}) we have
\begin{eqnarray}
P(N,p \approx p_c) \propto \exp\left\{-\frac{c_{p_c} N}{(\xi(p)/a)^D}\right\} \, .
\label{pnpc}
\end{eqnarray}
Therefore, from (\ref{nopc}) and (\ref{pnpc}),
the exponential part of the probability can be written in a
simple form
\begin{eqnarray}
P(N,p) \propto e^{-\kappa_p N^{\zeta_p}}
\label{pnpsimple}
\end{eqnarray}
where
\begin{eqnarray}
\kappa_{p\gg p_c} &\approx& \ln(1/c)
\nonumber
\\
\kappa_{p \approx p_c} &\approx&  (\xi(p)/a)^{-D} \approx |p-p_c|^{\nu D} \to 0 \, .
\label{kappap}
\end{eqnarray}

It is obvious from the above equations that the mean size of
the clusters diverges at $p=p_c$ and that for $p>p_c$ the cluster size
is determined by $\xi(p)$. In this paper we are mainly interested in
the physics of clusters in the paramagnetic case ($p>p_c$). For a given
concentration $c$ there are going to be clusters of all sizes but with
different probabilities. For instance, for a cubic system ($Z=6$), the
number of isolated atoms ($N=0$) becomes smaller than the number of
dimers ($N=1$) only when the concentration is larger than $c > 0.14$.
In understanding the response of such an inhomogeneous to external
probes we have to understand how a particular cluster responds to
these probes. We are going to assume that the clusters only couple
weakly to each other and that in first approximation they can be thought
of isolated and permeated by a paramagnetic matrix. The weak coupling
among the clusters, as we discuss at the end of the paper, can lead
to glassy states which do not show strong deviations from a Fermi liquid
ground state. In order to build up intuition about the physics of
the clusters we discuss the case of $1$ and $2$ magnetic atoms
in a paramagnetic matrix.
After the discussion it will become quite clear how a cluster of $N$
atoms should behave in the presence of a paramagnetic environment
(this is the $N$ impurity cluster Kondo effect).

\subsection{$1$ Magnetic Moment Kondo Effect}
\label{1mmke}

The one impurity Kondo effect should occur in the case of Kondo hole systems at
very low magnetic atom concentration and according to the argument given in
Subsection \ref{kondorder} also in the ligand systems in the ordered phase.
Our discussion now will follow very closely the approach
to the single impurity problem via bosonization techniques \cite{twolevel}.
In order to do so we reduce (\ref{hmf})
to an effective one dimensional problem via the bosonization technique.

The bosonization procedure follows three different steps. In the first step we trace out
the electrons far away from the Fermi surface exactly as in Section \ref{uceinter}. Since
there is just one magnetic moment in the problem no RKKY term is generated and the only effect
of this trace is to renormalize the Kondo couplings $J$.
Indeed, as shown by Anderson {\it et al.} \cite{yuval}
for the one impurity Kondo problem
the renormalization of the Kondo coupling is given by
\begin{eqnarray}
J_z^R = 8 v_F \delta(J_z)
\label{jzr}
\end{eqnarray}
where $\delta(J_z)$ is the phase shift of the electrons due to the scattering of a static
impurity (corresponding to the Ising component of (\ref{almosthere}))
which is given by
\begin{eqnarray}
\delta(J_z) = \arctan\left(\frac{\pi N(0) J_z}{4}\right) \, .
\label{djz}
\end{eqnarray}
On a second stage we linearize the electron dispersion close to the Fermi surface:
\begin{eqnarray}
E_{{\bf k},\sigma} = E_F + v_{F,\sigma} (|{\bf k}|-k_{F,\sigma}) \, ,
\label{linear}
\end{eqnarray}
where we are considering the generic case where the spin flavors can have
different Fermi velocities $v_{F,\sigma}$. In order to linearize the
problem we also have to introduce a momentum cut-off, $\Lambda$,
which is a non-universal constant (anisotropies in the shape of
the Fermi surface can be absorbed in $\Lambda$). The conduction
band Hamiltonian is written as
\begin{eqnarray}
H_{C} = \sum_{p,\sigma} v_{F,\sigma} p c^{\dag}_{p,\sigma} c_{p,\sigma}
\label{hc}
\end{eqnarray}
where $c_{p,\sigma}$ creates an electron with spin $\sigma$,
momentum $|{\bf k}| = p + k_F$ and angular momentum $l=0$.
Moreover, since we are treating the problem of a single impurity it
involves electrons moving in a single direction associated
with incoming or outgoing waves.
Thus, in writing (\ref{hc}) we have reduced the problem to an
effective one-dimensional problem. In order to do it we introduce
right, $R$, and left, $L$, moving electron operators
\begin{eqnarray}
\psi_{R(L),\sigma}(x) = \frac{1}{\sqrt{2 \pi}} \int_{-\Lambda}^{\Lambda}
dk c_{k \pm k_{F,\sigma},\sigma} e^{i k x}
\end{eqnarray}
which are used to express the electron operator as
\begin{eqnarray}
\psi_{\sigma}(x) = \psi_{R,\sigma}(x) e^{i k_{F,\sigma} x}
+ \psi_{L,\sigma}(x) e^{-i k_{F,\sigma} x} \, .
\end{eqnarray}
In any impurity problem the right and left moving operators
produce a redundant description of the problem since they
are actually equivalent to incoming or outgoing waves out of
the impurity. Therefore we have two options: either we work with
right and left movers in half of the line or we work in the full
line but impose the condition  $\psi_{R,\sigma}(x) = \psi_{L,\sigma}(-x)$.
Following tradition we choose the latter approach. Thus, from now on we drop the symbol $R$
from the problem and work with left movers only.
The left mover fermion can be bosonized as
\begin{eqnarray}
\psi_{\sigma}(x) = \frac{K_{\sigma}}{\sqrt{2 \pi a}}
e^{i \Phi_{\sigma}(x)}
\label{mandel}
\end{eqnarray}
where
\begin{eqnarray}
\Phi_{\sigma}(x) = \sum_{p>0} \sqrt{\frac{\pi}{p L}}
\left((b_p + \sigma a_p) e^{i \sigma p x} - h.c.\right)
\end{eqnarray}
and $K_{\sigma}$ is a factor which preserves
the correct commutation relations between electrons, that is,
$\{\psi_{\sigma}(x),\psi_{\sigma'}(y)\} = \delta(x-y)
\delta_{\sigma,\sigma'}$
and the bosons obey canonical commutation relations
$[a_k,a^{\dag}_p]=[b_k,b^{\dag}_p] = \delta_{p,k}$.

In what follows we are going to assume the simple case
in which $J_{a,b} = J_a \delta_{a,b}$ where $J_z < J_x <J_y$.
In terms of the boson operators, the Kondo Hamiltonian ({\ref{almosthere})
 becomes
\begin{eqnarray}
H &=& \overline{v}_F \sum_{p>0} p (a^{\dag}_p a_p + b^{\dag}_p b_p)
+ \delta v_F \sum_{p>0} p \left(b^{\dag}_p a_p+a^{\dag}_p b_p \right)
+J_z^R S^z \sum_{k>0} \sqrt{\frac{k}{\pi L}}
(a_k + a^{\dag}_k)
\nonumber
\\
&+& \frac{J_{+}}{4 \pi a} \left(S^+ K^{\dag}_{\downarrow} K_{\uparrow}
e^{\sum_{k>0} \sqrt{\frac{4 \pi}{k L}}
(a_k - a^{\dag}_k)} + h.c.\right)
\nonumber
\\
&+& \frac{J_{-}}{4 \pi a} \left(S^- K^{\dag}_{\downarrow} K_{\uparrow}
e^{\sum_{p>0} \sqrt{\frac{4 \pi}{k L}}
(a_k - a^{\dag}_k)} + h.c.\right)
\end{eqnarray}
where $\overline{v}_F = (v_{F,\uparrow}+v_{F,\downarrow})/2$ is the
average Fermi velocity,
$\delta v_F = (v_{F,\uparrow}-v_{F,\downarrow})/2$ is the mismatch
between Fermi velocities in different spin branches, $J_+ = (J_x+J_y)/2$ and
$J_-=(J_x-J_y)/2$. Notice that
because we are allowing for different Fermi velocities the charge
and spin degrees of freedom do not decouple from each other.
Moreover,
this Hamiltonian can be brought to a simpler form if one performs
a unitary transformation
\begin{eqnarray}
U = e^{S^z \sum_{k>0} \sqrt{\frac{\pi}{k L}}(a_k - a^{\dag}_k)}
\label{unitary}
\end{eqnarray}
which transforms the Hamiltonian to
\begin{eqnarray}
H' &=& U^{-1} H U = \overline{v}_F \sum_{p>0} p (a^{\dag}_p a_p + b^{\dag}_p b_p)
+ \delta v_F \sum_{p>0} p \left(b^{\dag}_p a_p+a^{\dag}_p b_p \right)
\nonumber
\\
&+& S^z \left[\left(J_z^R- \pi v_F \right) \sum_{k>0} \sqrt{\frac{k}{\pi L}}
(a_k + a^{\dag}_k) + \delta v_F \sum_{k>0} \sqrt{\frac{k}{\pi L}}
(b_k + b^{\dag}_k) \right]
\nonumber
\\
&+& \frac{J_{+}}{4 \pi a} \left(S^+ K^{\dag}_{\downarrow} K_{\uparrow}
+ S^- K^{\dag}_{\uparrow} K_{\downarrow} \right)
\nonumber
\\
&+& \frac{J_{-}}{4 \pi a} \left(S^- K^{\dag}_{\downarrow} K_{\uparrow}
e^{4 \sum_{p>0} \sqrt{\frac{\pi}{k L}}
(a_k - a^{\dag}_k)} + h.c.\right)
\, .
\end{eqnarray}
An important observation here
is that the unitary transformation {\it does not} affect $S^z$.
The anti-commutation factors can be rewritten in terms of spin
operators in which case we can rewrite \cite{twolevel}
\begin{eqnarray}
H' &=& \overline{v}_F \sum_{p>0} p (a^{\dag}_p a_p + b^{\dag}_p b_p) +
\delta v_F \sum_{p>0} p \left(b^{\dag}_p a_p+a^{\dag}_p b_p \right)
+ \frac{J_{+}}{2 \pi a} \sigma_x
\nonumber
\\
&+& \sigma_z \left[\left(J_z^R-\pi \overline{v}_F\right) \sum_{k>0} \sqrt{\frac{k}{\pi L}}
(a_k + a^{\dag}_k) + \delta v_F
\sum_{k>0} \sqrt{\frac{k}{\pi L}} (b_k + b^{\dag}_k) \right]
\nonumber
\\
&+&
\frac{J_{-}}{4 \pi a} \left(\sigma^+ e^{4 \sum_{p>0} \sqrt{\frac{\pi}{k L}}
(a_k - a^{\dag}_k)} + h.c.\right)
\label{2leve}
\end{eqnarray}
where $\sigma_x = (\sigma^++\sigma^-)/2$. Observe that (\ref{2leve})
describes the physics of a two level system coupled to a bosonic
environment \cite{twolevel}.

When $J^z\gg  2 v_F$ and $J_-=0$ (the limit of large uniaxial anisotropy) we
see from (\ref{jzr}) that $J^z_R \to \pi v_F$ ($\delta v_F=0$) and the Hamiltonian
reduces to
\begin{eqnarray}
H = \frac{J_{\perp}}{2 \pi a} \sigma_x +
v_F \sum_{p>0} p a^{\dag}_p a_p
\end{eqnarray}
with the decoupling of the spin degrees of freedom to the bosonic modes
($J_x=J_y=J_+=J_{\perp}$).
This is the dissipationless limit of the problem.
Observe that in this limit the eigenstates of the system are
eigenstates of $\sigma_x$, that is, the transverse field.

The physical interpretation in terms of the Kondo problem is also
very straightforward: in the limit of
$J^z\gg 2 v_F$ the electron forms a {\it virtual} bound state with
the localized spin. There are two states which are degenerate
for an antiferromagnetic coupling
\begin{eqnarray}
|+\rangle &=& |\Uparrow, \downarrow \rangle
\nonumber
\\
|-\rangle &=& |\Downarrow, \uparrow \rangle
\end{eqnarray}
which are the eigenstates of $\sigma^z$. This degeneracy is lifted
by the transverse field $\sigma^x$ and one ends up with two states
\begin{eqnarray}
|s\rangle &=& \frac{1}{\sqrt{2}} (|+\rangle - |-\rangle) = \frac{1}{\sqrt{2}} (|\Uparrow, \downarrow \rangle
- |\Downarrow, \uparrow \rangle)
\nonumber
\\
|t\rangle &=& \frac{1}{\sqrt{2}} (|+\rangle + |-\rangle) = \frac{1}{\sqrt{2}} (|\Uparrow, \downarrow \rangle
+ |\Downarrow, \uparrow \rangle)
\label{state1imp}
\end{eqnarray}
which are the singlet and triplet states.

Thus, we can define a tunneling splitting, $\Delta_0$, between the two magnetic states
and the coupling constant to
the bosonic bath, $\alpha$, which are given by
\begin{eqnarray}
\Delta_0 &=& \frac{J_{\perp}}{2 \pi a}
\nonumber
\\
\alpha &=& \left[1-\frac{J^z_R}{\pi v_F}\right]^2 =
\left[1-\frac{2}{\pi} \arctan\left(\frac{\pi N(0) J_z}{4}\right)\right]^2  \, .
\label{da}
\end{eqnarray}
For simplicity we will consider the case of uniaxial symmetry in which $J_-=0$
and $v_{F,\uparrow}=v_{F,\downarrow}$ so we
rewrite (\ref{2leve}) as
\begin{eqnarray}
H' = \Delta_0 \sigma_x - \pi v_F \sqrt{\alpha} \, \sigma_z \, \, \sum_{k>0} \sqrt{\frac{k}{\pi L}}
(a_k + a^{\dag}_k) + \sum_{k>0} v_F k a^{\dag}_k a_k \, .
\label{tls}
\end{eqnarray}
Observe that the magnetic moment flips at a rate given by $1/\Delta_0$. The bosons with
energy much larger than $\Delta_0$ (that is, with momentum close to $\Lambda$)
will follow adiabatically the motion of the spin and their effect is to renormalize the
fluctuation rate. The low energy bosons (that is, with $k\approx 0$)
are too slow and cannot follow the motion of the spin. The spin can
dissipate energy into this slow bosonic bath. In order to take into account
the effect of
the fast bosons into the motion of the spin we can
consider the effects of bosons
which live in a thin shell $\Lambda-d\Lambda<k<\Lambda$ and treat
their coupling to the spin in perturbation theory. It is simple to show
that in this case the bare tunneling splitting $\Delta_0$ is renormalized
to
\begin{eqnarray}
\Delta_R = E_c \left(\frac{\Delta_0}{E_c}\right)^{\frac{1}{1-\alpha}}
\label{dr}
\end{eqnarray}
where $\Delta_0 = \Delta(\Lambda_0)$ is the value of the tunneling at the bare value of
the coupling constant and $E_c \approx E_F$ is a high energy cut-off.
The above discussion is valid at zero temperature. At finite temperatures the renormalization
group flow has to stop at $max(T,\Delta_R)$ because after this point
there are no bosonic modes to renormalize anymore.
Thus, there is a crossover in the problem
when the temperature becomes of order $T \approx T_R = \Delta_R$ so that for temperatures
larger than $T_R$ the tunneling is suppressed and for temperatures smaller than $T_R$ the tunneling is
renormalized to $\Delta_R$. This crossover temperature can be associated with the {\it Kondo temperature}
as we discuss below.

It turns out that the physics of (\ref{tls}) is now understood and many physical
quantities can be computed exactly \cite{saleur}. For instance, it was found that,
\begin{eqnarray}
\langle \sigma_z(t) \rangle = \langle \sigma_z(0) \rangle \, \, e^{-\Gamma t} \cos(\Omega t)
\label{finalthing}
\end{eqnarray}
where
\begin{eqnarray}
\Gamma(\alpha,\Delta_0) &=& \frac{2 T_K}{ \pi} \sin^2\left(\frac{\pi \alpha}{2 (1-\alpha)}\right)
\nonumber
\\
\Omega(\alpha,\Delta_0) &=& \frac{T_K}{ \pi} \sin\left(\frac{\pi \alpha}{(1-\alpha)}\right) \Theta(1/2-\alpha)
\label{go}
\end{eqnarray}
where
\begin{eqnarray}
K_B T_K = \frac{T_R}{\alpha} = \frac{E_c}{\alpha} \left(\frac{\Delta_0}{E_c}\right)^{\frac{1}{1-\alpha}}
\label{tk}
\end{eqnarray}
is the actual Kondo temperature of the system which depends on a non-universal cut-off
energy scale $E_c$. Observe that the ratio
\begin{eqnarray}
\frac{\Omega}{\Gamma} = \cot\left(\frac{\pi \alpha}{2 (1-\alpha)}\right) \,
\label{ratio}
\end{eqnarray}
only depends on $\alpha$ and it is universal.
Moreover, it was shown that the zero temperature contribution of the magnetic moment to
the susceptibility is simply \cite{theo}
\begin{eqnarray}
\chi_{imp}(T=0) = \frac{1}{2 \pi \Delta_R} = \frac{1}{2 \pi \alpha T_K}
\label{chimp}
\end{eqnarray}
while the contribution to the specific heat is
\begin{eqnarray}
\frac{C_{imp}(T)}{T} = \frac{\pi}{3} \frac{\alpha}{\Delta_R} = \frac{\pi}{3}  \frac{1}{T_K} \, .
\label{cimpt}
\end{eqnarray}
The imaginary part of
the frequency dependent susceptibility is given by \cite{saleurunp}
\begin{eqnarray}
\Im\left[\chi_{imp}(\omega)\right] = -\frac{8 \Omega \Gamma}{\alpha} \sin\left(\frac{\pi \alpha}{2 (1-\alpha)}\right)
\, \frac{\omega}{\left(\omega^2-\Omega^2+\Gamma^2\right)^2 + 4 \Gamma^2 \Omega^2} \tanh(\frac{\beta \omega}{2})
\label{ichimp}
\end{eqnarray}
where $\beta = 1/T$. Notice that the Kramers-Kronig relation 
($\chi(0) = (1/\pi) \int d\omega'{\cal P}(1/\omega') \Im[\chi(\omega')]$), 
leads to (\ref{chimp}).
At high temperatures, $T\gg T_K$, we have, for instance,
\begin{eqnarray}
\chi_{imp}(T) &\approx& \frac{1}{4 T}
\nonumber
\\
C_{imp}(T) &\propto& \left(\frac{\Delta_R}{T}\right)^{2(1-\alpha)} \, .
\label{chicht}
\end{eqnarray}

The physics of the single impurity Kondo problem follows immediately
from the exact solution (\ref{finalthing}): the electron spin acts
as a transverse field on the local moment (see (\ref{tls})) which
flips (precesses) at a rate given by $\Omega$.
In the process of flipping the magnetic moment produces particle-hole
excitations in the Fermi sea which lead to an effective damping
of the flipping process which is given by $\Gamma$. Observe that
the flipping process is {\it underdamped} when $\alpha<1/3$
(since $\Omega(\alpha=1/3)=\Gamma(\alpha=1/3)$) and it is
{\it overdamped} when $1/2<\alpha<1/3$. Oscillations cease to
exist when $\alpha>1/2$. Going back to the Kondo problem we see
that the oscillations can be classified as: underdamped if
$N(0) J_z > 0.996$, overdamped if $0.63 > N(0) J_z > 0.996$
and damped if $N(0) J_z < 0.63$. Finally, just to make
connection with the SU(2) Kondo problem let us observe that
for $J_z,J_{\perp} \ll E_c$ the Kondo temperature looks very similar to
the SU(2) expression $T_K \approx E_c \exp\{-1/(N(0) J)\}$.
Indeed, from (\ref{tk}) we have
\begin{eqnarray}
T_K \approx E_c \exp\left\{-\frac{\ln[1/(N(0)J_{\perp})]}{N(0) J_z}\right\}
\label{tktls}
\end{eqnarray}
which is the form of the Kondo temperature for the anisotropic
Kondo problem. Observe that the Kondo temperature of an anisotropic
Kondo problem is not a single parameter quantity since it depends
on the Ising component $J_z$ and the XY component given by $J_{\perp}$.

\subsection{$2$ Magnetic Moments Kondo Effect}

The single impurity Kondo problem discussed in the last subsection
gives us a hint on how a cluster of coupled moments should behave,
that is, one would expect renormalizations of the tunneling
energy due to dressing of high energy particle-hole excitations
and dissipation due to the low energy particle-hole excitations.
It is however too simple because it does not contain the RKKY
coupling between spins. The next level of complexity is the two
impurity Kondo model \cite{tikm} which has the first features of a cluster
problem.

We consider the problem of two magnetic atoms at a distance ${\bf R}$
from each other interacting via an RKKY
interaction $\Gamma_{a,b} = \Gamma_a \delta_{a,b}$
in the presence of a metallic host,
that is, the so-called two impurity Kondo problem which is described by
(\ref{almosthere}):
\begin{eqnarray}
H_d &=& \sum_{{\bf k},\alpha} \epsilon_k c^{\dag}_{{\bf k},\alpha}
c_{{\bf k},\alpha} + \sum_{a} \Gamma^R_{a}(R) S_a({\bf R}/2)
S_a(- {\bf R}/2)
\nonumber
\\
&+& \sum_{{\bf k},{\bf k'},a}
J^R_{a} \left[e^{i({\bf k}-{\bf k'})\cdot {\bf R}/2}
c^{\dag}_{{\bf k},\alpha} \sigma^{a}_{\alpha,\beta}
c_{{\bf k'},\beta} S_a({\bf R}/2) + e^{-i({\bf k}-{\bf k'})\cdot {\bf R}/2}
c^{\dag}_{{\bf k},\alpha} \sigma^{a}_{\alpha,\beta}
c_{{\bf k'},\beta} S_a(-{\bf R}/2)\right]
\label{dimerhamil}
\end{eqnarray}
where the couplings are renormalized by the trace over high
energy degrees of freedom.

In order to study this problem we reduce it to an one-dimensional
problem by rewriting the electron operators as \cite{tikm}
\begin{eqnarray}
\psi_{j,\alpha}(k) = \frac{k}{\sqrt{2}} \int d\Omega
\left[\frac{1}{N_e(k)} \cos\left(\frac{{\bf k} \cdot {\bf R}}{2}\right)
+ i (-1)^j \frac{1}{N_o(k)} \sin\left(\frac{{\bf k} \cdot {\bf R}}{2}\right)
\right] c_{{\bf k},\alpha}
\label{psiexp}
\end{eqnarray}
where $j=1,2$ and $\Omega$ is the solid angle and
\begin{eqnarray}
N_{e(o)}(k) = \sqrt{1 \pm \frac{\sin(kR)}{kR}}
\end{eqnarray}
and $\{\psi^{\dag}_{j,\alpha}(k),\psi_{l,\beta}(k')\}=2 \pi \delta(k-k')
\delta_{\alpha,\beta} \delta_{j,l}$.
Furthermore, observe that all the momenta here are defined
in a thin shell around the Fermi surface. Thus, we can linearize the
band by writing $\epsilon_k = v_F(k_F-k)$ and rewrite the whole
Hamiltonian close to the Fermi surface as (see Appendix \ref{twoimp}):
\begin{eqnarray}
H &=& -i v_F \sum_{j,\alpha} \int dx \psi^{\dag}_{j,\alpha}(x)
\frac{\partial}{\partial x} \psi_{j,\alpha}(x) + \sum_{a} \Gamma^R_{a} S_{1,a} S_{2,a}
+ V \sum_{j,l,\alpha} \psi^{\dag}_{j,\alpha}(0) \tau^x_{j,l} \psi_{l,\alpha}(0)
\nonumber
\\
&+& \frac{v_F}{2} \sum_{a,j,l\alpha,\beta}
\left[J_{+,a}  \psi^{\dag}_{j,\alpha}(0)
\delta_{j,l} \sigma^a_{\alpha,\beta} \psi_{l,\beta}(0) S_{a} +
J_{m,a}  \psi^{\dag}_{j,\alpha}(0)
\tau^z_{j,l} \sigma^a_{\alpha,\beta} \psi_{l,\beta}(0) \delta S_{a}
\right.
\nonumber
\\
&+& \left. J_{-,a} \psi^{\dag}_{j,\alpha}(0)
\tau^x_{j,l} \sigma^a_{\alpha,\beta} \psi_{l,\beta}(0) S_{a}\right]
\label{1d2i}
\end{eqnarray}
where $\tau$ are Pauli matrices which act in the states $j=1,2$,
$J_{s,a}$ are exchange constants,
$\psi_{j,\alpha}(x)$ is the Fourier transform
of (\ref{psiexp}) and
\begin{eqnarray}
S_a &=& S_{1,a}+S_{2,a}
\nonumber
\\
\delta S_a &=& S_{1,a}-S_{2,a} \, .
\label{sds}
\end{eqnarray}
Moreover, we have explicitly introduced a new
coupling constant $V$ associated with impurity scattering and which breaks
the particle-hole symmetry.
This kind of term is unavoidable in any realistic model
of impurities and plays an important role in what follows.

Exactly as in the case of the single impurity given in (\ref{mandel}) we
bosonize the problem but take into account that besides $\uparrow$ and $\downarrow$
spin states we also have $j=1,2$. This leads to four types of bosonic fields, namely \cite{nfltikm},
\begin{eqnarray}
\phi_c(x) &=& (\phi_{1,\uparrow}(x) + \phi_{1,\downarrow}(x)  + \phi_{2,\uparrow}(x) + \phi_{2,\downarrow}(x))/2
\nonumber
\\
\phi_s(x) &=& (\phi_{1,\uparrow}(x) - \phi_{1,\downarrow}(x)  + \phi_{2,\uparrow}(x) - \phi_{2,\downarrow}(x))/2
\nonumber
\\
\phi_f(x) &=& (\phi_{1,\uparrow}(x) + \phi_{1,\downarrow}(x)  - \phi_{2,\uparrow}(x) - \phi_{2,\downarrow}(x))/2
\nonumber
\\
\phi_{sf}(x) &=& (\phi_{1,\uparrow}(x) - \phi_{1,\downarrow}(x)  - \phi_{2,\uparrow}(x) + \phi_{2,\downarrow}(x))/2
\label{newb}
\end{eqnarray}
which are associated with the {\it charge, spin, flavor and spin-flavor} currents of the problem.
And again, like in the one impurity Kondo problem, we perform a rotation in the spin space in order
to eliminate the spin currents. This can be accomplished with the unitary transformation
\begin{eqnarray}
U = e^{-i (S_{1,z}+S_{2,z}) \Phi_s(0)}
\end{eqnarray}
in which case the Hamiltonian becomes
\begin{eqnarray}
H &=& \frac{v_F}{2} \sum_{i=s,f,sf} \int dx \left[\Pi_i^2(x) + \left(\frac{\partial \phi_i}{\partial x}\right)^2
\right]  +  \tilde{\Gamma}_{z} S_{1,z} S_{2,z}
\nonumber
\\
&+& \Gamma_{\perp} (S_{1,x} S_{2,x} + S_{2,y} S_{1,y}) +
\frac{2 V}{\pi a} \cos(\Phi_{sf}(0)) \cos(\Phi_f(0)-\theta)
\nonumber
\\
&+&\frac{v_F \tilde{J}_{+,z}}{2 \pi} \frac{\partial \Phi_s(0)}{\partial x} S_{z}
+ \frac{v_F J_{+,\perp}}{\pi a} \cos(\Phi_{sf}(0)) S_{x}
\nonumber
\\
&+&\frac{v_F J_{m,z}}{2 \pi} \frac{\partial \Phi_{sf}(0)}{\partial x} \delta S_{z}
- \frac{v_F J_{m,\perp}}{\pi a} \sin(\Phi_{sf}(0)) \delta S_{y}
\nonumber
\\
&+& \frac{v_F J_{-,z}}{\pi a} \sin(\Phi_f(0)-\theta) \sin(\Phi_{sf}(0)) S_{z} +  \frac{v_F J_{-,\perp}}{\pi a} \cos(\Phi_f(0)-\theta) S_x
\label{hbosofinal}
\end{eqnarray}
where
\begin{eqnarray}
\Phi_{i}(x) &=& \sqrt{\pi} \left(\phi_{i}(x) - \int_{-\infty}^{x} dy \, \, \Pi_i(y)\right)
\nonumber
\\
\theta &=& \frac{1}{2} \int_{-\infty}^{+\infty} dx \, \,
\frac{\partial}{\partial x} (\Phi_{c}(x)-\Phi_{sf}(x))
\label{bigp}
\end{eqnarray}
and
\begin{eqnarray}
\tilde{\Gamma}_{z} &=& \Gamma^R_{z}-\frac{2 v_F}{\pi a} (J_{+,z}-\pi)
\nonumber
\\
\tilde{J}_{+,z} &=& J_{+,z} - 2 \pi
\label{reco}
\end{eqnarray}
are the new renormalized couplings.
Observe that (\ref{hbosofinal}) has strong similarities with
its one impurity counterpart (\ref{tls}) but the bosons have not decoupled
from
the transverse field terms because of the appearance of bosonic modes
$\Phi_{sf}$. The most studied case of this Hamiltonian is associated
with the NFL fixed point {\it which is not our main interest}
since it does not describe the cluster physics we are looking for.
The main problem with the NFL fixed point
is that it is unstable to the particle-hole symmetry breaking
operator $V$ defined in (\ref{hbosofinal}). This operator is
relevant under the renormalization group and drives
the system away from the NFL fixed point. In the presence of
scattering the system flows to a line of fixed points
which represent the different phase shifts.
This line of fixed points can be reached
in different ways as we discuss below

Let us consider the case
of $V \neq 0$.  As shown in refs. \cite{nfltikm} $V$ grows under the
RG and accordingly to (\ref{hbosofinal}) the value of the
$\Phi_{sf}$ and $\Phi_f$ fields {\it freeze} at values such
that
\begin{eqnarray}
\Phi_{sf}(0) = m \pi
\nonumber
\\
\Phi_f(0) = \pi n + \theta
\label{vneq}
\end{eqnarray}
where $m$ and $n$ are integers. By its definition in (\ref{bigp}) we see
that $\theta$ is purely topological and depends only on the
value of the fields at infinity, that is, depends on the boundary
conditions. Thus it is clear that $\theta$ is related to the
phase shift the electrons acquire by scattering from the
impurities. Therefore to each value of $\theta$ we have a different
fixed point and (\ref{vneq}) shows that in the absence of particle-hole
symmetry the system flows to a {\it line of fixed points} as
$\theta$ varies in between $0$ and $\pi/2$.
Using (\ref{vneq}) we find that (\ref{hbosofinal}) simplifies to:
\begin{eqnarray}
H &=&\tilde{\Gamma}_z S_{1,z} S_{2,z} + \Gamma_{\perp} \left(S_{1,x} S_{2,x} + S_{1,y} S_{2,y} \right) + \frac{v_F J_{\perp}}{\pi a} S_{x}
\nonumber
\\
&+& \frac{v_F}{2}  \sum_{i=s,sf} \int dx \left[\Pi_{i}^2(x) + \left(\frac{\partial \phi_{i}}{\partial x}\right)^2\right] +
\frac{v_F \tilde{J}_{+,z}}{2 \pi} \frac{\partial \Phi_{s}(0)}{\partial x} S_z
+ \frac{v_F J_{m,z}}{2 \pi} \frac{\partial \Phi_{sf}(0)}{\partial x} \delta S_{z}
\label{haf}
\end{eqnarray}
where $J_{\perp} = J_{+,\perp} + J_{-,\perp}$.
Notice that the flavor degree of freedom decouples in this limit.
This Hamiltonian has a form which is very similar to the two level
system (\ref{tls}) we have studied previously. The main difference is that
the spin operators for each impurity appear in a linear combination
and therefore are not naively related to the two level system problem.

The only trivial limit of the the Hamiltonian (\ref{haf}) is when
$J \gg \Gamma$ in which case the Kondo coupling has to be treated
first. The impurities decouple and we essentially have
two independent Kondo effects \cite{tikm}. We are interested in the
opposite limit since in the cluster the RKKY is stronger than the
Kondo effect.

In order to understand the
physics of Hamiltonian (\ref{haf}) one needs to consider the type of RKKY
interaction we are dealing with. Let us first consider the case where
$\tilde{J}_{+,z} = J_{m,z} =0$. The problem is purely magnetic since the bosons decouple from the
spins. We can diagonalize the magnetic Hamiltonian since
the it describes the simple problem of two interacting spins
in a transverse field. When $J_{\perp}=0$ the
system decouples into two triplets, $|\Uparrow,\Uparrow\rangle$ and $|\Downarrow,\Downarrow\rangle$
with $S_z =+1$ and $-1$,
respectively, and with energy $\tilde{\Gamma}_z/4$; a $S_z=0$ triplet,
$(|\Downarrow,\Uparrow\rangle+|\Uparrow,\Downarrow\rangle)/\sqrt{2}$, with energy
$-\tilde{\Gamma}_z/4 + \Gamma_{\perp}/4$; and a singlet
\begin{eqnarray}
|+,1\rangle = \frac{1}{\sqrt{2}} (|\Uparrow,\Downarrow\rangle
- |\Downarrow,\Uparrow\rangle) \, ,
\end{eqnarray}
with energy $-\tilde{\Gamma}_z/4 - \Gamma_{\perp}/4$.
On the one hand, the transverse field $J_{\perp}$
splits the degeneracy of the $|S_z|=1$ triplets down to
\begin{eqnarray}
|+,2 \rangle = \frac{1}{\sqrt{2}} (|\Downarrow,\Downarrow\rangle - |\Uparrow,\Uparrow\rangle)
\end{eqnarray}
with energy $\tilde{\Gamma}_z/4$ and (not normalized)
\begin{eqnarray}
|-,2\rangle = \frac{1}{\sqrt{2}} (|\Downarrow,\Downarrow\rangle + |\Uparrow,\Uparrow\rangle) +
\epsilon (|\Uparrow,\Downarrow\rangle - |\Downarrow,\Uparrow\rangle)
\end{eqnarray}
with energy
$\sqrt{(\tilde{\Gamma}_z/4-\Gamma_{\perp}/4)^2+\tilde{J}_{\perp}^2}+\Gamma_{\perp}/4$ where $\tilde{J}_{\perp} = v_F J_{\perp}/(2 \pi a)$.
The coefficient $\epsilon$ is a normalization coefficient which depends on the relation between
energy scales. On the other hand, while the singlet $|+,1\rangle$ is still an eigenstate
of the problem with energy $-\tilde{\Gamma}_z/4-\Gamma_{\perp}/4$ and
we get a new state
\begin{eqnarray}
|-,1\rangle = \frac{1}{\sqrt{2}} (|\Downarrow,\Uparrow\rangle-|\Uparrow,\Downarrow\rangle)
+ \epsilon (|\Uparrow,\Uparrow\rangle
+ |\Downarrow,\Downarrow\rangle)
\end{eqnarray}
with energy $-\sqrt{(\tilde{\Gamma}_z/4-\Gamma_{\perp}/4)^2+\tilde{J}_{m,\perp}^2}+\Gamma_{\perp}/4$. The splitting of the energy levels is shown in
the diagram Fig. \ref{splits}.

Let us consider first the case of antiferromagnetic coupling where 
$\tilde{\Gamma}_z>0$ and large ($\tilde{\Gamma}_z\gg \Gamma_{\perp},
\tilde{J}_{m,\perp}$). The lowest energy levels are
$|+,1\rangle$ and $|-,1\rangle$ which are separated in energy by $-\Gamma_{\perp}/2
+ 2 \tilde{J}_{\perp}^2/\tilde{\Gamma}_z \approx \tilde{J}_{\perp}^2/\tilde{\Gamma}_z$ (since
$\Gamma_{\perp} \approx J_{\perp}^2/E_F$ and $E_F \gg \tilde{\Gamma}_z$). Moreover, $\epsilon \approx \tilde{J}_{\perp}/\tilde{\Gamma}_z \ll 1$ and therefore
the main effect of the transverse field is to split the degeneracy
of the $|\Uparrow,\Downarrow\rangle$ and $|\Downarrow,\Uparrow\rangle$ states.
At low temperatures we just have to keep these two low lying states
(we always assume $T \ll  \tilde{\Gamma}_z$) and introduce
back the couplings $J_{+,z} = J_{m,z}$. We see from
(\ref{haf}) that $\Phi_s$ decouples from the spins since
\begin{eqnarray}
S_z |\sigma,-\sigma \rangle = (S_{1,z}+S_{2,z}) |\sigma,-\sigma\rangle = 0
\end{eqnarray}
while the coupling $J_{m,z}$
gets a renormalization of a factor of $2$ since
\begin{eqnarray}
\delta S_z |\sigma,-\sigma\rangle =  (S_{1,z}-S_{2,z}) |\sigma,-\sigma\rangle = 2 \sigma |\sigma,-\sigma\rangle
\end{eqnarray}
which just tells that the sub-Hilbert space is spanned by the states $|\Uparrow,\Downarrow\rangle$ and $|\Downarrow,\Uparrow\rangle$. The effective Hamiltonian can be written as
\begin{eqnarray}
H_{AF} = \frac{v_F}{2} \int dx \left[\Pi_{sf}^2(x) + \left(\frac{\partial \phi_{sf}}{\partial x}\right)^2\right] -
 2 \frac{v_F J_{m,z}}{2 \pi} \frac{\partial \Phi_{sf}(0)}{\partial x} \sigma_z
+ \frac{\tilde{J}_{\perp}^2}{\tilde{\Gamma}_z} \sigma_x
\label{afclus}
\end{eqnarray}
where
\begin{eqnarray}
\sigma_x |1,+\rangle &=& + |1,+\rangle
\nonumber
\\
\sigma_x |1,-\rangle &=& - |1,-\rangle
\nonumber
\\
\sigma_z |1,+\rangle &=& + |1,-\rangle
\nonumber
\\
\sigma_z |1,-\rangle &=& + |1,+\rangle \, .
\end{eqnarray}
Thus (\ref{afclus}) has essentially the same physics as the initial Hamiltonian and its physical meaning
is obvious, namely, it describes a dissipative two level system problem (\ref{tls}) where the two levels
represent the two states $|\Uparrow,\Downarrow\rangle$ and $|\Downarrow,\Uparrow\rangle$
of the magnetic moments, in order words, it describes a Kondo effect of the conduction band with the
antiferromagnetic cluster made out of two local moments. Observe that the heat bath for the
antiferromagnetic cluster is made out of the spin-flavor bosons. From the above Hamiltonian we
immediately conclude that the Kondo temperature of the antiferromagnetic cluster is given by
\begin{eqnarray}
T_K(J_{m,\perp},J_{m,z}) &=& \frac{E_c}{\alpha(J_{m,z})}
\left(\frac{v^2_F J_{\perp}^2}{4 \pi^2 a^2 \tilde{\Gamma}_z E_c}\right)^{\frac{1}{1-\alpha(J_{m,z})}}
\nonumber
\\
\alpha(J_{m,z}) &=& 4 \frac{J_{m,z}^2}{(2 \pi)^2} = (N(0) J^R_z)^2
\left[1-\left(\frac{\sin(k_F R)}{k_F R}\right)^2\right]\,
\label{aftk}
\end{eqnarray}
where we used (\ref{jexch}). Observe that we have extracted a factor of $4$
in front of $\alpha$ just to stress that $\alpha$ is coming from the
coupling of the staggered moment of two impurities and therefore
$\alpha$ scales like $2^2 \alpha$ since $\sqrt{\alpha} \to 2 \sqrt{\alpha}$
in contrast with the single impurity case.

In the ferromagnetic case the situation is reversed since the two lying states due
to the RKKY and transverse field are $|2,+\rangle$ and $|2,-\rangle$ which are split from each
other in energy by
$\sqrt{(\tilde{\Gamma}_z/4-\Gamma_{\perp}/4)^2 + \tilde{J}_{\perp}^2}+\Gamma_{\perp}/4-\tilde{\Gamma}_z/4 \approx \tilde{J}_{\perp}^2/\tilde{\Gamma}_z$. Moreover, we have
\begin{eqnarray}
S_z |\sigma,\sigma \rangle = (S_{1,z}+S_{2,z}) |\sigma,\sigma\rangle = 2 |\sigma,\sigma\rangle
\nonumber
\\
\delta S_z |\sigma,\sigma\rangle =  (S_{1,z}-S_{2,z}) |\sigma,\sigma\rangle = 0
\end{eqnarray}
and therefore from (\ref{haf}) the $\Phi_{sf}$ fields decouple in the ferromagnetic case and the effective
Hamiltonian in the sub-Hilbert space spanned by the triplet states  $|\Uparrow,\Uparrow\rangle$ and $|\Downarrow,\Downarrow\rangle$ reads,
\begin{eqnarray}
H_{F} = \frac{v_F}{2} \int dx \left[\Pi_{s}^2(x) + \left(\frac{\partial \phi_{s}}{\partial x}\right)^2\right] -
 2 \frac{v_F \tilde{J}_{+,z}}{2 \pi} \frac{\partial \Phi_{s}(0)}{\partial x} \sigma_z
+ \frac{\tilde{J}_{\perp}^2}{\tilde{\Gamma}_z} \sigma_x
\label{fclus}
\end{eqnarray}
where
\begin{eqnarray}
\sigma_x |2,+\rangle &=& + |2,+\rangle
\nonumber
\\
\sigma_x |2,-\rangle &=& - |2,-\rangle
\nonumber
\\
\sigma_z |2,+\rangle &=& + |2,-\rangle
\nonumber
\\
\sigma_z |2,-\rangle &=& + |2,+\rangle \, .
\end{eqnarray}
Notice again that (\ref{fclus}) describes a Kondo effect between the two low-lying states of a
ferromagnetic cluster where the heat bath is provided $\Phi_s$. The Kondo
temperature of the ferromagnetic cluster is given by
\begin{eqnarray}
T_K(J_{m,\perp},J_{+,z}) &=& \frac{E_c}{\alpha(J_{+,z})} \left(\frac{v^2_F J_{\perp}^2}{4 \pi^2 a^2 \tilde{\Gamma}_z E_c}\right)^{\frac{1}{1-\alpha(J_{+,z})}}
\nonumber
\\
\alpha(J_{+,z}) &=& 4 \frac{(\tilde{J}_{+,z})^2}{(2 \pi)^2} =
4 \left(1-\frac{N(0) J^R_z}{2}\right)^2
\label{ftk}
\end{eqnarray}
where we used (\ref{reco}) and (\ref{jexch}). Notice the close resemblance
of (\ref{ftk}) and the single impurity problem where $\alpha$ is given by
(\ref{da}). The only difference, like in the antiferromagnetic case, is
a number in front which is associated with the number of spins involved.
Indeed, the ferromagnetic problem is more directly related
to the usual Kondo effect than the antiferromagnetic case.
A direct comparison between the antiferromagnetic (\ref{aftk}) and
the ferromagnetic case (\ref{ftk}) reveals a basic difference between
the two problems: while dissipation scales like $(N(0) J_z)^2\ll 1$
in the antiferromagnetic case it scales like $1 - N(0) J_z \approx 1$
in the ferromagnetic case. Thus, for the same coupling constants the
ferromagnetic case has a stronger coupling to the bosonic bath.
The antiferromagnetic, on the other hand, is weakly coupled and therefore
dissipation is weaker. The same type of effect occurs in the problem
of tunneling of magnetic grains where dissipation is more important
for ferromagnetic than for antiferromagnetic granular systems \cite{stamp,nickolai,chud}.

The conclusion of the two-impurity Kondo problem is that the stable fixed points
of the system represent either a ferromagnetic or an antiferromagnetic cluster
undergoing a Kondo effect with the conduction band. The effective spin in this
case is associated with the low lying doublet generated by the RKKY interaction
which is split by the XY component of the Kondo coupling. Because the splitting of
the doublet requires the flip of two spins the transverse field scales like the $J_{\perp}^2$ in direct contrast with the single impurity case
where it is a linear function of the XY coupling. Moreover, the coupling to
the bath is also modified since the flip of two spins leads to the production
of more particle-hole excitations close to the Fermi surface. Although there
is no way to define long range order for the two impurity problem it is clear
that when the RKKY is dominant it is the ``order parameter'', magnetization
or staggered magnetization depending on the sign of the RKKY interaction, which
couples to the heat bath. This trend seems to be easily generalized to more
than two impurities, that is, although NFL fixed points are probably possible due
to specific symmetries of the problem, when symmetries are broken a simpler Kondo
effect of many moments is possible where the moments flip coherently in the
spin field generated by the Kondo coupling.

\subsection{$N$ Magnetic Moments - XYZ Magnetism}
\label{rkkytun}

Consider now the problem of $N$ magnetic moments forming a cluster
close to the QCP. If $N$ is large we can envisage this cluster as
a large magnetic grain. The ground
state of the grain is just the classical one, that is, it is the
fully ordered state for a ferromagnet ($|\Uparrow,\Uparrow,...,\Uparrow\rangle$
or $|\Downarrow,\Downarrow,...,\Downarrow\rangle$) or the Ne\'el state
for an antiferromagnet ($|\Uparrow,\Downarrow,\Uparrow,\Downarrow...\rangle$
or $|\Downarrow,\Uparrow,\Downarrow,\Uparrow...\rangle$).
Because of time reversal symmetry the ground state of a magnetic
system has to be at least double degenerate. Like in the examples of
the single impurity or two impurity the cluster can fluctuate quantum
mechanically between the two degenerate states in the absence of
an applied magnetic field which breaks explicitly the symmetry and
bias one of the configurations.

The tunneling between degenerate states can have many origins. In
the preceding sections we have discussed the tunneling due to the
XY coupling of the Kondo interaction to the magnetic moment. Another
more ``mundane'' source of interaction is the magnetic anisotropy
in XYZ magnets. This kind of anisotropy exists even in insulating
magnets and has to do with the interplay between spin-orbit and
crystal field effects. To understand
the origin of this anisotropy we can just look at the RKKY interaction
alone in (\ref{almosthere}). Observe that the RKKY interaction commutes
with ${\bf S}_T^2$ where ${\bf S}_T = \sum_{i=1}^N {\bf S}_i$ is the
total spin of the cluster. Thus, the eigenstates of the problem
can be classified accordingly to the eigenstates of ${\bf S}_T^2$,
that is, $S_T(S_T+1)$ where $S_T=0,1,2,...,N/2$ if $N$ is even or
$S_T=1/2,3/2,..,N/2$ if $N$ is odd. Because the cluster is in the
ordered state it will select $S_T$ accordingly to the interactions
($S_T=N/2$ for a ferromagnet and $S_T=0$ or $1/2$ for an antiferromagnet).
If the cluster is large it will behave like a
magnetic grain and the total spin operator  ${\bf S}_T$
behaves like a classical variable. Let us consider the simplest case
of a ferromagnetic cluster (the antiferromagnetic case is essentially
analogous) with $N$ atoms. Since the atoms are all locked together
we can describe their spins as classical variables:
\begin{eqnarray}
S_{z,i} &=& S_T \sin(\theta) \cos(\phi)
\nonumber
\\
S_{x,i} &=& S_T \sin(\theta) \sin(\phi)
\nonumber
\\
S_{y,i} &=& S_T \cos(\theta)
\label{szsxsy}
\end{eqnarray}
where $\theta$ is the angle the spins make with the $Y$ axis and
$\phi$ is the angle in the $X-Z$ plane.
The RKKY interaction between the spins is given by (\ref{almosthere})
with $\Gamma_{a,b}= \Gamma_z \delta_{a,b}$ with $|\Gamma_z|>|\Gamma_x|>|\Gamma_y|$
are the principal magnetic axis of the crystal. The
energy due to the RKKY interaction can be rewritten in terms of the angles as
\begin{eqnarray}
E(\theta,\phi) = N \left[ (\overline{\Gamma}_y-\overline{\Gamma_x}) \cos^2(\theta)
+ (\overline{\Gamma}_z - \overline{\Gamma}_x) \sin^2(\theta) \cos^2(\phi) + \overline{\Gamma}_x \right]
\label{etp}
\end{eqnarray}
where $\overline{\Gamma}_a = \sum_i \Gamma_a(i)$ is the average exchange
within the cluster. The energy (\ref{etp}) describes a magnet with
$Z$ easy axis and a $X-Z$ easy plane if
$0>\overline{\Gamma}_y > \overline{\Gamma_x}>\overline{\Gamma}_z$. The energy of the
cluster is minimized at $\phi=0,\pi$ and $\theta = \pi/2$ (that is,
all the spins point along the Z axis). It is usual to rewrite (\ref{etp}) as
\begin{eqnarray}
E(\theta,\phi) = N \left[K_{\perp} \cos^2(\theta) - K_{||} \sin^2(\theta)
\cos^2(\phi) \right]
\label{etpnice}
\end{eqnarray}
where $K_{\perp} = \overline{\Gamma}_y-\overline{\Gamma_x}>0$ and
$K_{||} =  \overline{\Gamma}_x - \overline{\Gamma}_z>0$ are the anisotropy energies
of the cluster (we have subtracted an unimportant constant from (\ref{etpnice})).

The dynamics of a cluster described by the energy (\ref{etpnice}) is
given by the well-known Landau-Lifshitz equations \cite{ll}:
\begin{eqnarray}
\frac{d {\bf S}_T}{d t} = -{\bf S}_T \times
\frac{\delta E}{\delta {\bf S}_T}
\end{eqnarray}
which, in terms of the angle variables, are
\begin{eqnarray}
\frac{d \phi}{d t} &=& - \frac{1}{S_T \sin \theta} \frac{\partial E}{\partial \theta}
\nonumber
\\
\frac{d \theta}{d t} &=& \frac{1}{S_T \sin \theta} \frac{\partial E}{
\partial \phi} \, .
\end{eqnarray}
Of course, such equations do not allow for any tunneling. To study tunneling
for such a magnetic grain we have to allow for solutions which are
not classical in nature. The simplest way to do it is by the path
integration method in imaginary time. The generating functional for the
magnetic grain can be written as \cite{chud}
\begin{eqnarray}
Z = \int d\mu[\phi] d\mu[\theta] e^{-{\cal S}_E}
\end{eqnarray}
where $S_E$ is the Euclidean action
\begin{eqnarray}
{\cal S}_E = \int d\tau \left\{i S_T [1 -\cos(\theta)] \frac{d \phi}{d \tau} + E(\theta,\phi)\right\} \, .
\label{setun}
\end{eqnarray}
The tunneling process is now described as an {\it instanton} solution
of the equations of motion for ${\cal S}_E$ which interpolate between the
minima of the potential. It can be shown
\cite{chud} that the tunneling energy is given by
\begin{eqnarray}
\Delta_A = \omega_0 |\cos(\pi S_T)| \exp\left\{- 2 S_T \ln\left[
\sqrt{1+K_{||}/K_{\perp}} + \sqrt{K_{||}/K_{\perp}}\right]\right\}
\label{deltaa}
\end{eqnarray}
where
\begin{eqnarray}
\omega_0 = 2 \sqrt{K_{||} (K_{||}+K_{\perp})}
\label{o0a}
\end{eqnarray}
is the attempt frequency of the cluster. Since $S_T \approx N/2$
we see that the tunneling splitting $\Delta$ can be written as
\begin{eqnarray}
\Delta = \omega_0 e^{-\gamma N}
\label{deln}
\end{eqnarray}
where
\begin{eqnarray}
\gamma &=& \ln\left[
\sqrt{1+K_{||}/K_{\perp}} + \sqrt{K_{||}/K_{\perp}}\right]
\nonumber
\\
&\approx& \frac{1}{2} \ln\left(\frac{K_{||}}{K_{\perp}}\right) \, .
\label{gammarkky}
\end{eqnarray}
Notice that the result (\ref{deln}) is what one would expect from
a WKB calculation for the tunneling splitting of $N$ atoms.
The factor of $\cos(\pi S_T)$
has to do with the Kramer's theorem: if $S_T$ is an integer (even number
of spins) the magnetization can tunnel and split the
degeneracy of the ground state; if $S_T$ is a half-integer (odd number of
electrons) tunneling is not allowed and the degeneracy of the ground state
remains. This term has its origin in the first term in the Euclidean
action (\ref{setun}) and it is topological in origin \cite{loss}.
Furthermore, the splitting is exponentially small in the number of spins
in the cluster since $S_T \propto N$ for a ferromagnet. Notice
that tunneling is only possible if $K_{\perp} \neq 0$ (that is,
$\overline{\Gamma}_x \neq \overline{\Gamma}_y$) which requires a magnetic
cluster with very low spin isotropy, that is, XYZ magnetism.
In an isotropic or uniaxial magnet (Heisenberg or XXZ, respectively) tunneling
is suppressed. As we show below, only the Kondo effect
can lead to tunneling.
We also observe
that at finite temperatures the system can be thermally activated
from one minimum to another. The process has the usual
Ahrenius factor $\exp\{-\beta N K_{||}\}$ related to the jump of
the system over an energy barrier of height $N K_{||}$. Comparing
this exponent with the one found in (\ref{deltaa}) we see that there
is a temperature $T_A$ above which thermal activation dominates
and below which quantum tunneling dominates. This temperature
is approximately given by
\begin{eqnarray}
T_A \approx \frac{N K_{||}}{2 S_T \ln\left[
\sqrt{1+K_{||}/K_{\perp}} + \sqrt{K_{||}/K_{\perp}}\right]} \, .
\label{ta}
\end{eqnarray}
All the arguments presented here can be easily generalized to the
case of antiferromagnetic grain \cite{chud}.

In a metallic substrate dissipation due to particle-hole excitations
in the conduction band plays an important role.
Dissipation can be introduced in the problem via (\ref{almosthere})
\cite{nickolai}. Since within the cluster the moments are locked together
we can rewrite (\ref{almosthere}) in Hamiltonian form as
\begin{eqnarray}
H = \Delta_A \sigma_x + \sum_{{\bf k},\alpha} \epsilon_k
c^{\dag}_{{\bf k},\alpha,0} c_{{\bf k},\alpha,0} +
\sum_{a,b} \overline{J}_{a,b} M_{a} \sigma_a \tau^b_{\alpha,\gamma}
\sum_{{\bf k},{\bf q}} F_{{\bf q}} c^{\dag}_{{\bf k}+{\bf q},\alpha,0}
c_{{\bf k},\gamma,0}
\label{nickhamil}
\end{eqnarray}
where $\overline{J}_{a,b}$ is the average exchange in the cluster,
$M_a$ is the order parameter ($S_a({\bf r}) = M_a \sigma_a \cos({\bf Q} \cdot {\bf r})$
where ${\bf Q}$ is the ordering vector in the cluster) and
\begin{eqnarray}
F_{{\bf q}} = \sum_{i=1}^N \cos({\bf Q} \cdot {\bf r}_i) e^{i {\bf q}
\cdot {\bf r}_i}
\label{fq}
\end{eqnarray}
is the form factor of the cluster.

The Hamiltonian (\ref{nickhamil}) describes the Kondo scattering
of the electrons by the cluster in the presence of a transverse field.
As shown in ref.\cite{nickolai} the transverse field is a relevant
perturbation in the problem and all the processes associated with
the XY component of the Kondo effect are irrelevant. In other words,
in the presence of tunneling due to the anisotropy in the RKKY interaction
the cluster flipping due to the Kondo effect does not play an important
role. We have seen, however, that the splitting due to the RKKY
interaction vanishes for Heisenberg or XXZ magnets and therefore the
Kondo scattering becomes relevant.
Thus, in the presence of $\Delta_A$ we just have to keep the
component of $M_a$ in the ordering direction (say, $z$).
Because the Kondo coupling does not play a role any longer it is possible
to apply perturbation theory to Hamiltonian (\ref{nickhamil}).
It is easy to see that (\ref{nickhamil}) maps
again into the dissipative
two level system (\ref{tls}) with $\Delta_0$ replaced by
$\Delta_A$ and the dissipative constant $\alpha$ is
given by a Fermi surface average \cite{nickolai}:
\begin{eqnarray}
\alpha &=& 2 (N(0)\overline{J}_z)^2 M_z^2 \int \frac{d{\bf r}}{V_c} \int \frac{d{\bf r'}}{V_c}
 \cos({\bf Q} \cdot {\bf r}) \, \cos({\bf Q} \cdot {\bf r'})
\left[\frac{\sin(k_F |{\bf r}-{\bf r'}|)}{k_F |{\bf r}-{\bf r'}|}\right]^2
\nonumber
\\
&=&  2 (N(0)\overline{J}_z)^2 M_z^2 \frac{1}{2 V_c} (\delta_{{\bf Q},0}+1)
\Re \left\{ \int d{\bf r} e^{i {\bf Q} \cdot {\bf r}}
\frac{\sin^2(k_F r)}{k^2_F r^2}  \right\}
\end{eqnarray}
where the integrals are performed in the volume $V_c$ of the cluster. Thus,
we introduce a cut-off term $e^{-r/L}$ (so that $V_c = 8 \pi L^3$)
and perform the integration exactly:
\begin{eqnarray}
\alpha &=& (N(0)\overline{J}_z)^2 M_z^2 \left[\frac{\delta_{{\bf Q},0}}{1+(2 k_F L)^2}
\right.
\nonumber
\\
&+& \left.
\frac{1}{4 L^3 k_F^2 Q} \left(\arctan(Q L) - \frac{\arctan[(Q+2 k_F)L]}{2}
-\frac{\arctan[(Q-2 k_F)L]}{2}\right)\right] \, .
\label{nickalpha}
\end{eqnarray}
The physical situation here is quite interesting. Although the XY coupling
of the Kondo effect does not play a role, the anisotropy of the RKKY
interaction introduces a transverse field. Thus, we can map back the
two level system problem to a pure Kondo effect of the cluster! The
only difference is that now in this ``fake'' Kondo effect the transverse
coupling (what we called $J_{\perp}$ in (\ref{da})) is related to $\Delta_A$.
The only difference from the original Kondo effect is that the transverse
component is proportional to the original exchange as $J^2$ instead of
$J$ as in the true Kondo effect. Thus, one has a cluster Kondo effect
with the RKKY coupling with a characteristic Kondo temperature
\begin{eqnarray}
T_K = \frac{E_F}{\alpha} \left(\frac{\Delta_A}{E_F}\right)^{\frac{1}{1-\alpha}} \, .
\label{fakeq}
\end{eqnarray}

Observe that for a ferromagnet ($Q=0$) the magnetization is $M_z = N/2$ and
in the large cluster size limit, $k_F L\gg 1$, we find \cite{nickolai}
\begin{eqnarray}
\alpha_F \approx \frac{1}{2 (3 \pi)^{2/3}}
\left(\frac{\rho_f}{\rho_e}\right)^{2/3} (N(0)\overline{J}_z)^2 N^{4/3}
\label{alphaf}
\end{eqnarray}
where we assumed the cluster to be homogeneous with $\rho_f = N/(8 \pi L^3)$
and $\rho_e$ is the electronic density given in (\ref{kf}).
Observe that dissipation
grows like a power law of the number of spins while the splitting
decreases exponentially with the number of spins.

In the antiferromagnetic case $Q \neq 0$, the {\it staggered}
magnetization is $M_z = N/2$, and from (\ref{nickalpha}) we find
that for large clusters ($Q L, 2 k_F L \gg 1$) we have
\begin{eqnarray}
\alpha_{AF} \approx (2 \pi N(0)\overline{J}_z)^2 \left(\frac{\rho_f}{k_F^2 Q}\right)
L^3 \left[\frac{\pi}{2} \Theta(2 k_F - Q) + \frac{(2 k_F)^2}{Q^2 - (2 k_F)^2}
\frac{1}{Q L}\right] + {\cal O}(1/L^3)
\label{alphaaf}
\end{eqnarray}
Thus, in the antiferromagnetic case $\alpha$ has a singularity at
$Q=2 k_F$ due to the Fermi surface effect. If $Q \leq 2 k_F$ the leading
order term is written
\begin{eqnarray}
\alpha_{AF} \approx \frac{\pi^2}{4} (N(0)\overline{J}_z)^2 \left(\frac{\rho_f}{k_F^2 Q}\right) N
\label{alphaqsmall}
\end{eqnarray}
and dissipation grows linearly with the number of atoms. Observe that
the dissipation is substantially smaller than in the
ferromagnetic case. For $Q>2 k_F$ we have
\begin{eqnarray}
\alpha_{AF} \approx 4 \pi^{4/3} (N(0)\overline{J}_z)^2 \left(\frac{\rho_f^{1/3}}{
Q}\right)^4 \frac{N^{2/3}}{1-(2 k_F/Q)^2}
\label{alphaqlarge}
\end{eqnarray}
which grows much slower with $N$ as the previous cases. From this
simple calculations we can conclude that the dissipation is much weaker
effect in antiferromagnetic clusters.

We can summarize these results as
\begin{eqnarray}
\alpha = \left(\frac{N}{N_c}\right)^{\varphi}
\label{alrkky}
\end{eqnarray}
where
\begin{eqnarray}
\varphi &=& 4/3
\nonumber
\\
N_c &=& \sqrt{6 \pi} \left(\frac{\rho_e}{\rho_f}\right)^{1/2}
\frac{1}{(N(0) J_z)^{3/2}}
\label{sumf}
\end{eqnarray}
for a ferromagnetic cluster,
\begin{eqnarray}
\varphi &=& 1
\nonumber
\\
N_c &=& \frac{4 k_F^2 Q}{\pi^2 \rho_f (N(0) J_z)^{2}}
\label{sumafqsmall}
\end{eqnarray}
for an antiferromagnetic cluster with $Q \leq 2 k_F$ and
\begin{eqnarray}
\varphi &=& 2/3
\nonumber
\\
N_c &=& \frac{(1-(2 k_F/Q)^2)^{3/2}}{8 \pi^2 (N(0) J_z)^{3}}
\left(\frac{Q}{\rho_f^{1/3}}\right)^6
\label{sumafqlarge}
\end{eqnarray}
for an antiferromagnetic cluster with $Q > 2 k_F$.
Notice that $N_c$
gives the critical number of spins in a given cluster
above which the Kondo effect ceases to occur because for $N>N_c$
we find $\alpha>1$ which corresponds to the {\it ferromagnetic}
Kondo coupling and therefore no Kondo effect. As we discussed before
this case is related to the cessation of tunneling and the freezing
of the cluster motion. Thus, the value of $N_c$ determines the
largest size of a cluster which can still tunnel in the presence
of a metallic environment when the anisotropy generated by the RKKY
interaction is the source of quantum tunneling.

We can estimate
the value of $N_c$ for typical values of the constants. We assume
$E_F \approx 10^4 K$, $\overline{J}_z \approx 300 K$ (
which corresponds to an ordering temperature of the order
$\overline{\Gamma}_z \approx \overline{J}^2_z/E_F \approx 10 K$). Moreover,
for U$^{+3.5}$Cu$^{+1}_4$Pd$^0$ we have $\rho_f \approx 1.2 \times
10^{-2} \AA^{-3}$ and $\rho_e\approx 8.7 \times 10^{-2} \AA^{-3}$
(which corresponds to $k_F \approx 1.4 \AA^{-1})$. Moreover, from
neutron scattering we have $Q \approx 0.8 \AA^{-1}$ \cite{ucupdn}.
In the ferromagnetic case we find $N_c \approx 2 \times 10^{3}$
atoms and for the antiferromagnetic case $N_c \approx 6 \times 10^{4}$
atoms! Thus, in the case of RKKY anisotropy a large number of
atoms can quantum tunnel at low temperatures. We are going to
see that the Kondo effect imposes much stronger restrictions on
these numbers.

\subsection{$N$ Magnetic Moments - XXZ and Heisenberg magnets}
\label{kondotun}

We describe in this section that when the system has XXZ or Heisenberg
symmetry ($\overline{\Gamma}_x = \overline{\Gamma}_y$)
the tunneling due to RKKY is suppressed. In this
case the only source of tunneling, as we have discussed for the two
impurity problem, is the Kondo effect itself. Again the XY
component of the Kondo effect acts as a local magnetic field that
flips the magnetic moment.
This Kondo effect, in perfect analogy with
the two-impurity Kondo problem, is due to the dissipative dynamics
of states, say, $|+\rangle = |\downarrow,\Uparrow;\downarrow,\Uparrow;...\rangle$
and $|-\rangle = |\uparrow,\Downarrow;\uparrow,\Downarrow;...\rangle$ in the
case of ferromagnetic coupling. The existence of quasi-degenerate low lying states
separated from higher energy states is guaranteed by the fact that the cluster
is effectively {\it within the ordered phase} and therefore states of the spins
can be related by a finite number of spin flips.
Now we can just
borrow the results from the previous sections for the response functions.
In particular, the cluster Kondo temperature is given by
\begin{eqnarray}
T_K(N) \approx \frac{E_F}{\alpha(N)} \left(\frac{\Delta_0(N)}{E_F}\right)^{1/(1-\alpha(N))}
\label{fakeqag}
\end{eqnarray}
where $\Delta_0(N)$ is the bare splitting between the two low lying states of the cluster
and $\alpha(N)$ is the dissipative coupling to the bath. Since $\Delta_0(N)$ is the splitting
between two states of the cluster where all the spins are flipped it has to scale like $(J_{\perp}/\Gamma^R_z)^N$
which corresponds to the energy required to flip $N$ spins. Furthermore, for a cluster
the dissipation comes from the fact that the ``order parameter'' flips and produces a
wake of particle-hole excitations close to the Fermi surface. It is clear that in this
case in (\ref{HN}) ``order parameter'' is an
extensive quantity which implies that $\alpha(N) \propto N^2$ in complete accordance
with the discussion of the two-impurity Kondo problem. Thus, we conclude that for a cluster
one has
\begin{eqnarray}
\Delta_0(N) \approx \Gamma^R_z \left(\frac{J_{\perp}}{\Gamma^R_z}\right)^N
= \Gamma^R_z e^{-N \ln(\Gamma^R_z/J_{\perp})}
\label{don}
\end{eqnarray}
which by direct comparison with (\ref{deln}) we find $\gamma \approx \ln(\Gamma_z^R/J_{\perp})$
and $ \omega_0 \approx \Gamma^R_z$. Notice from (\ref{reco}) that the
RKKY interaction is strongly renormalized and in general $\Gamma^R_z \gg J_{\perp}$.
Furthermore, for the antiferromagnetic case the coupling
of the cluster to the bath can be written as
\begin{eqnarray}
\alpha(N) \approx N^2 \left(\frac{J_z}{E_F}\right)^2 = \left(\frac{N}{N_c}\right)^2
\label{anaf}
\end{eqnarray}
while for a ferromagnetic cluster we expect (see (\ref{ftk}))
\begin{eqnarray}
\alpha(N) \approx N^2 \left(1-\frac{J_z}{E_F}\right)^2 =  \left(\frac{N}{N_c}\right)^2
\label{anf}
\end{eqnarray}
where $N_c \approx E_F/J_z$ ($\approx (1-\frac{J_z}{E_F})^{-1}$) for an
antiferromagnetic (ferromagnetic) cluster.
Observe that in a typical antiferromagnetic system $N_c \approx
100 - 1000$ spins while for a ferromagnetic cluster $N_c \approx 1 $.
We can immediately conclude that in the case of ferromagnetic clusters,
like in the case where tunneling is generated by anisotropy of the RKKY
interaction, dissipation will be so important that the cluster will freeze
at low temperatures. We expect that ordinary superparamagnetism
will occur - a situation much closer to a classical spin glass.
Moreover, observe that the Kondo temperature
drops exponentially with the number of spins in the cluster and therefore
for practical purposes
the Kondo temperature drops to vanishing small values before the
cluster size reaches $N_c$. Thus, the region where dissipation
occurs is rather narrow in the case of antiferromagnetic clusters.
The Kondo effect, therefore, imposes much stronger requirements in
the cluster size than the RKKY interaction. This is due to the fact
that the Kondo effect requires the collective flipping of the cluster
magnetization from electron scattering while the RKKY tunneling
is a product of lattice anisotropy. Which effect is the dominant one
in real systems
depends on the lattice structure and the spin-orbit coupling.
Furthermore, without dissipation the tunneling splitting vanishes
only when the number of spins in the cluster diverges. In the
presence of dissipation it vanishes for a finite number of spins given
by $N_c$.

\section{The Dissipative Quantum Droplet Model}

We have seen in the previous section that the problem of
a set of magnetic impurities interacting through a
conduction band has two main features: tunneling and
dissipation.  In a simple mean field like picture there are
three different energy scales in this problem: the ordering
temperature which scales like $T_c \propto N(0) J_z^2$,
the tunneling energy which is given by $ \Omega$ and
the damping energy which is given by $ \Gamma$ both
given in (\ref{go}).

In order to understand how these energy scales affect the Kondo lattice
consider a ligand system in the compositionally ordered phase. In this
case the exchange between atoms is small enough so that
the Kondo effect does not take place. As the system is
doped with a metallic atom with size smaller than the original
one the lattice {\it locally} contracts. The
local matrix elements have exponentially large values and
thus also the exchange $J_z$. Then a particular
region of the system can have an exchange parameter much larger than
the average exchange in the lattice. This is the situation
described by Hamiltonian (\ref{hmf}), for instance. If
the exchange is locally very large then in face of the
discussion of the single impurity Kondo problem it will
be given by a Hamiltonian like (\ref{tls}) where the
tunnelling splitting is replaced by $ \Omega$. Thus
in the limit of large magnetic anisotropy the effective
magnetic Hamiltonian which is generated from (\ref{almosthere})
is given by
\begin{eqnarray}
H_{eff} &=& \sum_{n,m} \Gamma_z({\bf r}_n-{\bf r}_m) S_z({\bf r}_n)
S_z({\bf r}_m) + \sum_n  \Omega({\bf r}_n) S_x({\bf r}_n)
\nonumber
\\
&+& \sum_{n,\alpha} S_z({\bf r}_n) \left(\lambda_{\alpha}({\bf r}_n) b_{\alpha}
+ \lambda^*_{\alpha}({\bf r}_n) b^{\dag}_{\alpha}\right)
+ \sum_{\alpha} \omega_{\alpha}  b^{\dag}_{\alpha} b_{\alpha}
\label{ansatz}
\end{eqnarray}
where $\alpha$ labels the relevant quantum numbers for the particle-hole
excitations and
$\lambda({\bf r}_n)$ is the local coupling constant of the spin
to the electronic bath. When $\lambda=0$ the bath decouples
from the spins and the problem is mapped into the transverse
field Ising model. We should stress at this point that
(\ref{ansatz}) can be only formally demonstrated for weak
magnetic moment concentration. For larger concentrations it
becomes an {\it ansatz} which has the correct features of
the problems we are discussing.

It is clear from the above that (\ref{ansatz}) will drive
the system through a quantum phase transition as $\Omega$ increases.
The simplest way to understand this effect is to rewrite the problem in
path integral form and use the Suzuki-Troter trick to map the $d$-dimensional
quantum problem (\ref{ansatz}) into a $d+1$ classical problem by breaking
the imaginary time direction into $N_{\tau}=\beta/\epsilon$ pieces.
The effective Hamiltonian associated with (\ref{ansatz}), after tracing out
the bosons degree of freedom, reads:
\begin{eqnarray}
H_e = \sum_{i=1}^{N_{\tau}} \sum_{n,m} \epsilon \Gamma_{n,m} S(i,n) S(i,n) +
\sum_{i=1}^{N_{\tau}} \sum_n J_n S(i,n) S(i+1,n)
+ \sum_{i \neq j=1}^{N_{\tau}} \sum_n \frac{\alpha_n}{8} \frac{S(i,n) S(j,n)}{(i-j)^2}
\label{he}
\end{eqnarray}
where $J_n = \ln[\tanh(\epsilon \Omega_n)]$ and $\alpha_n$ is the
dissipation constant of each spin (observe that in the limit of $\epsilon \to 0$ we have
$J_n \to \infty$, thus, the mapping into the classical problem is valid
when we take $\beta \to \infty$). The original problem
can now be thought as an anisotropic $d+1$ classical Ising model with
long range interactions in the imaginary time direction and short
range interactions in the space direction. In the
classical problem the sign of $\Gamma_n$ is irrelevant and for small $J_n$
the system will order magnetically (ferro, antiferro or spin-glass, depending
on $\Gamma_n$) in $d$ dimensions. Because the interactions are long-range in
the imaginary direction the system orders in this direction as well.
As in the classical case one assumes
that the lowest excitation energy above the classical configuration is a
droplet involving $N$ spins which can be reversed at some energy cost
$E(N)$ which scales with the size of the droplet as $N^{\theta}$.
Moreover, we will assume that the droplets are very diluted and do
not interact with each other in the spatial domain. It is clear
that the problem is equivalent to independent
droplets which are correlated in the imaginary time direction.
The effective action for this problem is obtained directly from (\ref{he})
by coarse-graining the spatial coordinate:
\begin{eqnarray}
H_D = \sum_{N} \left\{ \sum_{j=1}^{N_{\tau}}\left[ \epsilon E(N) S(\epsilon j)
+ K(N) S( \epsilon j) S(\epsilon(j+1)) \right] + \frac{\alpha(N)}{8}
\sum_{j \neq k}\frac{S(\epsilon j ) S(\epsilon k)}{(k-j)^2}  \right\}
\end{eqnarray}
where $S=-1$ corresponds to the ground state and $S=+1$ corresponds to the droplet
excitation present at some imaginary time. The problem of a droplet
at low energies is totally equivalent to a two level system problem.
This approach is an extension of the quantum droplet model proposed by
Thill and Huse \cite{thill} in the context of insulating magnets.
We should point out that the quantum droplet model gives essentially
the same results obtained numerically \cite{tfinum} and analytically
\cite{dfisher} for the random transverse field Ising model.

Notice that the energy barriers
involved in the imaginary time are given by $\Delta_0(N)$ which is the surface free energy
in $d+1$ dimensions.
It is obvious from the above
discussion that in real time the dynamics of the problem is described by
the Hamiltonian
\begin{eqnarray}
H = \sum_N \left(E(N) \sigma_z(N) + \Delta_0(N) \sigma_x(N) -\pi \sqrt{\alpha(N)}
\sigma_z(N) \sum_{k>0} \sqrt{\frac{k}{\pi L}} (b_k + b^{\dag}_k) +
\sum_{k>0} k b^{\dag}_k b_k \right)
\label{HN}
\end{eqnarray}
which is the analogue of the two level system Hamiltonian (\ref{tls}).
Observe that $\sigma_z$ is therefore related to magnetic order parameter of
the system and $\sigma_x$ is associated with the coherent flip of $N$ spins within
the droplet. In a ferro or antiferromagnetic state $E(N)$ is proportional
to the mean magnetic field of the spins which keep the droplet in the ordered
phase. In a spin glass this energy is distributed accordingly to some distribution
$P(E(N))$, say,
\begin{eqnarray}
P(E) = \frac{2}{\sqrt{\pi} \overline{\Gamma}} e^{-E^2/\overline{\Gamma}^2}
\end{eqnarray}
where $\overline{\Gamma} = \sqrt{\overline{\Gamma_{n,m}^2}}$ is the average
interaction within the droplet. Moreover, as it is well-known in
the context of the two-level system problem the bias $E(N)$ is a
relevant perturbation and will lead to ordering of the moments.
In the rest of the paper we will interested only in the
paramagnetic phase where $E(N)=0$ and the droplets are just
the clusters discussed in this paper.

The statistical problem is now very similar to the one discussed
in Section \ref{twolevel}. We have a distribution of
energy scales (or cluster Kondo temperatures) which have their origin
in a distribution of cluster sizes. Thus, contrary to the well-known
distribution of Kondo temperatures approach to NFL behavior \cite{ucupdnmr,miranda} the
effect is not of single impurity nature and the distribution is
{\it fixed} by percolation theory. Using (\ref{dr}), (\ref{don}) and
(\ref{anaf})
we see that the renormalized tunneling splitting is given by
\begin{eqnarray}
\Delta_R(N) = W e^{- \left(\frac{\gamma N + \ln(W/\omega_0)}{1-(N/N_c)^{\varphi}}\right)}
\label{drn}
\end{eqnarray}
where $W \approx E_F$. Usually it is not possible to invert the
$N$ as a function of $\Delta_R$ for a generic value of $\varphi$.
Thus, for practical purposes the averages over the distributions of
clusters have to be done with (\ref{Pn}) and  (\ref{fpx}):
\begin{eqnarray}
P(N) = \frac{N^{1-\theta} e^{- N/N_{\xi}}}{\Gamma(2-\theta) N_{\xi}^{2-\theta}} \, .
\label{pntheta}
\end{eqnarray}

We will consider first the example of the cluster Kondo effect
for $\varphi=2$ which serves as a good illustration of the effect
of dissipation in the cluster problem.
When $\varphi=2$, (\ref{drn}) can be inverted in order to give the
size of the cluster in terms of the splitting or fluctuation
time $\tau = 1/\Delta_R$ as
\begin{eqnarray}
\frac{N}{N_c} = \frac{1}{2 \ln(W/\Delta_R)} \left[\sqrt{(N_c \gamma)^2+ 4 \ln(W/\Delta_R) \ln(\omega_0/\Delta_R)}
- N_c \gamma\right] \, .
\label{ndr}
\end{eqnarray}
Notice that $N$ vanishes when $\Delta_R > \omega_0$ which corresponds
to the smallest cluster (that is, $N=1$ single Kondo impurity).
Therefore the probability of finding a cluster
with splitting $\Delta_R$ is given by
\begin{eqnarray}
P(\Delta_R) &=& P(N(\Delta_R)) \left|\frac{dN}{d\Delta_R}\right|
\nonumber
\\
&\propto&
\frac{\kappa_p N_c \left\{\ln(W/\Delta_R) \ln(W/\omega_0) + \frac{\gamma N_c}{2}
\left[\sqrt{(N_c \gamma)^2+ 4 \ln(W/\Delta_R) \ln(\omega_0/\Delta_R)}
- N_c \gamma\right] \right\}}{(1-e^{-N_c \kappa_p}) \Delta_R \ln^2(W/\Delta_R)
\sqrt{(N_c \gamma)^2+ 4 \ln(W/\Delta_R) \ln(\omega_0/\Delta_R)}}
\nonumber
\\
&\times& \exp\left\{-\kappa_p \frac{N_c}{2 \ln(W/\Delta_R)} \left[\sqrt{(N_c \gamma)^2+ 4 \ln(W/\Delta_R) \ln(\omega_0/\Delta_R)}
- N_c \gamma\right]\right\} \, .
\label{pdr}
\end{eqnarray}
Observe that when $N_c \to \infty$ we have
\begin{eqnarray}
P(\Delta) \propto \Delta^{\kappa_p/\gamma-1} \,
\label{pdelp}
\end{eqnarray}
which gives a power law distribution for the energy levels. If $\kappa_p/\gamma < 1$ the probability is divergent when $\Delta \to 0$.
As we are going
to see in the next section it is this power law that gives rise to quantum
Griffiths singularities.
However, for finite $N_c$ the power law is modified at very small
splittings. Indeed, when $\Delta_R \to 0$ we find:
\begin{eqnarray}
P(\Delta_R \to 0) \propto \frac{N_c \kappa_p (\ln(W/\omega_0)+ \gamma N_c)}{2(e^{N_c \kappa_p}-1)}
\frac{1}{\Delta_R \ln^2(\omega_0/\Delta_R)}
\label{psdr}
\end{eqnarray}
which, unlike (\ref{pdelp}), diverges logarithmically.
Thus, there is a crossover in the problem from the power law given by (\ref{pdelp}) for
$\Delta_R>\Delta^*$ to (\ref{psdr}) for $\Delta_R < \Delta^*$
where:
\begin{eqnarray}
\Delta^* = \sqrt{W \omega_0} \, \, \exp\left\{-\frac{1}{2} \sqrt{\ln^2(W/\omega_0)+ (\gamma N_c)^2}\right\} \, .
\label{dstar}
\end{eqnarray}
Thus, when $N_c \to \infty$ we have $\Delta^* \to 0$ and the power law behavior
dominates the entire range of energy scales.

\subsection{Magnetic properties: $T<T^*$}

Since we now have the distribution of energy scales it is possible
to calculate the physical properties of the clusters. For that we
need the response functions for individual clusters which are given
in Section \ref{twolevel}. For the Kondo effect there are no analytic
expressions for the physical properties which have to be usually calculated
numerically \cite{theo}. Here we will study only the asymptotic behavior.
The temperature crossovers,
however, can only be obtained numerically and in some special points
which are not of practical interest.

The magnetic susceptibility can be obtained directly from (\ref{chimp})
and (\ref{chicht}) and for $T \ll \Delta_R(N)$ is given by
\begin{eqnarray}
\chi(T,\Delta_R(N)) \approx \frac{1}{2 \pi \Delta_R(N)}
\label{lowchi}
\end{eqnarray}
while for $T \gg  \Delta_R(N)$ we have
\begin{eqnarray}
\chi(T,\Delta_R(N)) \approx \frac{1}{4 T} \, .
\label{highchi}
\end{eqnarray}
The average susceptibility is given by for $T < \omega_0$
\begin{eqnarray}
\overline{\chi}(T) &=& \int_0^{N_c} dN P(N) \chi(T,\Delta_R(N))
\nonumber
\\
&\approx& \int_0^{N^*(T)} dN \frac{P(N)}{2 \pi \Delta_R(N)} + \int_{N^*(T)}^{N_c} dN \frac{P(N)}{4 T}
\label{avechikd}
\end{eqnarray}
where $\Delta_R(N^*(T)) = T$ and we have approximated the integral by its two asymptotic pieces.
Since $\Delta_R \leq \omega_0$ in the high temperature limit
($T>\omega_0$) we have $\overline{\chi(T)} \approx 1/(4 T)$ and therefore one has the usual Curie behavior.

In what follows we will always assume that $T \ll \omega_0$ in which case $N^*(T) \approx N_c$
since $\Delta_R(N_c)=0$. Using (\ref{drn}) we easily find
\begin{eqnarray}
N^*(T) \approx N_c \left(1-\frac{\gamma N_c + \ln(W/\omega_0)}{\gamma N_c + \varphi \ln(\beta W)}\right) \, .
\label{nstar}
\end{eqnarray}
In this limit the second integral in (\ref{avechikd}) can be computed immediately
\begin{eqnarray}
\int_{N^*(T)}^{N_c} dN \frac{P(N)}{4 T} \approx \frac{
N^{2-\theta}_c
(\gamma N_c + \ln(W/\omega_0)) e^{-N_c/(\xi/a)^D}}{4 \Gamma(2-\theta) (\xi/a)^{2-\theta} T (\gamma N_c + \varphi \ln(\beta W))}
\label{avechi2}
\end{eqnarray}
and therefore for $T \ll T^* = W e^{-\gamma N_c/\varphi}$ we see that this term diverges like $1/(T \ln(W/T)$.
The first integral in (\ref{avechikd}) is more complicated but can be written in terms of exponential integrals
and we find that in the limit of $T \ll T^*$ it diverges like $1/(T \ln^2(W/T)$ and therefore is less
singular than the second term. Thus, we conclude that when $T \ll T^*$ the susceptibility diverges at zero
temperature as
\begin{eqnarray}
\overline{\chi}(T) \propto \frac{1}{T \ln(W/T)}
\label{ltavechi}
\end{eqnarray}
which has stronger divergence a the power law singularity
and it is independent on the value of $\varphi$.

In the opposite limit, that is $T \gg T^*$ we see that $N^*(T)$ is very small and given by
\begin{eqnarray}
N^*(T) \approx \frac{1}{\gamma} \ln(\beta \omega_0)
\end{eqnarray}
which is what we expect from the usual Griffiths-McCoy case. We conclude therefore that power
law behavior has to be observed at temperatures larger than $T^*$. In systems
where RKKY is the source of tunneling as we saw in
Subsection \ref{rkkytun} the value of $N_c$ is very
large and therefore $T^*$ is effectively zero. Thus power law behavior will
be predominant over all realistic low temperatures.
In systems where the Kondo effect is responsible
for tunneling, as described in Subsection \ref{kondotun} we can have $N_c \approx 10$
if $J_z \approx 1,000 \, K$ and
$\gamma \approx 2$ (for ordering temperatures of the order of
$\Gamma^R_z \approx 10 \, K$  and $J_{\perp} \approx 1 \, K$) and therefore
$T^* \approx 0.5 \, K$ which is a reasonable energy scale. Notice, however,
that $T^*$ is an exponential function of $N_c$ and therefore a change
to $J_z \approx 500 \, K$ (which is probably a better estimative
for the exchange in rare earths) leads to $T^* \approx 10^{-9} \, K$!

\subsection{Specific Heat: $T<T^*$}

Once again analytic expressions for the specific heat of a Kondo
system are not known. We will use the asymptotic forms given in
(\ref{cimpt}) and (\ref{chicht}). For $ T \ll  \Delta_R(N)$ we
have \cite{theo}
\begin{eqnarray}
\gamma(T,N) = \frac{C_V(T,N)}{T} \approx \frac{\pi}{3} \frac{\alpha(N)}{\Delta_R(N)}
\label{galow}
\end{eqnarray}
while for $ T \gg \Delta_R(N)$
\begin{eqnarray}
\gamma(T,N) \approx \left(\frac{\Delta_0}{E_c}\right)^2 \left(\frac{E_c}{T}\right)^{2 (1-\alpha(N))}
\label{gahigh}
\end{eqnarray}
where we used (\ref{dr}). Exactly as in the case of
the susceptibility we break the integral of the specific heat into
two pieces:
\begin{eqnarray}
\overline{\gamma}(T) \approx  \frac{\pi}{3}
\int_0^{N^*(T)} dN \frac{P(N) \alpha(N)}{\Delta_R(N)} +
\left(\frac{\Delta_0}{E_c}\right)^2 \int_{N^*(T)}^{N_c} dN P(N) \left(\frac{E_c}{T}\right)^{2 (1-\alpha(N))} \, .
\label{avega}
\end{eqnarray}
For $T \ll T^*$ the integrals can be evaluated as in the case of the
susceptibility. The main difference here is that the second integral in
(\ref{avega}) which gave the most divergent contribution in the case
of the susceptibility diverges only like $1/\ln(\beta W)$ because
$\alpha(N_c)=1$. Thus in the case of the specific heat the dominating
term is the first one which diverges like
\begin{eqnarray}
\overline{\gamma}(T) \propto \frac{1}{T \ln^2(W/T)}
\label{avegafinal}
\end{eqnarray}
which is a {\it weaker} divergence than the susceptibility. In the limit
of $T \gg T^*$ we recover the power law behavior characteristic of
Griffiths-McCoy singularities.
Other physical quantities can be directly calculated from the probability distribution
(\ref{pntheta}) and the dependence of $\Delta_R(N)$ on $N$ given in (\ref{drn}).

\section{$T>T^*$: quantum Griffiths singularities}
\label{clusternodissipation}

We have shown in the previous section that at temperatures smaller than
$T^*$ the cluster dynamics is essentially dissipative and the singularities
in the thermodynamic functions have logarithmic character. A reasonable
estimate of $T^*$ gives a very low temperature which in many systems is
experimentally not reachable. Thus, for most systems the physics in
a wide temperature range from $T^*$ to $T \approx \overline{\Gamma}$
where $\overline{\Gamma}$ is the average RKKY interaction in the system
the physics is non-dissipative. For temperatures above $\overline{\Gamma}$
the cluster does not exist as a well-defined object since temperature
fluctuations can excite isolated spins within the cluster.
In the temperature range $T^* < T < \overline{\Gamma}$ we can effectively treat the system as non-dissipative, that is, set
$\alpha=0$ (or equivalently, $N_c \to \infty$). In this crossover
temperature scale the physics is highly non-universal and depends strongly
from the distance from percolation threshold. Indeed, consider
the probability of finding a cluster with an energy splitting $\Delta$
which is given in (\ref{pdelp}).
Since there is a broad distribution of energy scales in disordered systems
it is often convenient to use a logarithmic scale so that
\begin{eqnarray}
P(\ln(\Delta)) \propto \Delta^{\lambda_p} \,
\label{plndel}
\end{eqnarray}
where
\begin{eqnarray}
\lambda_p = \frac{\kappa_p}{\gamma}
\label{lamb}
\end{eqnarray}
is an important exponent of the theory.
Equation (\ref{plndel}) is quite revealing if one considers the probability
of finding a spin in a cluster of size $N$ or excitation energy $\Delta$.
Since the probability is proportional to the size of the cluster one sees
that this probability is
\begin{eqnarray}
P_L(\ln(\Delta)) &\propto& N \Delta^{\kappa_p/\gamma} \propto L^D \Delta^{\kappa_p/\gamma}
\nonumber
\\
&\propto& (L \Delta^{1/Z_p})^D
\label{pldel}
\end{eqnarray}
where
\begin{eqnarray}
Z_p = \frac{D}{\lambda_p}
\label{Zp}
\end{eqnarray}
is the so-called dynamical exponent. Observe that $Z_p$ depends directly on $\lambda_p$
which is a non-universal quantity which depends on many microscopic details as
one can easily see from the definition (\ref{lamb}). Thus $Z_p$ varies
over the phase diagram. Indeed, using (\ref{kappap}) and (\ref{Zp})
\begin{eqnarray}
Z_{p\gg p_c} &\approx& \frac{\gamma D}{\ln(1/c)}
\nonumber
\\
Z_{p \approx p_c} &\approx& \gamma D (\xi(p)/a)^{D} \approx \frac{1}{|p-p_c|^{\nu D}} \to \infty
\label{Zpap}
\end{eqnarray}
which diverges at percolation threshold.
The physical meaning of the dynamical exponent
is clear from (\ref{pldel}): it defines the relationship between length and energy
scales in a correlated volume $L$ of the system.
The thermodynamic and response functions of the
system are directly related to $Z_p$.
Finally, we have to point out that the dynamical
exponent in the Griffiths phase has a different meaning than the one usually
used for clean systems \cite{hertz}
(thus we use the capital letter $Z_p$ for the dynamical
exponent in the Griffiths phase).
We should stress once again that the behavior of the system
for $T<T^*$ also corresponds to $Z \to \infty$ but unlike the
non-dissipative system $Z$ is divergent {\it away} from the
critical point since it is independent of $\kappa_p$ which vanishes
at $p=p_c$. Observe that even at this point (\ref{psdr}) is finite.

All the thermodynamic functions can be now calculated from the knowledge
of $\lambda_p$ or $Z_p$. Although we were able to find microscopic
expressions for $\gamma$ and $\omega_0$ the relationship between
$p$ and $Z_p$ depends on details of the percolation problem. Thus, another
way to face the problem here is to assume $\lambda_p$ as a phenomenological
parameter which can be varied as doping is varied and which vanishes
when $p \to p_c$. Thus, from now on we drop the subscript $p$ on $\lambda_p$.
In the rest of the section we give an explicit calculation for the
response functions in the crossover temperature range. Using
(\ref{pdelp}) and (\ref{deln}) we find that the normalized
distribution function (\ref{pdelp}) is given by
\begin{eqnarray}
P(\Delta) = \frac{[\ln(\omega_0/\Delta)]^{1-\theta}}{\Gamma[2-\theta]
(\gamma N_{\xi})^{2-\theta} \omega_0} \left(\frac{\Delta}{\omega_0}\right)^{\lambda-1} \Theta(\omega_0 - \Delta) \, .
\label{probpheno}
\end{eqnarray}
Notice that the Heavyside step function appears because the largest
value for the tunneling splitting is $\omega_0$.

The calculation of physical quantities is now rather simple because
we have just a two level system Hamiltonian to deal with. For generality
let us consider the problem of a magnetic field $H$ applied along the
easy axis of the cluster so that the effective cluster Hamiltonian
can be written as
\begin{eqnarray}
H_C = E_H \sigma_z + \Delta \sigma_x
\label{hclus}
\end{eqnarray}
where $E_H$ is the magnetic energy of the cluster. This magnetic energy
can be written as $E_H = m_c H$ where $m_c$ is the mean magnetic
moment of the clusters and it depends on the magnetic nature of
the cluster and the number of spins in the cluster.
For a ferromagnetic cluster we would have $m_c = g \mu_B S_T$. 
If the cluster is fully aligned
then $S_T=N/2$. For an antiferromagnetic cluster in a percolating lattice
we can have the following possibilities:
(1) the cluster has random fluctuations of the spins (uncompensated) in its
surface
in which case the net magnetization is proportional to $(N)^{1/D}$ (this
result can be obtained from the fact that the average size $R$ of cluster scales like
$R \propto N^{1/D}$ and therefore its area scales like $A \propto N^{2/D}$
leading to fluctuations or the order $\sqrt{A} \propto N^{1/D}$ as $N\gg 1$); (2) the cluster
has uncompensated spins in its volume so that the net magnetization proportional to
$(N)^{1/2}$.
In general we write $m_c = \mu_B q N^{\phi}$ where $q$
gives the magnitude of the average moment in the cluster (for a spin
glass it is the Edwards-Anderson parameter) and
$\phi =1 (1/2)$ for ferromagnets (antiferromagnets and spin glasses).
Thus, using (\ref{deln}) we can write
\begin{eqnarray}
E_H(\Delta) = q \mu_B \left[\frac{1}{\gamma} \ln\left(\frac{\omega_0}{\Delta}\right)\right]^{\phi} H \, .
\end{eqnarray} 
Notice, however, that the dependence of $E_H$ on $\Delta$ is rather weak and
since $\Delta$ is cut-off from above by $\omega_0$, $H$ or $T$, we can safely
replace the expression above by
\begin{eqnarray}
E_H(H,T) \approx q \mu_B \left[\frac{1}{\gamma} \ln\left(\frac{\omega_0}{
max(H,T)}\right)\right]^{\phi} H \, .
\end{eqnarray} 
In doing so we disregard higher logarithmic corrections in the expressions
given above.

The cluster magnetization is obtained from (\ref{hclus}):
\begin{eqnarray}
M(\Delta,H,T) = \frac{\mu_B^2 E_H}{\sqrt{E_H^2 + \Delta^2}}
\tanh \left(\beta \sqrt{E_H^2 + \Delta^2}\right) \, ,
\label{mdh}
\end{eqnarray}
the static magnetic susceptibility can obtained directly from (\ref{mdh}):
\begin{eqnarray}
\chi_0(\Delta,H,T) = \mu_B^3 \gamma \left[
\frac{\beta E_H}{E_H^2 + \Delta^2}
sech^2 \left(\beta \sqrt{E_H^2 + \Delta^2}\right)
+ \frac{\Delta^2}{(E_H^2 + \Delta^2)^{3/2}}
\tanh \left(\beta \sqrt{E_H^2 + \Delta^2}\right)\right] \, ,
\label{chi0}
\end{eqnarray}
the imaginary part of the frequency dependent susceptibility
for $\omega>0$ is (see Appendix (\ref{wsus})):
\begin{eqnarray}
\Im\left[\chi(\omega,\Delta,H,T)\right] = 2 \pi \frac{\Delta^2}{(E_H^2 + \Delta^2)}
\tanh\left(\beta \sqrt{E_H^2 + \Delta^2}\right)\delta(\omega-2 \sqrt{E_H^2 + \Delta^2}) \, ,
\label{ichiw}
\end{eqnarray}
and the specific heat is given by:
\begin{eqnarray}
C_V(\Delta,H,T) = \beta^2 (E_H^2 + \Delta^2) sech^2 \left( \beta \sqrt{E_H^2 + \Delta^2} \right) \, .
\label{cv}
\end{eqnarray}
Any physical constant can now be calculated directly from the expressions above and
the distribution of tunneling splittings given by (\ref{probpheno}).

\subsubsection{Magnetic Response}

From (\ref{mdh}) and (\ref{probpheno}) we see that the average magnetization
is given by
\begin{eqnarray}
\overline{M}(H,T) &\propto& H
 \int_0^{\omega_0} d\Delta  \frac{\Delta^{\lambda-1}}{\sqrt{E_H^2 + \Delta^2}}
\tanh \left(\beta \sqrt{E_H^2 + \Delta^2}\right) [\ln{\omega_0 \over \Delta}]^{1-\theta+\phi}
\nonumber
\\
&\propto& H \beta^{1-\lambda}
\int_{\beta E_H}^{\beta \sqrt{E_H^2+\omega_0^2}} dx
\frac{\tanh(x)}{\left(x^2-(\beta E_H)^2\right)^{1-\lambda/2}}
[\ln{\beta \omega_0\over \sqrt{x^2-(\beta E_H)^2}}]^{1-\theta+\phi} \, .
\label{avemag}
\end{eqnarray}
Observe that the magnetization has a scaling form
\begin{eqnarray}
\overline{M}(H,T) = \frac{H}{T^{1-\lambda}} f_{\lambda}\left(\frac{H}{T},\frac{T}{\omega_0}\right)
\label{mscal}
\end{eqnarray}
where $f_{\lambda}(x,y)$ is a simple scaling function.

Let us consider first the high temperature case ($T\gg  \omega_0, E_H$) in which case
the integral simplifies to
\begin{eqnarray}
\overline{M}\approx \frac{E_H}{T}
\label{hightm}
\end{eqnarray}
which is just the usual Curie behavior. Observe that the high temperature results have to be
understood with a little grain of salt since the cluster decomposes when the temperature
becomes of the order of the characteristic coupling between spins (which is of order of
ordering temperature of the pure system, $T_c$).
At high temperatures one has indeed paramagnetic behavior like the one described by
(\ref{hightm}) but the Curie constant can be quite different from the one we would get
from the cluster calculation since it comes from individual atoms. Thus, high temperature
behavior in the cluster picture only makes sense if $T_c\gg T\gg \omega_0$ which puts
a strong constraint on the temperature range.
Moreover, at high temperatures the clusters can be thermally activated
over the barrier and the two level system description is not complete.
We therefore focus on the case where $\omega_0\gg T, E_H$. Notice that in this
case the scaling function in (\ref{mscal}) depends only on the ratio of $H/T$.

Let us consider the case where $\omega_0\gg T, E_H$.
If $T\gg  E_H$ (low field limit) (\ref{avemag}) can be safely replaced by
\begin{eqnarray}
\overline{M} \propto \frac{(\ln{\beta \omega_0})^{1-\theta+\phi}}{T^{1-\lambda}} H\, .
\label{avemaglt}
\end{eqnarray}
Observe that the magnetization is linear in the field and diverges
at low temperatures as $T^{\lambda-1}$ for $\lambda<1$.
Thus, the scaling function is $f_{\lambda}(x,0) \approx 1$ when $x \to 0$.

On the other hand, if $\omega_0\gg E_H\gg T$ (high field limit) the integral has a different behavior,
namely,
\begin{eqnarray}
\overline{M} &\propto& H (\beta)^{1-\lambda}
\int_{\beta E_H}^{\infty} dx
\frac{\left[\ln{\beta \omega_0\over \sqrt{x^2-(\beta E_H)^2}}\right]^{
1-\theta+\phi}}
{\left(x^2-(\beta E_H)^2\right)^{1-\lambda/2}}
\nonumber
\\
&\propto& H^\lambda \left(\ln{\omega_0\over H}\right)^{1-\theta+\phi}
\label{avemaghf}
\end{eqnarray}
which shows that the magnetization scales like $H^{\lambda}$ and therefore the susceptibility
behaves like $H^{\lambda-1}$. Thus, the scaling function behaves like $f_{\lambda}(x,0)
\to x^{\lambda-1}$ when $x \to \infty$. The crossover from (\ref{avemaghf}) to (\ref{avemaglt})
occurs when $E_H \approx T$.

The imaginary part of the average frequency dependent susceptibility can be easily
calculated from (\ref{probpheno}) and (\ref{ichiw}) since the only contribution comes
from the Dirac delta function:
\begin{eqnarray}
Im[\chi(\omega,E_H,T)] \propto \left(\omega^2-(2 E_H)^2\right)^{\lambda/2}
\frac{\tanh\left(\frac{\beta \omega}{2}\right)}{\omega}
\left(\ln{2 \omega_0\over\sqrt{\omega^2-(2 E_H)^2}}\right)^{1-\theta+\phi}\Theta(\omega-2 E_H) \, .
\label{ichiwave}
\end{eqnarray}
This function can be measured in momentum integrated neutron scattering experiments since it is
the local response function. Observe that for $E_H=0$ it reduces to
\begin{eqnarray}
Im[\chi(\omega,0,T)] \propto
\frac{\tanh\left(\frac{\beta \omega}{2}\right)}{\omega^{1-\lambda}}
\left(\ln{2 \omega_0\over\omega}\right)^{1-\theta+\phi}
\label{finalichiw}
\end{eqnarray}
which is rather simple function. The fact that this function vanishes for $\omega<2 E_H$ has to do
with the blocking of the tunneling by a magnetic field which biases one
specific magnetic configuration. Observe that the application of a magnetic field suppresses the quantum
fluctuations and therefore destroys the NFL behavior caused by the clusters.

Another important quantity for the characterization of a Griffiths singularity is the nonlinear
magnetic susceptibility which can be calculated from (\ref{mscal}) as the third derivative
of the magnetization. Due to the scaling behavior it is obvious that the leading behavior is
\begin{eqnarray}
\chi_3(T) \propto \frac{1}{T^{3-\lambda}}
\label{chi3}
\end{eqnarray}
which therefore diverges more strongly than the linear susceptibility.

In certain experiments, like the Knight shift $K(T)$ measurement in
NMR, it is possible to observe the width of
the distribution of energy scales by studying the distribution of local susceptibilities
in the system. Indeed, we can write:
\begin{eqnarray}
\frac{\delta \chi(T)}{\chi} = D \frac{\delta K(T)}{K}
\end{eqnarray}
where $D$ is a constant which depends on the range of the interactions.
For long range interactions $D=1$ while for short range interactions
one has $D>1$. $\delta \chi(T) = \sqrt{\overline{\chi^2}(T) -
\overline{\chi}(T)^2}$ can be calculated directly from
(\ref{chi0}) and (\ref{probpheno})
\begin{eqnarray}
\frac{\delta \chi(T)}{\overline{\chi}} \propto \frac{1}{T^{\lambda/2}}
\end{eqnarray}
for $T \to 0$. This result shows that as the temperature is lowered
the width of the distribution of susceptibilities grows and eventually
diverges at $T=0$. In fact any moment of the susceptibility can
be calculated from (\ref{chi0}) and (\ref{probpheno}).

\subsubsection{Neutron Scattering}

Consider now the problem of scattering of neutrons by the magnetic
clusters discussed here. The differential cross-section for
neutron scattering is given by the usual expression
\begin{eqnarray}
\frac{d^2 \sigma}{d \Omega d\omega} = N \left(\frac{\gamma e^2}{m_N c^2}\right)^2
\frac{k'}{k} S({\bf q},\omega)
\end{eqnarray}
where ${\bf q}={\bf k}-{\bf k'}$ is the momentum transfer and
$\omega = ((k')^2-k^2)/(2 m_N)$ the energy transfer
in the scattering. $S({\bf q},\omega)$ is the dynamical form factor
which can be written as
\begin{eqnarray}
S({\bf q},\omega) = \frac{F^2({\bf q})}{N \pi} \sum_{i,j}
\int_{-\infty}^{+\infty} dt e^{-i (\omega t-{\bf q} \cdot({\bf r}_i-{\bf r}_j)}
\langle S_i(0) S_j(t) \rangle \, ,
\label{sqm}
\end{eqnarray}
where $F(\bf{q})$ is the nuclear form factor.
It is usual to rewrite the spin-spin correlation function as
$\langle S_i(0) S_j(t) \rangle = (\langle S_i(0) S_j(t) \rangle
- \langle S_i \rangle \langle S_j \rangle) + \langle S_i \rangle \langle S_j \rangle$ so that we can divide the dynamical form factor into a static
and a dynamic part: $S({\bf q},\omega) = S_S({\bf q}) \delta(\omega)
+ S_D({\bf q},\omega)$ where
\begin{eqnarray}
 S_S({\bf q}) = \frac{2 F^2({\bf q})}{N}\sum_{i,j}
e^{i{\bf q} \cdot({\bf r}_i-{\bf r}_j)} \langle S_i \rangle \langle S_j \rangle
\end{eqnarray}
gives the static response. We are only interested in the dynamic response
which can be rewritten in terms of the imaginary part of the susceptibility
as
\begin{eqnarray}
S_D({\bf q},\omega) = \frac{2 F^2({\bf q})}{N \pi}
\left[\frac{1}{1-e^{-\beta  \omega}}\right]
\Im\left\{\chi({\bf q},\omega)\right\} \, .
\label{sdqw}
\end{eqnarray}

In the previous subsections we have calculated the $q=0$ component of
$\Im\left\{\chi({\bf q},\omega)\right\}$ but for neutron scattering we
need a more detailed description. In order to do that we again consider
the clusters as independent scatters. Moreover, since the cluster
scatters as a whole we write for a cluster of $N$ atoms that
\begin{eqnarray}
\Im\left\{\chi_N({\bf q},\omega)\right\} = F^2_N(q)
\Im\left\{\chi(\omega,\Delta(N),T)\right\}
\label{ichin}
\end{eqnarray}
where $F^2_N(q)$ is the form factor of the cluster and $\Im\left\{\chi(\omega,\Delta(N),T)\right\}$ is given in (\ref{ichiw}). The main theoretical
problem is to calculate the cluster form factor since in a percolating
lattice clusters have different sizes and shapes.
We observe however, that if we are sufficiently close
to the ordering vector $q=Q$ we can write
\begin{eqnarray}
F_N(q) = e^{-R^2(N) [(q-Q)^2-Q^2]}
\label{fnq}
\end{eqnarray}
where $R(N)$ is the gyration radius of the cluster and is roughly given
by $N^{1/D} a$ (we normalized the form factor such that $F_N(q=0)=1$).
Substituting (\ref{fnq}) and (\ref{ichiw}) into
(\ref{ichin}) and averaging the result with (\ref{probpheno}) we
easily find
\begin{eqnarray}
\Im\left\{\chi({\bf q},\omega)\right\} \propto
\frac{\left[\ln(2 \omega_0/\omega)\right]^{1-\theta} \tanh(\beta \omega/2)}{
(\omega/\omega_0)^{1-\lambda}} \exp\left\{- \left[\ln(2 \omega_0/\omega)/\gamma \right]^{2/D}
[(q-Q)^2-Q^2] a^2 \right\}
\label{neutrons}
\end{eqnarray}
where $c$ is a constant of order of unit. Notice that for $q=0$ we obtain
(\ref{finalichiw}). Also notice that
$\int d{\bf q} \Im\left\{\chi({\bf q},\omega)\right\} \approx
\Im\left\{\chi({\bf q}=0,\omega)\right\}$ apart from logarithmic corrections.

A first direct consequence of the Griffiths phase is the
weak interplay between $q$ and $\omega$ which are essentially decoupled.
The scattering is peaked around the ordering wave-vector $Q$ and
the width of the peak is given by
\begin{eqnarray}
\delta q(\omega) \approx \frac{1}{a \left[\ln(2 \omega_0/\omega)/\gamma\right]^{1/D}}
\label{dqw}
\end{eqnarray}
which is independent of temperature and weakly dependent on the frequency.
Actually only measurements over many decades of frequency can detect
any significant changes in the width of the peak. Notice, moreover,
that the width of the peak does {\it not} depend on the distance from
the QCP which is given by $\lambda \propto \xi^D(p)$ and therefore
we predict the width of the peak to be essentially independent on
the concentration. All these results, however, have a limitation which
is the assumption of the gaussian form factor given in (\ref{fnq}).

\subsubsection{Specific Heat}

An important function in the context of NFL physics is the low temperature specific
heat which is given by
\begin{eqnarray}
\overline{C}_V(H,T) &\propto& \beta^2
\int_0^{\omega_0} d\Delta \Delta^{\lambda-1} (E_H^2 + \Delta^2)
sech^2 \left( \beta \sqrt{E_H^2 + \Delta^2} \right)
\left[\ln{\omega_0\over \Delta}\right]^{1-\theta}
\nonumber
\\
&\propto& \beta^{-\lambda}
\int_{\beta E_H}^{\beta \sqrt{E_H^2 + \omega_0^2}} dx \frac{x^3 sech^2(x)}
{\left(x^2-(\beta E_H)^2\right)^{1-\lambda/2}}
\left[\ln{\beta \omega_0\over \sqrt{x^2-(\beta E_H)^2}}\right]^{1-\theta}
\label{avecv}
\end{eqnarray}
which has the scaling form:
\begin{eqnarray}
\overline{C}_V(H,T) = T^{\lambda} g_{\lambda}\left(\frac{H}{T},\frac{T}{\omega_0}\right) \, .
\label{avecvsca}
\end{eqnarray}

Again we will consider only the case of $T, E_H  \ll  \omega_0$. At very low fields, $E_H \ll T$,
we see that (\ref{avecv}) reduces to:
\begin{eqnarray}
\overline{C}_V(H,T) \propto T^{\lambda} (\ln{\beta \omega_0})^{1-\theta} \,.
\label{dvlowt}
\end{eqnarray}
Observe that in this
case the specific heat coefficient diverges at low temperatures:
\begin{eqnarray}
\gamma(T) = \frac{\overline{C}_V(H,T)}{T}
\propto \frac{1}{T^{1-\lambda}} (\ln{\beta \omega_0})^{1-\theta}
\label{gcv}
\end{eqnarray}
with the same leading exponent as the susceptibility.
Notice that the scaling function is such that
$g_{\lambda}(x \to 0) \to 1$.

On the other hand at high fields, $E_H \gg  T$, we have as leading behavior:
\begin{eqnarray}
\overline{C}_V(H,T) &\propto& \frac{(\beta H)^{2+\lambda/2}}{(\beta \omega_0)^\lambda} e^{-\beta H}
\nonumber
\\
&\propto& \frac{H^{2+\lambda/2}}{T^{2-\lambda}} \exp\left\{-\frac{H}{T}\right\}
\label{avecvhf}
\end{eqnarray}
which has a Sckotty anomaly due to the field. All these results are summarized
 in Fig.\ref{sumclus}.
These results are all valid for temperatures above $T^*$. Below $T^*$
we expect the crossover discussed in the previous section and a deviation
from power law behavior.

\section{Conclusions}

The problem of the effect of disorder in magnetic phase transitions is probably
one of the most important problems of modern condensed matter. The intrinsic
complexity of this problem is related to the fact that at very low temperatures
close to a QCP statistical fluctuations due to disorder and quantum fluctuations
due to quantum mechanics are intrinsically linked to each other. This is
especially true for the alloys of U and Ce because of the existence of the Kondo
effect.

We have presented a unified picture of Kondo hole and ligand systems and the role
played by alloying in the physical properties of these systems. Starting from
the Kondo lattice model we have shown that it is possible to write down an
effective quantum action (\ref{almosthere}) which contains both the RKKY and the Kondo effect by
tracing energy degrees of freedom outside of a shell of momentum $\Lambda$ around
the Fermi surface. This treatment
allows for a direct comparison between the effects of RKKY and Kondo and reproduces,
at the Hamiltonian level, the famous Doniach argument. We have shown, however,
that alloying leads to a broad distribution of exchange couplings and therefore
to a very inhomogeneous environment. This should be contrasted with old approaches
to this problem where the concept of {\it chemical pressure} has been introduced.
Just briefly, this concept proposes that chemical substitution is equivalent to
applying pressure to the system: if one substitutes a smaller atom by a larger one
then the lattice expands and it is equivalent to negative pressure; if we substitute
a larger atom by a smaller one the lattice contracts and therefore it is equivalent
to positive pressure. The next step on the chemical pressure argument is to assert
that the exchange $J$ over the entire lattice changes uniformly under pressure
and the system can undergo a phase transition from an ordered state to a Kondo
compensated state as shown by $T_N$ in Fig.\ref{donifig}. From our point
of view chemical pressure and applied pressure have different
effects on the physics of these systems.

A simple example of the large difference between pressure and chemical
substitution
can be captured in specific heat measurements in CeCu$_6$ under pressure
\cite{fisher} and with substitution of Ce by Th in Ce$_{1-x}$Th$_x$Cu$_6$
\cite{kim}. From the experimental  data on CeCu$_6$ one finds that the inverse of $\gamma$
can be fitted by a linear relation with the applied pressure, $P$:
\begin{equation}
\frac{\gamma_0}{\gamma(P)} = 1+1.017 \frac{P}{P_{max}}
\label{gammap}
\end{equation}
where $\gamma_0 = 1.67$ J mole$^{-1}$ K$^{-2}$ is the zero pressure
result and $P_{max} = 8.8$ kbar is the maximum applied pressure in
the experiment.
Moreover, the unit cell volume in pure CeCu$_6$ at zero applied pressure
is ${\cal V}_0 = 420.6 \AA^3$ while at
maximum applied pressure, $P_{max}$, of $8.8$ kbar the volume
is $416.5 \AA^3$. Assuming a linear relation between the volume and
the pressure (a constant bulk modulus) one finds,
\begin{eqnarray}
\frac{{\cal V}(P)}{{\cal V}_0} = 1 - \beta \frac{P}{P_{max}}
\label{pressure}
\end{eqnarray}
where $\beta = 9.7 \times 10^{-3}$.
In the case of the doped samples it is experimentally
established that
pure ThCu$_6$ has a unit cell volume of $417.1 \AA^3$.
Since the contraction of the lattice seems to be homogeneous
and isotropic with doping
one finds that the volume as a function of
doping can be written as,
\begin{eqnarray}
\frac{{\cal V}(c)}{{\cal V}_0} = 1 - \alpha c
\label{doping}
\end{eqnarray}
where $\alpha = 8.3 \times 10^{-3}$. The main idea
behind the ``chemical pressure'' theory is to assert that (\ref{pressure}) and
(\ref{doping}) are equivalent to each other, that is, there is a one-to-one
correspondence between an applied pressure and doping the system.
If one assumes that this is indeed the case
we can find the correspondence between applied pressure and doping:
\begin{equation}
\frac{P}{P_{max}}= \frac{\alpha}{\beta} c.
\label{relation}
\end{equation}
Using (\ref{relation}) together with (\ref{gammap}) one finds that
the ``chemical pressure" theory yields to the following relation,
\begin{equation}
\left[\frac{\gamma_0}{\gamma(c)}\right]_P = 1 + \zeta_P c
\label{gammad}
\end{equation}
where $\zeta_{P} = 0.858$. Now one has
to compare (\ref{gammad}) to the actual dependence of $\gamma$
on doping. Since we are interested in the low doping regime
one uses the experimental data up to $c=0.4$ which can be well
fitted by a straight line of the form \cite{kim},
\begin{equation}
\frac{\gamma_0}{\gamma(c)} = 1 + \zeta_D c
\label{gammar}
\end{equation}
where $\zeta_{D} = 0.249$. We must stress that (\ref{gammar})
is obtained from an {\it extrapolation} of the data to the
zero temperature limit. A more careful look at the data, however,
seems to indicate a divergence of the specific heat exponent in
Ce$_{1-x}$Th$_x$Cu$_6$. We will however, insist using the extrapolation
just to check the identity between chemical pressure and real pressure.
We see the clear difference between the actual doped sample (\ref{gammar})
and the calculation based on the ``chemical pressure" argument, (\ref{gammad}),
for the same value of doping. We find that $\zeta_{P}/\zeta_{D} \approx 3.4$
and therefore a quite large difference between pressure and doping.
The conclusion is that there is more
than a change in the lattice constants or the volume of the unit cell
when a system is doped. Actually, from our arguments we actually
expect that the ration $\zeta_D/\zeta_P$ to be temperature
dependent!

In this paper we have proposed that the cut-off $\Lambda$ which determines
the energetic crossover between Kondo effect and RKKY coupling is directly
related to the number of compensated moments in the system which is given by 
(\ref{lkf}). Thus, $\Lambda$
has to be determined in a self-consistent way from the disorder and characteristic
energy scales in the problem. We argue that this scale is not determined by
the Kondo temperature alone but also by the RKKY coupling between moments.
Since the RKKY coupling involves states well deep into the Fermi surface
$\Lambda$ can be rather large and of the order of the Fermi momentum.
As a secondary consequence of this discussion we show that the exhaustion
paradox proposed by Nozi\`eres does not apply in the concentrated limit
of the Kondo lattice because it fails to take into account the high energy
degrees of freedom associated with the RKKY interaction.

Within the percolation theory of the Kondo lattice we show that close to
the QCP strong response from quantum fluctuation magnetic clusters gives
singular contributions at low temperatures which are known to be
Griffiths-McCoy singularities. We have presented a phenomenological
approach for this problem which allows the calculation of the various
physical quantities of interest such as magnetic susceptibilities (as
given in (\ref{ltavechi}) and (\ref{avemaglt})),
dynamical form factors for neutron scattering (given in (\ref{neutrons})) 
and thermodynamic quantities (such as the specific heat in (\ref{avegafinal}) 
and (\ref{dvlowt})).
The integrals involved are rather complicated and sometimes one has to
perform them numerically. Moreover, we have shown that when a magnetic
field is applied to the system there is a recovery of the ordinary
behavior due to the quenching of the clusters in the presence of
the magnetic field (see (\ref{avecvhf})). 
We have also discussed the various scalings in the
presence of a magnetic field which are given in (\ref{mscal}) and (\ref{avecvsca}).

We have also reviewed the single impurity Kondo problem and have shown
that in the anisotropic case, which is relevant for the case of
spin-orbit coupling, it can be interpreted as the dissipative dynamics
of a two level system in a transverse field. We have shown that while
the XY component of the Kondo coupling acts as a transverse field on
the magnetic spin, the Ising component gives rise to dissipation due
to the production of particle-hole excitations at the Fermi surface.
Using exact results for the dissipative two level system we have shown
that the physical quantities differ slightly from the non-dissipative
two level system by the presence of the parameter $\alpha$ that controls
the coupling of the two level system to the electronic heat bath.
We have also revisited the two impurity Kondo problem and have shown
that the stable fixed points are related to the coherent flipping motion
of the spins as a single degree of freedom (as shown in (\ref{afclus}) and
(\ref{fclus})). This is a clear sign of
cluster behavior, that is, coherent quantum mechanical motion.
In a cluster of $N$ atoms we have identified two mechanisms for
quantum tunneling: in XYZ magnets the anisotropy of the RKKY interaction
generated coherent cluster tunneling (with a cluster Kondo temperature
given in (\ref{fakeq})) while in XXZ or Heisenberg systems the Kondo effect
generates the tunneling (with a cluster Kondo temperature given in (\ref{fakeqag})). In both cases the tunneling splitting is
an exponential function of the number of atoms in the cluster as shown in
(\ref{deln}). This
calculation allows us to relate the microscopic quantities in
the Hamiltonian (\ref{almosthere}) and the phenomenological parameters
in Section \ref{clusternodissipation}. We have shown that there
is a cluster Kondo effect which is parameterized also by a dissipative
constant $\alpha(N) = (N/N_c)^{\varphi}$ where $\varphi$ is an exponent
which depends on the type of coupling with the particle-hole continuum
(see (\ref{alrkky}))
and $N_c$ (given in (\ref{sumf})-(\ref{sumafqlarge}) and (\ref{anaf})-(\ref{anf})) is the largest size of the cluster for which the Kondo
effect ceases to exist and the cluster are frozen. Thus, the Kondo
temperature of the cluster vanishes at $N_c$. In this respect
our picture is quite close to the Kondo disorder picture proposed
for the explanation of NFL in these systems \cite{ucupdnmr,miranda}.
We have to point out, however, that the microscopic origin of both
pictures is quite different since in our scenario the system has
to be close to the magnetic phase transition and the Kondo temperature
is defined for an extensive object which is determined by the laws
of percolation theory.

Based on these results we proposed a dissipative droplet model for
the magnetic behavior in disordered metallic magnets which is described by
the action (\ref{HN}). This picture
is the extension of the quantum droplet model proposed by Thill and
Huse \cite{thill} for the case of insulating magnets. We have shown
that in this model there is a crossover temperature
$T^* \approx E_F e^{-\gamma N_c/\varphi}$ above which the results
of the Griffiths-McCoy phase are still valid with power law behavior
in many physical quantities (see (\ref{dstar})). 
Below $T^*$ the behavior of the
physical quantities is modified by dissipation and
a new behavior emerges where the magnetic susceptibility and specific
heat diverge stronger than power law and do not scale together
as shown in (\ref{ltavechi}) and (\ref{avegafinal}).
Actually the magnetic susceptibility has a stronger divergence than the
specific heat coefficient. 
This could perhaps explain why the exponents $\lambda$ found from
magnetic susceptibility measurements are systematically smaller than the ones
found in the specific heat data \cite{marcio,hilbert}.
We have estimated, however, that $T^*$ is rather small. 
Thus, power laws as given in (\ref{avemaglt})
and (\ref{dvlowt}) should dominate the crossover behavior at low temperatures. 
We have calculated the thermodynamic properties of the clusters alone but
it is clear that in a real system one has to add also the contributions coming
from the paramagnetic environment (Fermi liquid or heavy Fermi liquid) which
are given by (\ref{xipauli}) and (\ref{cvfermi}), for instance. Since
the response functions we have calculated are singular as $T \to 0$ they
dominate the response at low enough temperatures.

Finally, we should mention that our picture is based
on two main features: the short range character of the disorder
and the complete decoupling of the clusters. On the one hand,
if the disorder
is long range correlated then stronger divergences of the physical
quantities are possible \cite{rieger}. On the other hand, if
the cluster are not decoupled but there is a residual
interaction between them it is possible that the system reaches
a spin-glass state at lower temperatures.
The nature of the spin glass state depends on the dissipation.
If the freezing temperature is of order of $T^*$ then the dissipation
will kill quantum coherence and a classical spin glass state is
formed. If $T^*$ is smaller than the freezing temperature then a
quantum spin glass ground state is possible \cite{thill}.
The existence of such phases can only be decided on experimental basis.

Thus, in this paper we put forward the theoretical basis for
Griffiths-McCoy singularities in the paramagnetic region close to
the QCP for magnetic order of U and Ce intermetallics. We show
that strong temperature dependence in the physical quantities is
expected at low temperatures due to the quantum mechanical response
of magnetic clusters. We believe that our picture is in agreement
with a large number of experiments in these systems \cite{miraflores,marcio,hilbert}.
It would be interesting to investigate further the existence of
a even lower temperature regime where dissipation plays a decisive
role in the physics of these systems.

We would like to acknowledge E.~Abrahams, I.~Affleck, M.~Aronson, 
G.~Castilla, T.~Costi, D.~Cox, I.~Dzyaloshinsky, 
R.~Egger, F.~Guinea, J.~Kaufman, 
H.~v.~L\"ohneysen, D.~MacLaughlin, B.~Maple, N.~Prokof'ev, R.~Ramazashvili,
H.~Rieger, H.~Saleur, S.~Senthil, G.~Stewart and A.~Tsvelik for 
illuminating comments and discussions. A.~H.~C.~N. acknowledges partial 
support from the Alfred P.~Sloan foundation
and a Los Alamos CULAR grant under the auspices of the U.~S.
Department of Energy.

\newpage

\appendix{}

\section{Perturbation Theory}
\label{pertrg}

Our starting point is (\ref{lsplit}) where we
integrate
out the electrons in regions $\Omega_1$ and $\Omega_2$ and obtain an effective action for the electrons residing in
region $\Omega_0$:
\begin{eqnarray}
Z= \int D{\bf S}(n,t) D\overline{\psi}_0({\bf r},t) D\psi_0({\bf r},t) e^{i S_{eff}({\bf S},\overline{\psi}_0,\psi_0)}
\end{eqnarray}
where
\begin{eqnarray}
S_{eff} = S_0 -i tr \ln(G^{-1}) -i \sum_{{\bf r},{\bf r'}} \int dt \int dt' \sum_{\alpha,\alpha'}
\sum_{\iota,\iota'} \overline{\eta}_{\alpha,\iota}({\bf r},t)
G^{\iota,\iota'}_{\alpha,\alpha'}({\bf r},t;{\bf r'},t')
\eta_{\alpha',\iota'}({\bf r'},t')
\label{seff}
\end{eqnarray}
where $S_0$ is (\ref{s0psi}) for the electrons in region $\Omega_0$ only. Furthermore,
\begin{eqnarray}
\eta_{\alpha,\iota}({\bf r},t) = \sum_{a,b}J_{a,b}({\bf r}) S^a({\bf r},t) \tau^b_{\alpha,\beta}
\psi_{\beta,0}({\bf r},t)
\label{etaai} \, ,
\end{eqnarray}
and
\begin{eqnarray}
\langle {\bf r},t,\iota,\alpha | G^{-1} |{\bf r'},t',\iota',\alpha) =
\left(i \frac{\partial}{\partial t} + \mu
+\epsilon_{\alpha}(-i \nabla) \right) \delta_{{\bf r},{\bf r'}} \delta(t-t') \delta_{\iota,\iota'}
\delta_{\alpha,\alpha'}
\nonumber
\\
+ \sigma^x_{\iota,\iota'}
\sum_{a,b} J_{a,b}({\bf r}) S^a({\bf r},t) \tau^b_{\alpha,\beta} \delta(t-t') \delta_{{\bf r},{\bf r'}}
\label{gm1}
\end{eqnarray}
where $\sigma^x_{\iota,\iota'}$ is a Pauli matrix which works in the subspace of states $\Omega_1$ and $\Omega_2$. Moreover,
\begin{eqnarray}
J_{a,b}({\bf r}) = \sum_n J_{a,b}(n) \delta_{{\bf r},{\bf r}_n} \, .
\end{eqnarray}

It is convenient to define the bare Green's function as the solution of the $J=0$
problem, that is,
\begin{eqnarray}
\left(i \frac{\partial}{\partial t} + \mu
+\epsilon_{\alpha}(-i \nabla) \right) G_{0,\alpha}({\bf r}-{\bf r'},t-t')
= \delta(t-t') \delta_{{\bf r},{\bf r'}} \, .
\label{goai}
\end{eqnarray}
Notice that the {\it time ordered} Green's function can be trivially obtained
from the above equation and reads,
\begin{eqnarray}
G_{0,\alpha}({\bf k},\omega) = \frac{1}{\omega+\mu-\epsilon_{\alpha}({\bf k})-
\frac{i}{\tau} sgn(\epsilon_{\alpha}-\mu)}
\end{eqnarray}
where we have introduced the electron relaxation time $\tau$ due to weak disorder.
Thus, in the regions of momentum space introduced in Section (\ref{uceinter}) we have
\begin{eqnarray}
G_{0,\alpha,1}({\bf k},\omega) =
\frac{\Theta(k_{F,\alpha}-\Lambda-k)}{\omega+\mu-\epsilon_{\alpha}({\bf k})+
\frac{i}{\tau}}
\nonumber
\\
G_{0,\alpha,2}({\bf k},\omega) = \frac{\Theta(k-k_{F,\alpha}-\Lambda)}{\omega+\mu-\epsilon_{\alpha}({\bf k})-
\frac{i}{\tau}} \, .
\label{g012}
\end{eqnarray}

Notice that the trace over the fast modes
has generated two kinds of terms in (\ref{seff}):
new interactions between the $\Omega_0$ electrons and the spins
(through $\eta$) and spin-spin interactions through $G$. In what follows we will consider
perturbation theory of $J$ for these fast modes.
First let us consider the new interaction terms between electrons and spins. Since $\eta$
is already order $J$ we can consider $G \approx G_0$ in (\ref{seff}) and rewrite
\begin{eqnarray}
I = \sum_{{\bf r},{\bf r'}} \int dt \int dt' \sum_{\alpha,\alpha'}
\sum_{\iota,\iota'} \overline{\eta}_{\alpha,\iota}({\bf r},t)
G^{\iota,\iota'}_{\alpha,\alpha'}({\bf r},t;{\bf r'},t')
\eta_{\alpha',\iota'}({\bf r'},t')
\nonumber
\end{eqnarray}
\begin{eqnarray}
I = \sum_{{\bf r},{\bf r'}} \int dt \int dt'
\sum_{\gamma,\delta}
\overline{\psi}_{\gamma,0}({\bf r},t)
\left[\sum_{a,b,c,d} \sum_{\alpha,\beta}\sum_{\iota,\iota'}
J_{a,b}({\bf r}) J_{c,d}({\bf r'}) S^a({\bf r},t) S^c({\bf r'},t')
\right.
\nonumber
\\
\left. \tau^b_{\alpha,\gamma} \tau^d_{\delta,\beta}
G^{\iota,\iota'}_{\alpha,\alpha'}({\bf r},t;{\bf r'},t')
\right]
\psi_{\delta,0}({\bf r'},t') \, .
\label{gen1}
\end{eqnarray}
Thus, in order to evaluate $I$ one needs an expression for $G$. From (\ref{gm1})
we have (symbolically),
\begin{eqnarray}
G^{-1} &=& G_0^{-1} + {\cal J} = G_0^{-1} \left(1 + G_0 {\cal J}\right)
\nonumber
\\
G &=& \left(1 + G_0 {\cal J}\right)^{-1} G_0
\nonumber
\\
&\approx& G_0 + G_0 {\cal J} G_0 + h.o.t.
\end{eqnarray}
where ${\cal J}$ is the second term in (\ref{gm1}).

To leading order one has,
\begin{eqnarray}
I &\approx& \sum_{{\bf r},{\bf r'}} \int dt \int dt'
\sum_{\gamma,\delta} \overline{\psi}_{\gamma,0}({\bf r},t) \psi_{\delta,0}({\bf r'},t')
\sum_{a,b,c,d} J_{a,b}({\bf r}) J_{c,d}({\bf r'})
S^a({\bf r},t) S^c({\bf r'},t')
\nonumber
\\
&\times& \sum_{\alpha} \tau^b_{\alpha,\gamma} \tau^d_{\delta,\alpha}
\left(\sum_{\iota} G_{0,\alpha,\iota}({\bf r}-{\bf r'},t-t') \right)
\label{int}
\end{eqnarray}
which is a electron spin-flip scattering process involving two different local
moments. Observe that this interaction is mediated by the single particle
Green's function $G_0$ which can be evaluated from (\ref{g012}). After a trivial frequency
integral we find
\begin{eqnarray}
\sum_{\iota} G_{0,\alpha,\iota}({\bf r},t) &=& \frac{i e^{-|t|/\tau}}{\rho_L}
\int \frac{d{\bf k}}{(2 \pi)^3} e^{- i ({\bf k} \cdot {\bf r}-(\epsilon_{{\bf k}}-\mu)t)}
\left[\Theta(k-k_{F,\alpha}-\Lambda)\Theta(t)
\right.
\nonumber
\\
&-& \left. \Theta(k_{F,\alpha}-\Lambda-k) \Theta(-t)\right]
\end{eqnarray}
where $\rho_L=N/V$ is the lattice density. Since we are only interested in the
asymptotic behavior of the system we observe that when $t \to \infty$ the integral
is dominated by the
saddle point
at $\mu=\epsilon_{\alpha}(k_{F,\alpha})$ but the $\Theta$ functions are only finite outside
this region, thus, the above integral has very fast oscillations given null contribution.
In order to see this result explicitly let us expand around the saddle point, that is,
we change variables:
\begin{eqnarray}
{\bf k}= k_{F,\alpha} + q \, \, {\bf u}_{F,\alpha}
\end{eqnarray}
where ${\bf u}_{F,\alpha}$ is a vector perpendicular to the Fermi surface (that is,
in the direction of the Fermi velocity). Thus we can write
\begin{eqnarray}
\epsilon_{{\bf k}} &\approx& \mu + v_{F,\alpha} q
\nonumber
\\
k &\approx& k_F + q
\end{eqnarray}
and the integral above becomes
\begin{eqnarray}
\sum_{\iota} G_{0,\alpha,\iota}({\bf r},t) &\approx& \frac{i e^{-|t|/\tau}}{\rho_L}
\sum_{{\bf k}_{F,\alpha}} \frac{k_{F,\alpha}^2}{(2 \pi)^3} \int_{-\infty}^{+\infty} dq
e^{- i [({\bf k}_{F,\alpha} + q {\bf u}_{F,\alpha})\cdot {\bf r} - q v_{F,\alpha} t ]}
\left[\Theta(q-\Lambda) \theta(t)
\right.
\nonumber
\\
&-& \left. \Theta(-\Lambda-q) \Theta(-t)\right] e^{-\nu |q|}
\end{eqnarray}
where we introduced an infinitesimal converging factor $\nu \to 0$. The integrals
are now trivial and give
\begin{eqnarray}
\sum_{\iota} G_{0,\alpha,\iota}({\bf r},t) &\approx& \sum_{{\bf k}_{F,\alpha}}
\frac{k_{F,\alpha}^2}{8 \pi^3 \rho_L} e^{i {\bf k}_{F,\alpha} \cdot {\bf r}}
\frac{e^{-i \Lambda ({\bf u}_{F,\alpha} \cdot {\bf r}- v_{F,\alpha} t) sgn(t)}}{
{\bf u}_{F,\alpha} \cdot {\bf r}- v_{F,\alpha} t - i \nu sgn(t)}
\end{eqnarray}
which shows that at long times the Green's function oscillates with frequency
$v_F \Lambda$ and therefore averages out to zero.

We now look at the the first term in (\ref{seff}) and write
\begin{eqnarray}
tr\ln(G^{-1}_0 + {\cal J}) &=& tr \ln [G^{-1}_0 (1+G_0  {\cal J})] =
tr \ln G^{-1}_0 + tr \ln (1+G_0  {\cal J})
\nonumber
\\
&\approx& tr\ln G^{-1}_0 + tr(G_0 {\cal J}) - \frac{1}{2} tr(G_0 {\cal J} G_0 {\cal J})
+ h.o.t.
\label{gexp}
\end{eqnarray}
In order to simplify the notation we will use a simple
dummy index $i=({\bf r},t,\alpha,\iota)$ for the matrix elements.
Observe that $tr(G_0 {\cal J}) = \sum_{i,j}
G_{0,i} \delta_{i,j} {\cal J}_{i,j} = \sum_i G_{0,i} {\cal J}_{i,i} = 0$ because
$\sigma^x_{\iota,\iota}=0$ in (\ref{gm1}). The second order term in (\ref{gexp}) becomes
\begin{eqnarray}
tr(G_0 {\cal J} G_0 {\cal J}) &=& \sum_{i,j,k,l} G_{0,i} \delta_{i,j} {\cal J}_{j,k}
G_{0,k} \delta_{k,l} {\cal J}_{k,i}
\nonumber
\\
&=& \sum_{i,k} \sum_{\alpha,\gamma} G_{0,i} {\cal J}_{i,k} G_{0,k} {\cal J}_{k,i} \, .
\end{eqnarray}
Using (\ref{gm1}) we find explicitly
\begin{eqnarray}
tr(G_0 {\cal J} G_0 {\cal J}) = \int d{\bf r} \int d{\bf r'} \int dt \int dt'
\sum_{a,b,c,d} J_{a,b}({\bf r}) J_{c,d}({\bf r'}) S^a({\bf r},t) S^c({\bf r'},t')
\Pi_{b,d}({\bf r}-{\bf r'},t-t')
\label{rkkyns}
\end{eqnarray}
where
\begin{eqnarray}
\Pi_{b,d}({\bf r},t) = 2
\sum_{\iota,\iota'} \sum_{\alpha,\gamma} \tau^b_{\alpha,\gamma} \tau^d_{\gamma,\alpha}
G_{0,\alpha,i}({\bf r},t) \sigma^x_{\iota,\iota'} G_{0,\gamma,\iota'}(-{\bf r},-t)
\label{pibdrs}
\end{eqnarray}
is the polarization function of the electron band. It
can be rewritten in momentum space as
\begin{eqnarray}
\Pi_{b,d}({\bf p},\Omega) = \frac{1}{2} \sum_{\alpha,\gamma} \sum_{\iota,\iota'}
\tau^b_{\alpha,\gamma} \tau^d_{\gamma,\alpha} \sigma^x_{\iota,\iota'}
 \int \frac{d\omega}{2 \pi} \int \frac{d{\bf k}}{(2 \pi)^d}
G_{0,\alpha,\iota}({\bf k},\omega) G_{0,\gamma,\iota'}({\bf k}+{\bf p},\omega+\Omega)
\label{pibdms}
\end{eqnarray}
which can be evaluated once the conduction band is known. Thus, to second order
the integration of the fast degree of freedom generates an interaction among
the spins in different sites. One further approximation which is usually
used is to take the static limit in (\ref{rkkyns}). This approximation is
good if one looks at times scales which are much larger than the scattering
time of the electrons between moments which is $\ell/v_{F,\alpha}$ where $\ell$
is the distance between moments and $v_{F,\alpha}$ is the Fermi velocity for
the $\alpha$ branch. We replace
$\Pi_{b,d}({\bf p},\omega)$ by $\Re[\Pi_{b,d}({\bf p},0)]$
(since $\Im[\Pi_{b,d}({\bf p},0)]=0$). In this limit the second order contribution is the RKKY interaction
between localized spins (which depends on the cut-off $\Lambda$)
which can be written as
\begin{eqnarray}
H_{RKKY}(\Lambda) = \sum_{m,n} \sum_{a,b} \Gamma_{a,b}({\bf r}_n-{\bf r}_m,\Lambda)
S^a({\bf r}_n) S^b({\bf r}_m)
\label{hrrkyl}
\end{eqnarray}
where
\begin{eqnarray}
\Gamma_{a,b}({\bf r},\Lambda) =
- \sum_{c,d} J_{a,c}  J_{b,d} \int d{\bf q} e^{i {\bf q} \cdot {\bf r}}
\Re[\Pi_{c,d}({\bf q},\Omega=0,\Lambda)] \, .
\label{exrkky}
\end{eqnarray}
The integral in (\ref{exrkky}) can be written directly from (\ref{pibdms}) and reads
\begin{eqnarray}
&\sum_{\iota,\iota'}& \sigma^x_{\iota,\iota'}  \int d{\bf k'}
e^{i ({\bf k}-{\bf k'}) \cdot {\bf r}}
\int \frac{d\omega}{2 \pi} \int \frac{d{\bf k}}{(2 \pi)^d}
G_{0,\alpha,\iota}({\bf k},\omega) G_{0,\gamma,\iota'}({\bf k'},\omega) =
\nonumber
\\
&=& \int d{\bf k'} \int \frac{d{\bf k}}{(2 \pi)^d} e^{i ({\bf k}-{\bf k'}) \cdot {\bf r}}
\frac{d\omega}{2 \pi} \left(
\frac{\Theta(k_{F,\alpha}-\Lambda-k) \Theta(k'-k_{F,\gamma}-\Lambda)}{
(\omega+\mu-\epsilon_{\alpha}(k)-i/\tau) (\omega+\mu-\epsilon_{\gamma}(k')+i/\tau)} \right.
\nonumber
\\
&+&  \left. \frac{\Theta(k-k_{F,\alpha}-\Lambda) \Theta(k_{F,\gamma}-\Lambda-k')}{
(\omega+\mu-\epsilon_{\alpha}(k)+i/\tau) (\omega+\mu-\epsilon_{\gamma}(k')-i/\tau)}\right)
\nonumber
\\
&=& i \int d{\bf k'} \int \frac{d{\bf k}}{(2 \pi)^d} e^{i ({\bf k}-{\bf k'}) \cdot {\bf r}}
\left(\frac{\Theta(k_{F,\alpha}-\Lambda-k) \Theta(k'-k_{F,\gamma}-\Lambda)}{
\epsilon_{\gamma}(k')-\epsilon_{\alpha}(k)+2i/\tau} \right.
\nonumber
\\
&+& \left. \frac{\Theta(k-k_{F,\alpha}-\Lambda) \Theta(k_{F,\gamma}-\Lambda-k')}{
\epsilon_{\alpha}(k)-\epsilon_{\gamma}(k')+2i/\tau} \right)
\label{incomplete}
\end{eqnarray}

Let us consider the special case where the dispersion are the same in both
spin branches and the Fermi surface is assumed to be spherical. In this
case the bare Green's function does not depend on the spin label and
$\Pi_{b,d}({\bf r},t) = \delta_{b,d} \Pi^0({\bf r},t)$. We
can write (\ref{incomplete}) as
\begin{eqnarray}
{\cal I}(r,\Lambda) &=& i \int d{\bf k'} \int \frac{d{\bf k}}{(2 \pi)^3} e^{i ({\bf k}-{\bf k'}) \cdot {\bf r}}
\left(\frac{\Theta(k_{F}-\Lambda-k) \Theta(k'-k_{F}-\Lambda)}{
\epsilon(k')-\epsilon(k)+2i/\tau}
\right.
\nonumber
\\
&+& \left.
\frac{\Theta(k-k_{F}-\Lambda) \Theta(k_{F}-\Lambda-k')}{
\epsilon(k)-\epsilon(k')+2i/\tau} \right) =
\nonumber
\\
&=& \int d{\bf k'} \int \frac{d{\bf k}}{(2 \pi)^3}
\frac{\cos[({\bf k}-{\bf k'}) \cdot {\bf r}]}{
\epsilon(k')-\epsilon(k)+2i/\tau}
\Theta(k_{F}-\Lambda-k) \Theta(k'-k_{F}-\Lambda)
\nonumber
\\
&=& \frac{2}{\pi r^2} \int_{0}^{k_F-\Lambda} dk  \int_{k_F + \Lambda}^{\infty} dk'
\frac{k k' \sin(k r) \sin(k' r)}{\epsilon(k')-\epsilon(k)+2i/\tau}
\label{rkkycomplete}
\end{eqnarray}

When $\Lambda \to 0$ and $E_F \gg  1/\tau$ we find the usual result \cite{mattis}
\begin{eqnarray}
{\cal I}(r,0)= \frac{4 m k_F^2}{\pi r^2} e^{-r/\ell} {\cal F}_0(2 k_F r)
\label{rkkyplain}
\end{eqnarray}
where
\begin{eqnarray}
\ell = \frac{k_F \tau}{m} = v_F \tau
\end{eqnarray}
is the electron mean-free path and
\begin{eqnarray}
{\cal F}_0(x) = \frac{\pi}{4} \left[\frac{\sin(x)-x \cos(x)}{x^2}\right]
\label{fx}
\end{eqnarray}
which leads to the final expression for the RKKY interaction in $d=3$.

Let us now consider the corrections to the RKKY interaction due to a finite
$\Lambda/k_F$. In all the cases here we will consider the situation in which
$k_F \geq  \Lambda \gg  1/\tau$.
From (\ref{rkkycomplete}) we see that the correction has the form
\begin{eqnarray}
\delta {\cal I}(r,\Lambda) = \frac{4 m k_F^2}{\pi r^2} e^{-r/\ell} 
{\cal F}(2 k_F r,\Lambda/k_F)
\label{rkkydres}
\end{eqnarray}
where, ${\cal F} =  {\cal F}_1+ {\cal F}_2 + {\cal F}_3$,
with 
\begin{eqnarray}
{\cal F}_1(2 x,y) &=& - \int_0^{1} dz \int_{1}^{1+y} dw
\frac{z w \sin(x z) \sin(x w)}{z^2-w^2+i \eta}
\nonumber
\\
{\cal F}_2(2 x,y) &=& - \int_{1-y}^{1} dz \int_{1}^{\infty} dw
\frac{z w \sin(x z) \sin(x w)}{z^2-w^2+i \eta}
\nonumber
\\
{\cal F}_3(2 x,y) &=& \int_{1-y}^{1} dz \int_{1}^{1+y} dw
\frac{z w \sin(x z) \sin(x w)}{z^2-w^2+i \eta}
\end{eqnarray}
where $\eta \to 0$. All the integrals can be evaluated after a very
tedious algebra. The final result reads
\begin{eqnarray}
{\cal F}(x,y) &=& \frac{1}{4 x^2} \left\{\pi \cos(x)-(\pi - 2xy) \sin(x)-
2 x \sin(xy) -2 x^2 y (Ci(x)-Ci(xy)) \right.
\nonumber
\\
&+& \cos[(1+y)x] \left[(Ci(x)-Ci(xy))-(1+y)x (Si(x)-Si(xy))\right]
\nonumber
\\
&+&  \sin[(1+y)x] \left[(Si(x)-Si(xy))+(1+y)x (Ci(x)-Ci(xy))\right]
\nonumber
\\
&-& \cos[(1-y)x] \left[(Ci(x)-Ci(xy))-(1-y)x (Si(x)+Si(xy))\right]
\nonumber
\\
&-& \left. \sin[(1-y)x] \left[(Si(x)+Si(xy))+(1-y)x (Ci(x)-Ci(xy))\right]
\right\}
\label{rkkylamb}
\end{eqnarray}
where
\begin{eqnarray}
Ci(x) &=& - \int_x^{\infty} dt \frac{\cos(t)}{t}
\nonumber
\\
Si(x) &=& - \int_x^{\infty} dt \frac{\sin(t)}{t}
\end{eqnarray}
are cosine and sine integrals. A plot of ${\cal F}(x,y)$ 
is shown in Fig.\ref{newrkky}. Thus, the actual RKKY interaction in
the presence of a finite $\Lambda/k_F$ is given by the sum of (\ref{rkkyplain})
plus (\ref{rkkydres}). Observe that besides the usual $2 k_F$ oscillations
a finite $\Lambda/k_F$ produces other oscillations at $2 \Lambda$.
Thus, the structure of the RKKY interaction is more complicated. 

At very short distances, $2 k_F r \ll 1$, we have
\begin{eqnarray}
{\cal I}(r,\Lambda) \approx \frac{2 m k_F^3}{3 r} \left(1-\frac{\Lambda}{k_F}
\right)^3
\end{eqnarray}
which, as in the case of the usual RKKY interaction, leads to ferromagnetic
coupling. The main difference, however is that the slope of the RKKY is
reduced by the factor $(1-\Lambda/k_F)^3$ and therefore the strength of the
ferromagnetic coupling is reduced as $\Lambda$ increases. 
Furthermore, while the first zero of the RKKY interaction for $\Lambda=0$ 
occurs for $k_F r \approx 2.2467$; when $\Lambda \neq 0$ the first zero is
reduced as shown in Fig.\ref{reduction} and saturates at $k_F r \approx 1$
when $\Lambda \approx k_F$. Overall the effect of a finite $\Lambda$ is to
increase the range of the antiferromagnetic coupling for small to moderate
$k_F r$. The component near the Fermi energy contributes a largely ferromagnetic component at these $k_F r$. Note that overall the contributions to RKKY
from the energy region away from the Fermi surface is at least comparable
to that for near the Fermi surface.
At intermediate distances, that is, $\Lambda^{-1} \gg r > k_F^{-1}$, 
the RKKY interaction can be written as
\begin{eqnarray}
{\cal I}(r,\Lambda) &\approx& {\cal I}(r,0) + \frac{2 m k_F \Lambda}{r^2}
\sin(k_F r) \left[-\pi \cos(k_F r) - 2(1-{\cal C}+ Ci(2 k_F r)-\ln(2 \Lambda r))
\sin(k_F r)  
\right.
\nonumber
\\
&+& \left. 2 \cos(k_F r) (Si(2 k_F r)+\pi/2) \right]
\end{eqnarray}
where ${\cal C} \approx 0.577216$ is the Euler constant.
We see from these expression that the correction to the $\Lambda=0$ case 
vanishes as $(\Lambda/k_F) \ln(\Lambda/k_F)$ as $\Lambda/k_F \to 0$.
The most striking difference between the behavior of the RKKY
interaction with $\Lambda/k_F \neq 0$ is it behavior at large distances.
It is very simple to show that for $r>>k_F^{-1},\Lambda^{-1}$ one has
\begin{eqnarray}
{\cal I}(r,\Lambda) \approx - \frac{m k_F^3}{2 \Lambda \pi r^4} \left[1-\left(\frac{
\Lambda}{k_F}\right)^2\right] \left(\cos(2 k_F r)+\cos(2 \Lambda r)\right)
\end{eqnarray}
which decays like $1/r^4$ instead of the usual $1/r^3$. Thus, for finite
$\Lambda$ the RKKY interaction has a shorter range.

\section{Mean field theory of itinerant ferromagnetism in the Kondo lattice}
\label{ferromf}

In this appendix we will discuss the physics of $H_{MF}$ in (\ref{hmf}).
Assuming that the magnetic order is homogeneous we can write
\begin{eqnarray}
H_{MF} = \sum_{{\bf k},\sigma} \epsilon_{\sigma}({\bf k}) c^{\dag}_{{\bf k},\sigma}
c_{{\bf k},\sigma}
+ J_z M_z \sum_{i=1}^{N_s} (n_{i,\uparrow}-n_{i,\downarrow})
+ J_z m_z \sum_{i=1}^{N_s} S^z_i
\label{hmf0}
\end{eqnarray}
where
\begin{eqnarray}
M_z &=& \frac{1}{N_s} \sum_{i=1}^{N_s}  \langle S^z_i \rangle
\nonumber
\\
m_z &=& \frac{1}{N_s} \sum_{i=1}^{N_s}  \left(\langle n_{i,\uparrow} \rangle -
\langle n_{i,\downarrow} \rangle \right)
\end{eqnarray}
where $N_s$ is the number of magnetic atoms.
As usual we can now perform a Fourier transform of the electron
operator
\begin{eqnarray}
c_{n,\sigma} = \frac{1}{\sqrt{N}} \sum_{{\bf k}} e^{i {\bf k} \cdot {\bf r}_n}
c_{{\bf k},\sigma}
\end{eqnarray}
where $N$ is the total number of atoms. We rewrite (\ref{hmf0})
as
\begin{eqnarray}
H_{MF} &=& \sum_{{\bf k}} \left[ \left(\epsilon_{\uparrow}({\bf k})
+ J_z c M_z \right) c^{\dag}_{{\bf k},\uparrow} c_{{\bf k},\uparrow}
+ \left(\epsilon_{\downarrow}({\bf k})
- J_z c M_z \right) c^{\dag}_{{\bf k},\downarrow} c_{{\bf k},\downarrow} \right]
\nonumber
\\
&+& J_z m_z \sum_i S^z_i
\end{eqnarray}
which brings the Hamiltonian to diagonal form ($c=N_s/N$). In the ground state we have
\begin{eqnarray}
\langle n_{{\bf k},\sigma} \rangle = \Theta(k_{F,\sigma}-k)
\end{eqnarray}
and therefore
\begin{eqnarray}
\langle n_{\sigma} \rangle &=&  \frac{1}{N_s} \sum_{i=1}^{N_s} \langle n_{i,\sigma} \rangle
= \frac{1}{N} \sum_{{\bf k}} \langle n_{{\bf k},\sigma} \rangle
\nonumber
\\
&=& \frac{1}{\rho_L} \frac{k_{F,\sigma}^3}{6 \pi^2}
\label{ns}
\end{eqnarray}
where $\rho_L = N/V$ is the lattice density. On the other hand, the total number of
electrons is
\begin{eqnarray}
N_e &=& \sum_{i=1}^N \sum_{\sigma} \langle n_{i,\sigma} \rangle =
\sum_{{\bf k}} \sum_{\sigma} \langle n_{{\bf k},\sigma} \rangle
\nonumber
\\
&=& V \sum_{\sigma} \frac{k_{F,\sigma}^3}{6 \pi^2}
\label{nev}
\end{eqnarray}
and thus the number of electrons per lattice site is
\begin{eqnarray}
\overline{n} = \frac{N_e}{N} = \sum_{\sigma} \langle n_{\sigma} \rangle
\end{eqnarray}
moreover
\begin{eqnarray}
m_z = n_{\uparrow} - n_{\downarrow}
\end{eqnarray}
from which we conclude
\begin{eqnarray}
n_{\uparrow}  &=& \frac{\overline{n} + m_z}{2}
\nonumber
\\
n_{\downarrow}  &=& \frac{\overline{n} - m_z}{2} \, .
\label{nund}
\end{eqnarray}

Since the up and down spins are in thermal equilibrium they must have
the same chemical potential, that is,
\begin{eqnarray}
\mu = \frac{k_{F,\uparrow}^2}{2 m^*} + J_z M_z c
= \frac{k_{F,\downarrow}^2}{2 m^*} - J_z M_z c
\end{eqnarray}
which can be rewritten with the help of (\ref{ns}) as
\begin{eqnarray}
(\langle n_{\uparrow} \rangle)^{2/3} - (\langle n_{\downarrow} \rangle)^{2/3}
= -\frac{4 m^* J_z M_z c}{(6 \pi^2 \rho_L)^{2/3}}  \, .
\end{eqnarray}
Using (\ref{nund}) we find
\begin{eqnarray}
(\overline{n} + m_z)^{2/3} - (\overline{n}-m_z)^{2/3}
= -\frac{4 m^* J_z M_z c}{(3 \pi^2 \rho_L)^{2/3}}
\end{eqnarray}
which has to be solved for $m_z$. If one is interested in the critical temperature
$T_c$ of the problem which is much smaller than $\mu$ it is sufficient to consider
the case of $m_z\ll \overline{n}$ in which case we find
\begin{eqnarray}
m_z \approx - \frac{3 m^* J_z M_z c (\overline{n})^{1/3}}{(3 \pi^2 \rho_L)^{2/3}}  \, .
\label{mz}
\end{eqnarray}
On the other hand, from (\ref{hmf0}) we have
\begin{eqnarray}
M_z = - S B_S\left(\frac{\beta S J_z c m_z}{2}\right) \, .
\label{Mz}
\end{eqnarray}
The solution of the problem is given by the set (\ref{mz}) and (\ref{Mz}).
In particular, substituting (\ref{mz}) into (\ref{Mz})
\begin{eqnarray}
M_z = S B_S\left(\frac{3 S m^* \beta J^2_z c^2 (\overline{n})^{1/3}}{
2 (3 \pi^2 \rho_L)^{2/3}} M_z\right) \, .
\end{eqnarray}
At $T \to T_c^-$ we have $M_z \to 0$ and therefore we find
\begin{eqnarray}
T_c &=& \frac{S (S+1) m^* J^2_z c^2 (\overline{n})^{1/3}}{
2 (3 \pi^2 \rho_L)^{2/3}}
\nonumber
\\
&=& \frac{S (S+1)}{4} \frac{c^2 \overline{n} J_z^2}{E_F}
= \frac{S (S+1)}{6 \rho_L} N(0) J_z^2
\end{eqnarray}
which agrees with (\ref{gammaab}).

\section{Frequency dependent susceptibility of a two level system}
\label{wsus}

Let us consider the response of a system described by the
Hamiltonian (\ref{hclus}) to an oscillating magnetic field
of frequency $\omega$. In linear response this is given
by the retarded spin-spin correlation function averaged over the
thermal ensemble
\begin{eqnarray}
\chi(t) &=&
-i \Theta(t) \langle [\delta \sigma_z(0),\delta \sigma_z(t)] \rangle
= -i \frac{\Theta(t)}{Z} tr \left(e^{-\beta H}
[\delta \sigma_z, e^{i H t} \delta \sigma_z e^{-i H t}]\right)
\nonumber
\\
&=&-i \frac{\Theta(t)}{Z}
\sum_{n,m} \left(e^{-\beta E_n} -e^{-\beta E_m}\right) e^{i (E_n-E_m) t}
|\langle n|\delta \sigma_z| m \rangle|^2
\label{pst}
\end{eqnarray}
where $H|n\rangle = E_n|n\rangle$ and $Z=\sum_n e^{-\beta E_n}$
and $\delta \sigma_z = \sigma_z - \langle \sigma_z \rangle$.
A simple Fourier transform gives
\begin{eqnarray}
\chi(\omega) &=& -\frac{1}{Z} \sum_{n,m} \left(e^{-\beta E_n} -e^{-\beta E_m}\right)
\frac{|\langle n|\delta \sigma_z| m \rangle|^2}{\omega + (E_n-E_m) - i \epsilon}
\nonumber
\\
 &=& -\frac{1}{Z} \sum_{n,m} \left(e^{-\beta E_n} -e^{-\beta E_m}\right)
\frac{|\langle n|\delta \sigma_z| m \rangle|^2}{E_n-E_m}\left(1-\frac{\omega}{\omega + (E_n-E_m) - i \epsilon}\right)
\nonumber
\\
&=&
\chi(0)+ \frac{\omega}{Z}  \sum_{n,m} \left(e^{-\beta E_n} -e^{-\beta E_m}\right)
\frac{|\langle n|\delta \sigma_z| m \rangle|^2}{(E_n-E_m)(\omega + (E_n-E_m) - i \epsilon)}
\label{psw}
\end{eqnarray}
where $\epsilon \to 0$ and
\begin{eqnarray}
\chi(0) =  -\frac{1}{Z} \sum_{n,m} \left(e^{-\beta E_n} -e^{-\beta E_m}\right)
\frac{|\langle n|\delta \sigma_z| m \rangle|^2}{E_n-E_m}
\label{ps0}
\end{eqnarray}
is the static susceptibility.

The calculation is very simple because we are dealing with a two-level system problem.
Indeed the Hamiltonian (\ref{hclus}) can be diagonalized by a unitary transformation
\begin{eqnarray}
U &=& e^{i \theta \sigma_y}
\nonumber
\\
\tan(2 \theta) &=& \frac{\Delta}{E_H}
\label{utl}
\end{eqnarray}
which transforms
\begin{eqnarray}
U H_C U^{-1}&=& \sqrt{E_H^2 + \Delta^2} \, \, \tau_z
\nonumber
\\
U \sigma_z U^{-1} &=& \frac{1}{\sqrt{E_H^2 + \Delta^2}} \left[E_H \tau_z - \Delta \tau_x\right]
\nonumber
\\
U \sigma_x U^{-1} &=& \frac{1}{\sqrt{E_H^2 + \Delta^2}} \left[\Delta \tau_z + E_H \tau_x\right] \, .
\label{hctrans}
\end{eqnarray}
Moreover, in the new basis of the Pauli matrices $\tau$ with eigenstates $|\pm\rangle$ and eigenenergies
$\pm\sqrt{E_H^2 + \Delta^2}$, respectively,  we can easily show that
\begin{eqnarray}
\langle \sigma_z \rangle &=& \frac{E_H}{\sqrt{E_H^2 + \Delta^2}} \tanh\left(\beta \sqrt{E_H^2 + \Delta^2}\right)
\nonumber
\\
\langle \pm |\sigma_z | \pm \rangle &=& \pm \frac{E_H}{\sqrt{E_H^2 + \Delta^2}}
\nonumber
\\
\langle \pm |\sigma_z | \mp \rangle &=& - \frac{\Delta}{\sqrt{E_H^2 + \Delta^2}} \, .
\label{me}
\end{eqnarray}
Substituting (\ref{me}) into (\ref{ps0}) we easily
find
\begin{eqnarray}
\chi(0) = \frac{\beta E_H}{E_H^2 + \Delta^2}
sech^2 \left(\beta \sqrt{E_H^2 + \Delta^2}\right)
+ \frac{\Delta^2}{(E_H^2 + \Delta^2)^{3/2}}
\tanh \left(\beta \sqrt{E_H^2 + \Delta^2}\right)
\end{eqnarray}
which is the result (\ref{chi0}). The dynamic part can be obtained in analogous way and it gives
\begin{eqnarray}
\chi(\omega) &=& \chi(0) - \frac{\beta E_H^2}{E_H^2 + \Delta^2} sech^2\left(\beta \sqrt{E_H^2 + \Delta^2}\right)
\frac{\omega}{\omega-i\epsilon}
\nonumber
\\
&-&
\frac{\Delta^2}{(E_H^2 + \Delta^2)^{3/2}}\tanh\left(\beta \sqrt{E_H^2 + \Delta^2}\right)
\left[\frac{\omega}{\omega-2 \sqrt{E_H^2 + \Delta^2} - i\epsilon}
+ \frac{\omega}{\omega+2 \sqrt{E_H^2 + \Delta^2} - i\epsilon}\right]
\label{chiw}
\end{eqnarray}
from which we can extract the real and imaginary parts:
\begin{eqnarray}
\Re\left[\chi(\omega)\right] &=& \chi(0) - \frac{\beta E_H^2}{E_H^2 + \Delta^2}
sech^2\left(\beta \sqrt{E_H^2 + \Delta^2}\right) \frac{\omega^2}{\omega^2+\epsilon^2}
\nonumber
\\
&-&
\frac{\Delta^2}{2 (E_H^2 + \Delta^2)^{3/2}}
\tanh\left(\beta \sqrt{E_H^2 + \Delta^2}\right)
\left[\frac{\omega (\omega-2 \sqrt{E_H^2 + \Delta^2})}{(\omega-2 \sqrt{E_H^2 + \Delta^2})^2 + \epsilon^2}
\right.
\nonumber
\\
&+& \left. \frac{\omega(\omega+2 \sqrt{E_H^2 + \Delta^2})}{(\omega+2 \sqrt{E_H^2 + \Delta^2})^2 + \epsilon^2}\right]
\nonumber
\\
\Im\left[\chi(\omega)\right] &=& 2 \pi \frac{\Delta^2}{(E_H^2 + \Delta^2)}
\tanh\left(\beta \sqrt{E_H^2 + \Delta^2}\right) \left[\delta(\omega-2 \sqrt{E_H^2 + \Delta^2})
- \delta(\omega+2 \sqrt{E_H^2 + \Delta^2})\right]
\label{richiw}
\end{eqnarray}
which are used in the calculations.

\section{Two impurity calculations}
\label{twoimp}

Starting from (\ref{dimerhamil}) and using the expansion
(\ref{psiexp})the
Hamiltonian is written as
\begin{eqnarray}
H &=&
\sum_{j,\alpha} \int \frac{dk}{2 \pi} \epsilon_k \psi^{\dag}_{j,\alpha}(k)
\psi_{j,\alpha}(k) +
\sum_{a} \Gamma^R_{a} S_{1,a} S_{2,a}
\nonumber
\\
&+& \frac{v_F}{2} \sum_{a,j,l,\alpha,\beta}
\int \frac{dk}{2 \pi} \int \frac{dk'}{2 \pi}
\left[J_{+,a}(k,k')  \psi^{\dag}_{j,\alpha}(k)
\delta_{j,l} \sigma^a_{\alpha,\beta} \psi_{l,\beta}(k')
( S_{1,a} + S_{2,a} )
\right.
\nonumber
\\
&-& \left.
J_{m,a}(k,k') \psi^{\dag}_{j,\alpha}(k)
\tau^z_{j,l} \sigma^a_{\alpha,\beta} \psi_{l,\beta}(k') ( S_{1,a} - S_{2,a} )
+ J_{-,a}(k,k') \psi^{\dag}_{j,\alpha}(k)
\tau^x_{j,l} \sigma^a_{\alpha,\beta} \psi_{l,\beta}(k') ( S_{1,a} + S_{2,a} )
\right.
\nonumber
\\
&-& \left.
i J_{t,a}(k,k') \psi^{\dag}_{j,\alpha}(k)
\tau^y_{j,l} \sigma^a_{\alpha,\beta} \psi_{l,\beta}(k') ( S_{1,a} - S_{2,a} )
\right]
\label{hbl}
\end{eqnarray}
where $\tau^a$ are Pauli matrices which act on the subspace of $j=1,2$,
and
\begin{eqnarray}
J_{\pm,a} &=& \frac{k k' J^R_a}{4 \pi v_F} \left(N_e(k) N_e(k') \pm N_o(k) N_o(k')\right)
\nonumber
\\
J_{m(t),a} &=& \frac{k k' J^R_a}{4 \pi v_F} \left(N_e(k) N_o(k') \pm N_o(k) N_e(k')\right)
\label{js}
\end{eqnarray}
are the exchange constants. Observe that all the momenta here are defined
in a thin shell around the Fermi surface. Thus, we can linearize the
band by writing $\epsilon_k = v_F(k_F-k)$ and expand the above exchange
constants to leading order in $k$ in which case we get
\begin{eqnarray}
J_{+,a}(k,k') &\approx& \pi N(0) J^R_a
\nonumber
\\
J_{-,a}(k,k') &\approx& \pi N(0) J^R_a \frac{\sin(k_FR)}{k_FR}
\nonumber
\\
J_{m,a}(k,k') &\approx& \pi N(0) J^R_a \sqrt{1-\left(\frac{\sin(k_FR)}{k_FR}\right)^2}
\nonumber
\\
J_{t,a}(k,k') &\approx& 0 \, .
\label{jexch}
\end{eqnarray}
in which case there is no momentum dependence in the coupling
constants of (\ref{hbl}) which transforms it to an impurity problem
if we define the Fourier transform:
\begin{eqnarray}
\psi_{j,\alpha}(x) = \int_{-\infty}^{+\infty} \frac{dk}{2 \pi} e^{i k x} \psi_{j,\alpha}(k)
\end{eqnarray}
in which case (\ref{hbl}) becomes turns into (\ref{1d2i}).

An important term we have not included is due to ordinary
scattering of electrons at the impurity. This term is just
the electrostatic coupling which is given by
\begin{eqnarray}
H_V &=&  \sum_{\alpha} \left(V({\bf R}/2) \psi^{\dag}_{\alpha}({\bf R}/2) \psi_{\alpha}({\bf R}/2)
+V(-{\bf R}/2) \psi^{\dag}_{\alpha}(-{\bf R}/2) \psi_{\alpha}(-{\bf R}/2) \right)
\nonumber
\\
&=& \sum_{\alpha} \sum_{{\bf k},{\bf k'}} \left(V({\bf R}/2) e^{-i({\bf k}-{\bf k'}) \cdot {\bf R}/2}
 + V(- {\bf R}/2) e^{i({\bf k}-{\bf k'}) \cdot {\bf R}/2}\right)
c^{\dag} _{{\bf k},\alpha} c_{{\bf k'},\alpha}
\nonumber
\\
&=& \sum_{\alpha} \int \frac{dk}{2 \pi} \int \frac{dk'}{2 \pi}
\left[V_e(k,k') \psi^{\dag}_{1,\alpha}(k) \psi_{1,\alpha}(k')
+ V_o(k,k') \psi^{\dag}_{2,\alpha}(k) \psi_{2,\alpha}(k')
\right.
\nonumber
\\
&+& \left. V_{eo}(k,k') \left(\psi^{\dag}_{1,\alpha}(k) \psi_{2,\alpha}(k') + \psi^{\dag}_{2,\alpha}(k) \psi_{1,\alpha}(k')
\right)\right]
\end{eqnarray}
where we used (\ref{psiexp}). As previously, assuming that the
electron momenta is close to the Fermi surface we can rewrite the
above term as
\begin{eqnarray}
H_V &=& \sum_{\alpha} \left[V_1  \psi^{\dag}_{1,\alpha}(x=0) \psi_{1,\alpha}(x=0)
+ V_2  \psi^{\dag}_{1,\alpha}(x=0) \psi_{1,\alpha}(x=0)
\right.
\nonumber
\\
&+& \left. V \left(\psi^{\dag}_{1,\alpha}(x=0) \psi_{2,\alpha}(x=0) + \psi^{\dag}_{2,\alpha}(x=0) \psi_{1,\alpha}(x=0)
\right)\right] \, .
\label{vs}
\end{eqnarray}

The bosonization procedure goes like the one described for the single
impurity problem in Subsection \ref{1mmke}. The only difference is that
we are going to construct the factors $K_{\sigma}$ explicitly in
terms of the fermion fields. We define
\begin{eqnarray}
K_{1,\sigma} &=& \exp\left\{i \pi \int dx \psi^{\dag}_{1,\uparrow} \psi_{1,\uparrow}\right\}
\nonumber
\\
K_{2,\sigma} &=& \exp\left\{i \pi \int dx \left[\sum_{\sigma}
\psi^{\dag}_{1,\sigma} \psi_{1,\sigma}\right] + \psi^{\dag}_{2,\uparrow} \psi_{2,\uparrow}\right\}
\end{eqnarray}
which can be rewritten in terms of the bosonic fields if we use that
\begin{eqnarray}
\psi^{\dag}_{i,\sigma} \psi_{i,\sigma} = \frac{1}{2 \pi} \frac{\partial \Phi_{i,\sigma}}{\partial x} \, .
\end{eqnarray}
It is easy to show that (\ref{1d2i}) reads (for simplicity we work
again with the problem of uniaxial symmetry)
\begin{eqnarray}
H &=& \frac{v_F}{2} \sum_{i=c,s,f,sf} \int dx \left[\Pi_i^2(x) + \left(\frac{\partial \phi_i}{\partial x}\right)^2
\right]  +  \sum_{a} \Gamma_{a} S_{1,a} S_{2,a} + \frac{2 V}{\pi a}
\cos(\Phi_{sf}(0)) \cos(\Phi_f(0)-\theta)
\nonumber
\\
&+& \frac{v_F J_{+,z}}{2 \pi} \frac{\partial \Phi_s(0)}{\partial x}(S_{1,z} + S_{2,z})
\nonumber
\\
&+& \frac{v_F J_{+,\perp}}{\pi a} \cos(\Phi_{sf}(0))
\left(\cos(\Phi_s(0)) (S_{1,x}+S_{2,x}) - \sin(\Phi_s(0)) (S_{1,y}+S_{2,y})\right)
\nonumber
\\
&+& \frac{v_F J_{m,z}}{2 \pi} \frac{\partial \Phi_{sf}(0)}{\partial x}(S_{1,z} - S_{2,z})
\nonumber
\\
&-& \frac{v_F J_{m,\perp}}{\pi a} \sin(\Phi_{sf}(0))
\left(\sin(\Phi_s(0)) (S_{1,x}-S_{2,x}) + \cos(\Phi_s(0)) (S_{1,y}-S_{2,y})\right)
\nonumber
\\
&+& \frac{v_F J_{-,z}}{\pi a} \sin(\Phi_{sf}(0)) \sin(\Phi_f(0)-\theta)
(S_{1,z} + S_{2,z})
\nonumber
\\
&+& \frac{v_F J_{-,\perp}}{\pi a} \cos(\Phi_f(0)-\theta)
\left(\cos(\Phi_s(0)) (S_{1,x}+S_{2,x}) - \sin(\Phi_s(0)) (S_{1,y}+S_{2,y})\right)
\end{eqnarray}
which can be rewritten in a more economical format:
\begin{eqnarray}
H &=& \frac{v_F}{2} \sum_{i=c,s,f,sf} \int dx \left[\Pi_i^2(x) + \left(\frac{\partial \phi_i}{\partial x}\right)^2
\right]  +  \sum_{a} \Gamma_{a} S_{1,a} S_{2,a}
\nonumber
\\
&+& \frac{2 V}{\pi a}
\cos(\Phi_{sf}(0)) \cos(\Phi_f(0)-\theta)
\nonumber
\\
&+&\frac{v_F J_{+,z}}{2 \pi} \frac{\partial \Phi_s(0)}{\partial x}(S_{1,z} + S_{2,z})
\nonumber
\\
&+& \frac{v_F J_{+,\perp}}{2 \pi a} \cos(\Phi_{sf}(0))
\left(e^{i \Phi_s(0)} (S_{1,+}+S_{2,+}) + e^{-i \Phi_s(0)} (S_{1,-}+S_{2,-})\right)
\nonumber
\\
&+& \frac{v_F J_{m,z}}{2 \pi} \frac{\partial \Phi_{sf}(0)}{\partial x}(S_{1,z} - S_{2,z})
\nonumber
\\
&+& i \frac{v_F J_{m,\perp}}{2 \pi a} \sin(\Phi_{sf}(0))
\left(e^{i \Phi_s(0)} (S_{1,+}+S_{2,+}) - e^{-i \Phi_s(0)} (S_{1,-}+S_{2,-})\right)
\nonumber
\\
&+&\frac{v_F J_{-,z}}{\pi a} \sin(\Phi_f(0)-\theta) \sin(\Phi_{sf}(0))
(S_{1,z} + S_{2,z})
\nonumber
\\
&+& \frac{v_F J_{-,\perp}}{2 \pi a} \cos(\Phi_f(0)-\theta)
\left(e^{i \Phi_s(0)} (S_{1,+}+S_{2,+}) +  e^{-i \Phi_s(0)} (S_{1,-}+S_{2,-})\right)
\label{hboso1}
\end{eqnarray}
where $S_{i,\pm} = S_{i,x} \pm i S_{i,y}$ and $\theta$ is defined in
(\ref{bigp}). Notice that the charge degrees of freedom have
decoupled entirely and can be disregarded and 
that the diagonal terms in (\ref{vs}) can be written as
gradients of the boson operators and therefore can be trivially
absorbed into a shift of the bosons and do not appear here.
After the unitary transformation:
$
U = e^{-i (S_{1,z}+S_{2,z}) \Phi_s(0)}
$
 the Hamiltonian becomes (\ref{hbosofinal}).

\newpage

\begin{figure}
\epsfysize10cm
\hspace{1cm}
\epsfbox{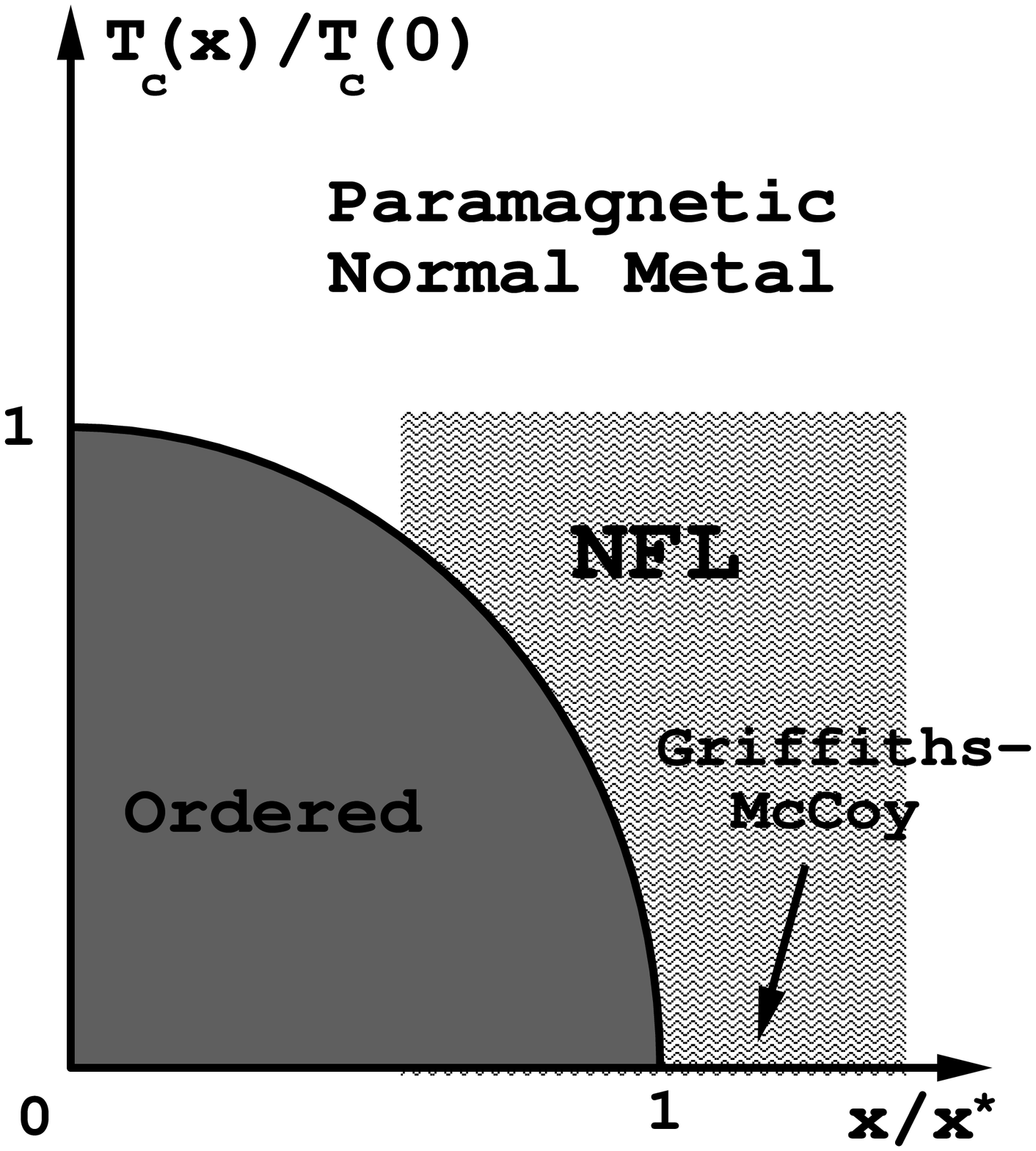}
\caption{Typical phase diagram of the systems studied with an ordered magnetic
phase and an anomalous metallic phase close to the QCP. The Griffiths-McCoy
singularities appear close to the QCP.}
\label{phdi}
\end{figure}

\begin{figure}
\epsfysize10cm
\hspace{1cm}
\epsfbox{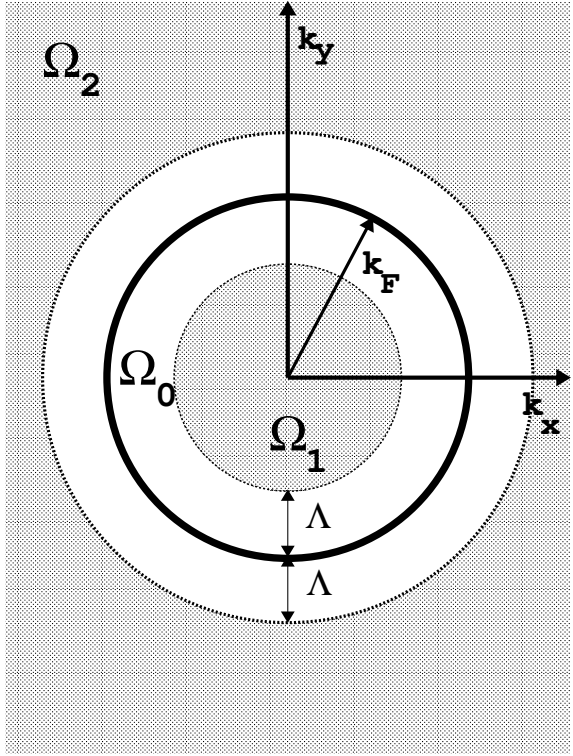 }
\caption{Fermi Surface and its regions of energy.}
\label{fs}
\end{figure}

\begin{figure}
\epsfysize15cm
\hspace{1cm}
\epsfbox{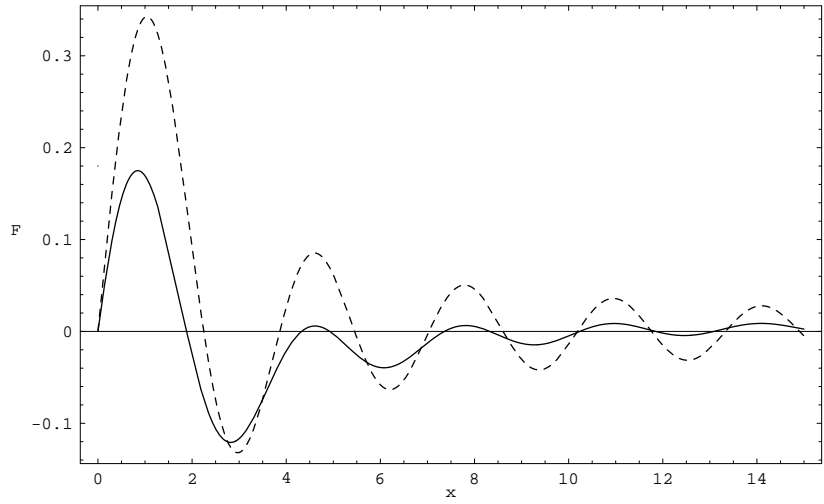}
\caption{Plot of ${\cal F}(2 x,y)$ which appears in the expression for the RKKY
interaction (\ref{gammaab}): dashed line is the usual RKKY ($\Lambda=0$);
continuous line $\Lambda/k_F =0.1$.}
\label{newrkky}
\end{figure}

\begin{figure}
\epsfysize15cm
\hspace{1cm}
\epsfbox{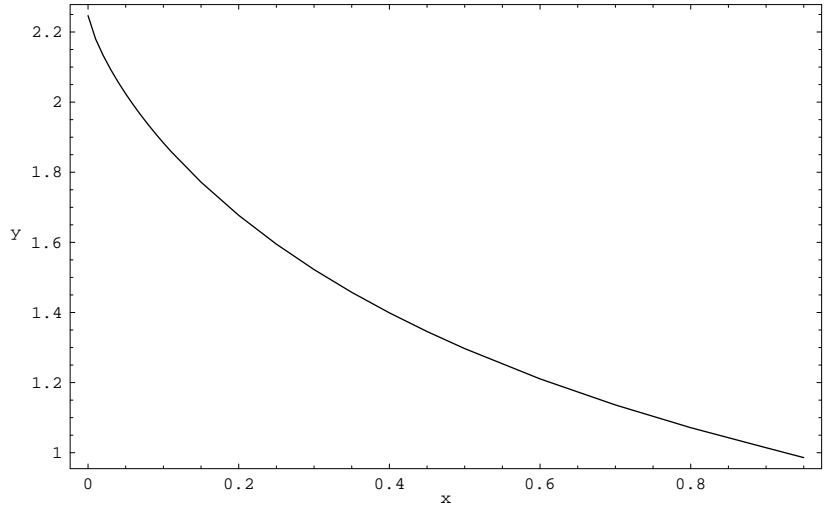}
\caption{Plot of the first zero of the RKKY interaction, $y=k_F r_0$, 
as a function of $x=\Lambda/k_F$.}
\label{reduction}
\end{figure}

\begin{figure}
\epsfysize10cm
\hspace{1cm}
\epsfbox{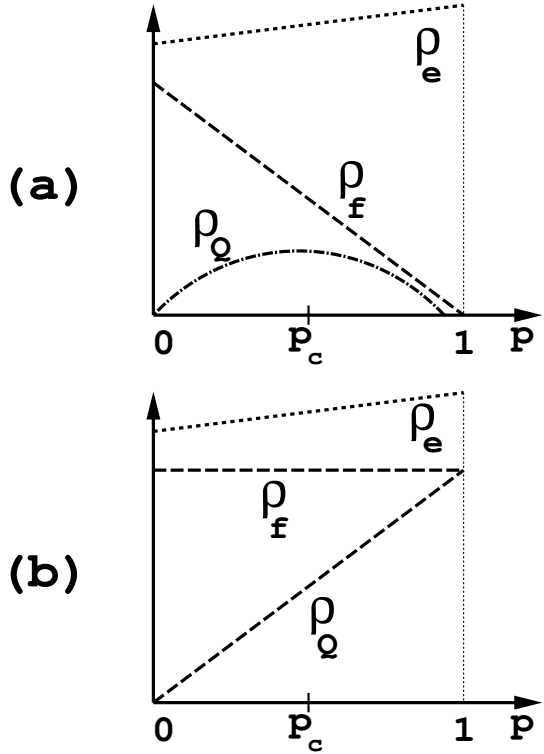}
\caption{Percolation problem in a (a) Kondo hole system and (b) Ligand system. $\rho_e$ is the
electronic density; $\rho_f$ is the density of magnetic moments; $\rho_Q$ is the density
of compensated moments; $p$ is the percolation parameter and $p_c$ is the percolation threshold.}
\label{perco}
\end{figure}

\begin{figure}
\epsfxsize=6cm
\epsfysize=6cm
\hspace{1cm}
\epsfbox{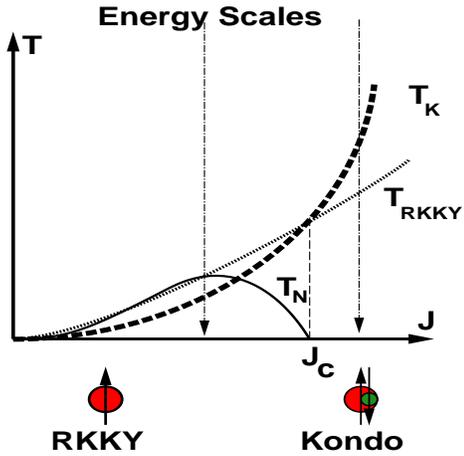}
\caption{Doniach phase diagram: long dashed line is the Kondo temperature, $T_K$;
short dashed line is the RKKY temperature,$T_{RKKY}$; the continuous line is the
ordering temperature $T_N$.}
\label{donifig}
\end{figure}

\begin{figure}
\epsfysize10cm
\hspace{1cm}
\epsfbox{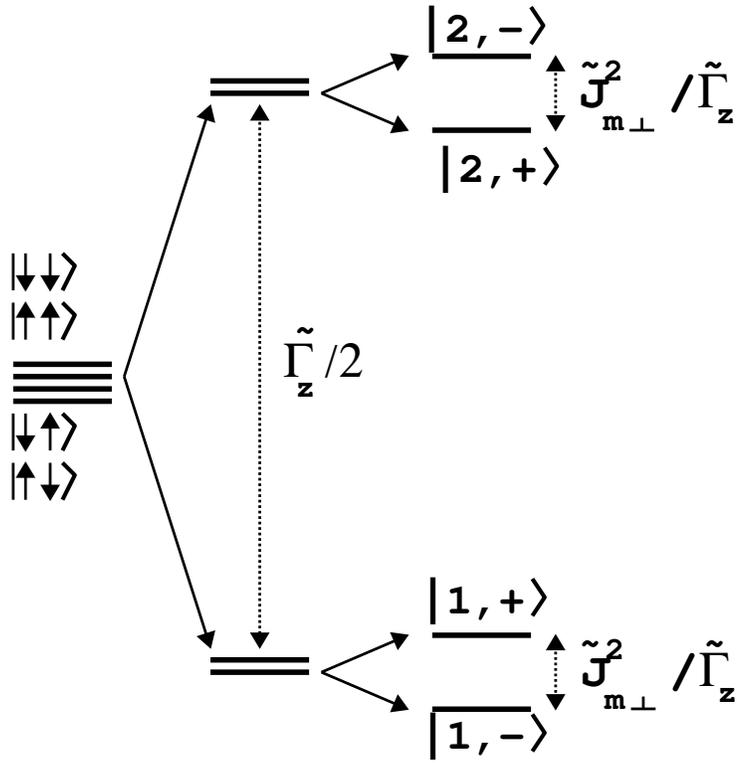}
\caption{Energy level splitting in the two-impurity Kondo effect.}
\label{splits}
\end{figure}

\begin{figure}
\epsfysize10cm
\hspace{1cm}
\epsfbox{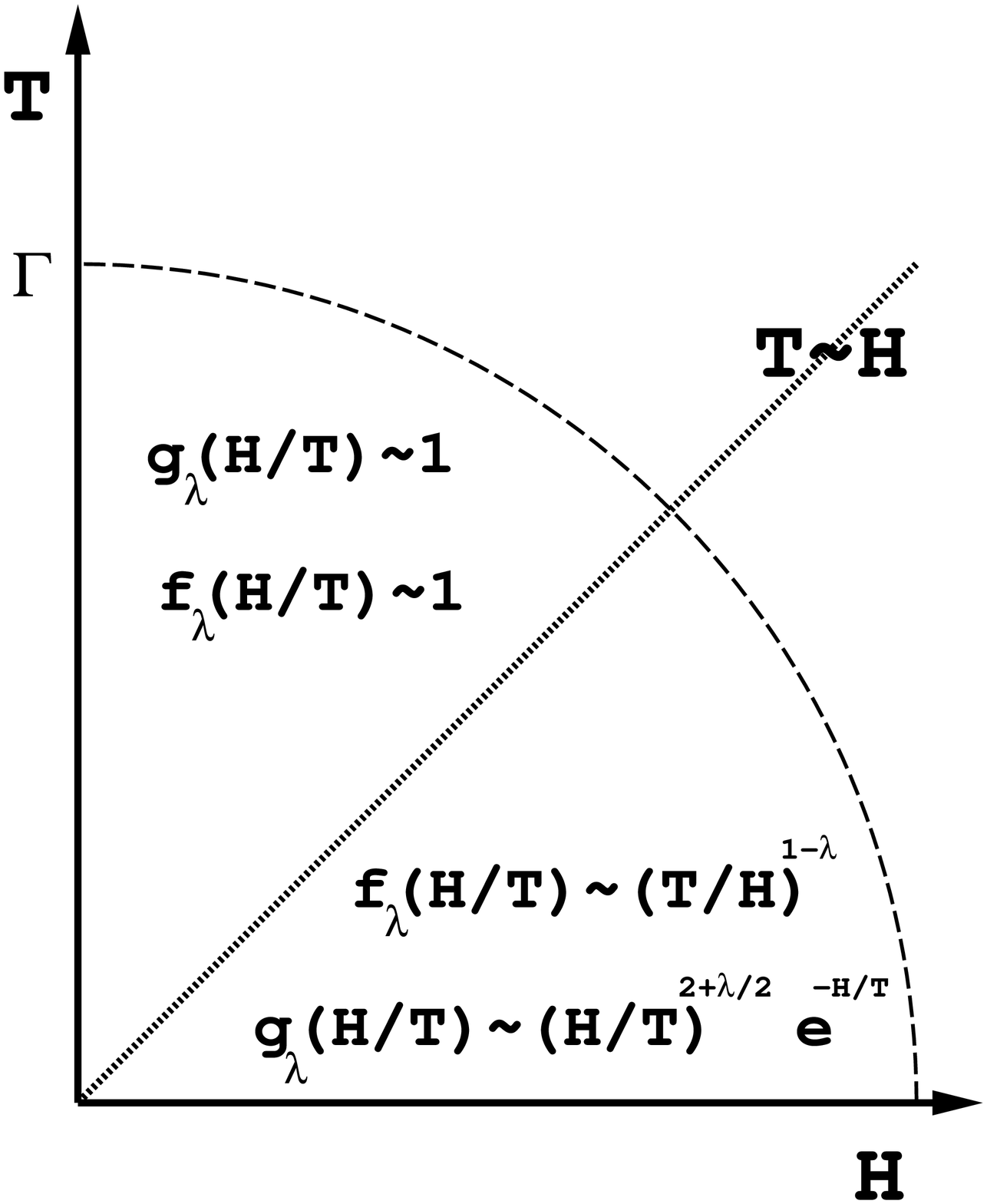}
\caption{Behavior of the scaling functions of the magnetization, $f_{\lambda}(H/T)$,
given in (\ref{mscal}) and the specific heat, $g_{\lambda}(H/T)$, given by (\ref{avecvsca}),
in the $H \times T$ phase diagram. The line $T \sim H$ marks the crossover
in behavior and $\Gamma$ is the value of the RKKY interaction above which
the cluster does not exist.}
\label{sumclus}
\end{figure}

\end{document}